%% file: long.tex
\documentstyle[preprint,tighten,aps]{revtex}
\begin{document}
\draft
\title{``Confined Coherence'' in Strongly Correlated, Anisotropic Metals}
\author{David G. Clarke}

\address{IRC in Superconductivity and Cavendish Laboratory, 
University of Cambridge,\\
Cambridge, CB3 0HE, United Kingdom \\}
\author{S. P. Strong}
\address{
NEC Research Institute, 4 Independence Way,
Princeton, NJ, 08540, U.S.A.\\}
%\date{\today}
\date{July 8, 1996}
\maketitle
\begin{abstract}
We present a detailed discussion of both theoretical and
experimental evidence in favour of the existence of states
of ``confined coherence'' in metals of sufficiently high
anisotropy and with sufficiently strong correlations.
The defining property of such a state is
that single electron coherence is confined to lower dimensional
subspaces (planes or chains) so that it is impossible
to observe interference effects between histories which
involve electrons moving between these
subspaces.
The most dramatic experimental manifestation of such a state is
the coexistence of
incoherent, non-metallic transport in one or two directions
(transverse to the lower dimensional subspaces)
with coherent transport in at least one other
direction (within the subspaces).
The magnitude
of the Fermi surface warping due to transverse 
(inter-subspace) momentum
plays the role of
an order parameter (in a state of confined coherence,
this order parameter vanishes) and the effect
can occur in a pure system at zero temperture.
Our theoretical approach is to treat 
an anisotropic 2D (3D) electronic system as a collection
of 1D (2D) electron liquids coupled by weak interliquid
single particle hopping. We find that
a necessary condition for the
destruction of coherent interliquid transport
is that the intraliquid state be a non-Fermi liquid.
We present a very detailed discussion of coupled 1D
Luttinger liquids and the reasons for
believing in the existence of a
phase of confined coherence in that model. This 
provides a paradigm for incoherent transport between
weakly coupled 2D non-Fermi liquids, the case relevant
to the experiments of which we are aware. 
Specifically, anomalous transport data in the (normal state of the) cuprate
superconductors and in the low temperature, metallic state
of the highly anisotropic organic
conductor (TMTSF)$_2$PF$_6$
cannot be understood within a
Fermi liquid framework, and, we argue, the only plausible way
to understand that transport is in terms of a state
of confined coherence.

\end{abstract}

%\narrowtext

\input intro.sec

\input tls.sec

\input fermions.sec

\input smalla.sec

\input wen.sec

\input luttliq.sec

\input interp.sec

\input otheroperators.sec

\input caxis.sec

\input organics.sec

\input concl.sec

We would like to thank P.~W.~Anderson for his comments and
continued encouragement, especially in the early stages of this work.
We have benefited greatly from interactions with P.~Chaikin, G.~Danner,
W.~Kang
and K.~Chashechkina regarding their anomalous experimental results for 
(TMTSF)$_2$PF$_6$.
A.~MacKenzie, N.~Hussey, S.~Tajima, T.~Timusk, S.~Uchida and J.~Wheatley
provided helpful
comments concerning the cuprates. We thank A.~MacKenzie for bringing the physics
of Sr$_2$RuO$_4$ to our attention, and V.~Yakovenko for sending us 
unpublished results of ``hot spot'' calculations. We would also like to
thank W.~Bialek, P.~Coleman, G.~Lonzarich and K.~Sch\"{o}nhammer for stimulating
discussions and interest in our work.

D.~G.~C. gratefully acknowledges the support of a Research Fellowship from 
St.~Catharine's College, Cambridge.

\newpage

\input appendix.sec

%We are grateful to P.~W.~Anderson for numerous helpful discussions on this
%and related issues.

%\samepage
\newpage
% now the references. delete or change fake bibitem. delete next three
%   lines and directly read in your .bbl file if you use bibtex.

%%%%%%%%%%%%%%%%% FIGURES %%%%%%%%%%%%%%%%%%%%%%%%%%

%%%%%%%%%%%%%%%%% wen.sec %%%%%%%%%%%%%%%%%%%%%%%%%%

\begin{figure}
%wenFS.eps
\caption{ Behavior of the poles of the Green's function 
for $k=k_F$ in
the approximation discussed in the text.
There is no physically sensible solution for
$\alpha > 1/4$ since the poles move into
physically inaccessible regions as
$\alpha \rightarrow 1/4$.
}
\label{fig:wenFs}
\end{figure}

\begin{figure}
%wennonFS.eps
\caption{ Behavior of the poles of the Green's function 
for $k \neq k_F$ in
the approximation discussed in the text.
There is a physically allowed solution for
$t_{\perp} > 0$ and
$\alpha > 1/4$, however, for
$\alpha > 1/3$ the alloed pole is shifted to energies
with a lower, not higher real part.
In addition,
the pole for $t_{\perp} < 0$
does not exist for any $\alpha > 1/4$
or for $k$ (measured from $k_F$) too large, as discussed in the text.
Instead a new pole appears on the real axis
with unphysical properties, as discussed in the text.
}
\label{fig:wennonFS}
\end{figure}

\begin{figure}
%wennonFSsc.eps
\caption{ Behavior of the poles of the Green's function 
for $k \neq k_F$ and spin charge separation included in
the approximation discussed in the text.
There is a physically allowed pole for
$t_{\perp} >0$ and $\alpha > 1/4$
only for sufficiently small
$k$ (measured from $k_F$) and the pole lies to the left of
$v_{\rho} k$ for $\alpha > 1/6$.
The pole for $t_{\perp} < 0$ 
which is continuously connected to the pole
for $\alpha = 0$ $v_{\sigma} = v_{\rho}$
is not
shown, but behaves essentially as in
the spinless case,
while the other pole for
$t_{\perp} < 0$ disperses along the chains like
$v_{\sigma} k$.
}
\label{fig:wennonFSsc}
\end{figure}

%%%%%%%%%%%%%%%%%% luttliq.sec %%%%%%%%%%%%%%%%%%%%%%%%

\begin{figure}
%spectral.eps
\caption{Electron spectral function (right moving part, $\rho_+(q,\omega)$
where $q\equiv k-k_F$) 
in a spin-charge separated Luttinger liquid (from J.~Voit, Ref. \protect\cite{spectral}).
The exponent $\gamma_{\rho}$ in the figure is the same as $\alpha$ in our notation.
}
\label{fig:spectral1}
\end{figure}

\begin{figure}
\caption{The interliquid hopping spectral function for {\em spinless\/} Luttinger
liquids for various values of 
$\alpha$. Here $\omega_l=(v_c-v)\Delta k$
and $\omega_u$ is the ultraviolet cutoff of order $v/a$. The plots do not
include the weak power law cutoff dependent prefactor. Note that
for $\alpha=0$,
$A_{12}(\omega)\propto \delta(\omega)$.}
\label{fig: inter_spectral_spinless}
\end{figure}

\begin{figure}
\caption{The interliquid hopping spectral function for {\em spinny\/} Luttinger
liquids, for various values of 
$\alpha$. Here $\omega_l=(v_s-v)\Delta k$, $\omega_i=(v_c-v)\Delta k$
and $\omega_u$ is the ultraviolet cutoff of order $v/a$. The plots do not
include the weak power law cutoff dependent prefactors. The vertical
arrow is the $\alpha=0$ spectral function, 
$A_{12}(\omega)\propto \delta(\omega)$.}
\label{fig: inter_spectral_spinny}
\end{figure}

%%%%%%%%%%%%%%%%% caxis.sec %%%%%%%%%%%%%%%%%%%%%%%%%%

\begin{figure}
\caption{Temperature dependence of the in-plane (upper panel) and
inter-plane (lower panel) resistivity for single crystals of
La$_{2-x}$Sr$_x$CuO$_4$ with various compositions in the metallic phase
(from Ref. \protect\cite{uchida_dc}).}
\label{fig: uchida_dc}
\end{figure}

\begin{figure}
\caption{In-plane ($\rho_{ab}$, closed circles) and interlayer
($\rho_{c}$, open squares) electrical resistivity of Sr$_2$RuO$_4$
plotted against the square of the temperature. The solid lines represent the
fits below 25 K: $\rho=\rho_0+AT^2$ 
(from Y.~Maeno {\em et al.\/}, preprint (1995)).}
\label{fig: ruthenate}
\end{figure}

\begin{figure}
\caption{C-axis optical conductivity spectra below 2.0 $eV$ for
La$_{2-x}$Sr$_x$CuO$_4$. Expanded spectra in the low-energy region are shown
in the inset
(from Ref. \protect\cite{uchida_ac}).}
\label{fig: uchida_ac}
\end{figure}

\begin{figure}
\caption{(a)c-axis reflectivity of
YBa$_2$Cu$_3$O$_{6+x}$ for several $x$ at room temperature.
The inset shows the a-axis reflectivity.
(b) Real part of the c-axis optical conductivity of YBa$_2$Cu$_3$O$_{6+x}$.
The inset shows the a-axis conductivity.
(from Ref. \protect\cite{cooper}).}
\label{fig: ybco_ac(cooper)}
\end{figure}

\begin{figure}
\caption{The temperature dependence of the c-axis conductivity spectra for
YBa$_2$Cu$_3$O$_{6+x}$ for various $x$.
(from Ref. \protect\cite{tajima_ac}).}
\label{fig: ybco_ac}
\end{figure}

\begin{figure}
\caption{Real and imaginary parts of the electronic dynamical conductivity
of fully oxygenated YBa$_2$Cu$_3$O$_{7-\delta}$ crystals at different
temperatures. The electronic conductivity is
estimated by subtracting the phononic contribution, which was fitted by 
five Lorentz oscillators.
(from Ref. \protect\cite{schutzmann}).}
\label{fig: schutz}
\end{figure}

%%%%%%%%%%%%%%%%% organics.sec %%%%%%%%%%%%%%%%%%%%%%%

\begin{figure}
%rxxraw.eps
\caption{ Resistance along the most conducting direction
(in milliohms) as a function of magnetic field strength
and orientation.  Field was rotated in the $bc$ plane
and angle $\Theta$ is defined so that
$\pm$90 degrees  coincide with the $b$ direction.
%Data are
%taken from Ref. \protect\cite{woowon}
}
\label{fig:woowon}
\end{figure}

\begin{figure}
%rzzraw.eps
\caption{ Resistance along the least conducting direction
(in ohms) as a function of magnetic 
orientation.  Field 
strenght was 4 Tesla.  The field was rotated in the $bc$ plane
and angle $\Theta$ is defined so that
$\pm$100 gradians coincide with the $b$ direction.
}
\label{fig:guybc}
\end{figure}

\begin{figure}
%dip_cmp.eps
\caption{ Resistance along the most conducting direction
(in milliohms) as a function of magnetic field strength
and orientation.  Field was ramped up for fixed orientation
in the $bc$ plane near to magic angles.
The angle $\Theta$ is defined so that
$\pm$90 degrees  coincides with the $b$ direction.
%Data are
%taken from Ref. \protect\cite{woowon}.
As explained in the text, after the destruction of
superconductivity, the angle dependence is pronounced
in the central dip, but emerges more slowly for
the off-center dip.
}
\label{fig:diphierarchy}
\end{figure}

\begin{figure}
%scaling.eps
\caption{ Log of the resistance along the most conducting direction
(in milliohms) as a function of 
log of the component of the magnetic field
perpendicular to the $ab$ plane.
As explained in the text, the data are expected to
scale away from the magic angle dips.
}
\label{fig:scalingxx}
\end{figure}

\begin{figure}
%rzllpf.eps
\caption{ Log of the resistance along the least conducting direction
(in ohms) as a function of 
log (base 10) of the component of the magnetic field (in Tesla)
perpendicular to the $ab$ plane.
As explained in the text, the data are expected to
scale away from the magic angle dips.
%Raw data were provided by Danner, {\it et al.}.
}
\label{fig:scalingzz}
\end{figure}

\begin{figure}
%acdata.eps
\caption{ Resonances in conducitivity in the least conducting direction
for (TMTSF)$_2$ClO$_4$
as a function of field orientation.
Magnetic fields of various strengths were rotated in the
$ac$ plane and the resistance (in ohms) is plotted as
a function of angle in the $ac$ plane, measured from
$\hat{a}$.
}
\label{fig:aclebed}
\end{figure}

\begin{figure}
%crvr.eps
\caption{ 
Calculated resonances in conductivity in the least conducting direction
for (TMTSF)$_2$ClO$_4$
as a function of field orientation.
Magnetic fields of various strengths were rotated in the
$ac$ plane and the resistance (in ohms) is plotted as
a function of angle in the $ac$ plane, measured from
$\hat{a}$.  
%Adapted from Ref. \protect\cite{batman}
}
\label{fig:crvr}
\end{figure}

\begin{figure}
%rzpvr.eps
\caption{
Measured resonances in conductivity in the least conducting direction
for (TMTSF)$_2$PF$_6$
as a function of field orientation.
Magnetic fields of various strengths were rotated in the
$ac$ plane and the resistance (in ohms) is plotted as
a function of angle in the $ac$ plane, measured from
$\hat{a}$. 
%Data provided by Danner, {\it et al.}
}
\label{fig:rzpvr}
\end{figure}

\begin{figure}
%dpfwcl.eps
\caption{
Comparison of the measured resonances in 
the second derivative of the
conductivity in the least conducting direction
for (TMTSF)$_2$ClO$_4$ and (TMTSF)$_2$PF$_6$
as a function of field orientation.
Angle in the $ac$ plane is measured from
$\hat{a}$.  PF$_6$ data are
offset for clarity.
%Adapted from Ref. \protect\cite{batman}
}
\label{fig:dbatmandtheta}
\end{figure}

\begin{figure}
%rzpvrhy.eps
\caption{
Replacement of the Danner resonances in
the 
conductivity in the least conducting direction
for (TMTSF)$_2$PF$_6$ with Lebed magic angle effects.
Angle in the $ac$ plane is measured from
$\hat{a}$.  
%Adapted from Ref. \protect\cite{batman}
}
\label{fig:rzpvrhy}
\end{figure}

\begin{figure}
%drphy.eps
\caption{
Replacement of the Danner resonances in
the second derivative of the
conductivity in the least conducting direction
for (TMTSF)$_2$PF$_6$ with Lebed magic angle effects.
Angle in the $ac$ plane is measured from
$\hat{a}$.  
%Adapted from Ref. \protect\cite{batman}
}
\label{fig:drphy}
\end{figure}

\begin{figure}
%cryvr.eps
\caption{
Theoretical behavior of the Danner resonances 
for fields out of the $ac$ plane.
Materials values are those used
%in \protect\cite{batman}
for (TMTSF)$_2$ClO$_4$.
Notice that the main resonances are little
affected by fields along the $b$ direction of up to 0.6 Tesla.
}
\label{fig:cryvr}
\end{figure}

%%%%%%%%%%%%%%%%% concl.sec %%%%%%%%%%%%%%%%%%%%%%%%%%

\begin{figure}
%RG2chains.eps
\vspace*{0.3cm}
\caption{ Schematic renormalization group flows for
the two chain problem.  It differs
from previous proposals in that the flows in $t_{\perp}$
away from the decoupled chains fixed point may
either flow into or approach closely a fixed point where there
is finite, incoherent hopping between the chains.  Which
of those occurs depends on the stability of the
incoherent fixed point to two body hopping between the
chains and other perturbation generated by $t_{\perp}$.
}
\label{fig:RG2chains}
\end{figure}

\begin{figure}
%RGmanychains.eps
\vspace*{0.3cm}
\caption{
Schematic renormalization group flows for
the many chain problem.  It differs
from previous proposals in that the flows in $t_{\perp}$
away from the decoupled chains fixed point may
either flow into or approach closely a fixed point where there
is finite, incoherent hopping between the chains. Experimentally,
it appears that the incoherent fixed point is stable 
for (TMTSF)$_2$PF$_6$
in the presence of a magnetic field.  It may be more generally
stable, but in the case of (TMTSF)$_2$PF$_6$, it
is unstable to superconductivity in zero magnetic field.
}
\label{fig:RGmanychains}
\end{figure}

%%%%%%%%%%%%%%%%% END OF FIGS %%%%%%%%%%%%%%%%%%%%%%%%%%%%%%%%

\end{document}

%% file: intro.sec
\section{Introduction}
\label{sec:intro}
\subsection{Preamble}

Motivated primarily by the difficulties faced in understanding the 
normal state of the high-temperature superconducting cuprates (HTSC's)
within the framework of Fermi liquid theory (FLT), there has been a
great deal of theoretical interest in the possible
existence of ``metallic'' states of matter controlled by
non-Fermi liquid (NFL) fixed points in two,
and possibly even three, dimensions.
%steve, I kept the above line to acknowledge the MFL guys
As has been known for many years, FLT is never an appropriate
description of a one dimensional (1D) metal; rather, the generic
paradigm to use is that of the Luttinger liquid \cite{duncan}, a quite specific
type of NFL. 
In two and three dimensions
it is widely accepted that, at least for sufficiently weak
electron-electron interactions and generic
Fermi surface shapes, the free Fermi gas is a stable fixed point.
%with respect to all but the BCS instabilities.
This belief is based on the stability of generic Fermi surfaces
in the conventional 
diagrammatic perturbation theory 
\cite{landau} and the renormalization group \cite{shankar}
to everything except the usual BCS instability.
However, in two dimensions there are some unresolved questions
about the applicability of the starting point
\cite{philshift,casmetz} and non-Fermi liquid
behavior for weak interactions is somewhat controversial.
Certainly, for strong interactions
and/or Fermi surfaces with strong nesting properties
the ``NFL'' spin and charge density wave states
represent alternative, strong coupling fixed points,
and there is a great deal of interest
at present in the  possible existence of
other strong coupling fixed points, especially in 
two dimensions.

Whatever the case for 1D and 2D models, 
bulk materials are never purely 1D or 2D systems, {\em e.g.\/}
the so-called quasi-1D organic conductors have interchain hopping
integrals typically of order a few hundred degrees,
and one might be led to the view that any metallic state
existing at sufficiently low temperature
will be three dimensional and
therefore
a Fermi liquid, provided that the
interactions are not too strong.
There are two commonly accepted ways 
around this argument, one being
that sufficiently strong interaction
can render the hopping between chains
irrelevant, or at least less relevant than
other interchains couplings, the other
being that for Fermi surfaces with special
shapes the interactions are never weak.
The former is not known to occur in any
actual material, while the latter is
not uncommon and results in the known
charge and spin density wave materials.
The purpose of this paper is to review
a third proposal \cite{prl,magic,prl2,lees2} for avoiding the
low temperature crossover to a three
dimensional Fermi liquid.

Our motivation for doing this is quite simple:
we are interested in trying to explain experiments 
we believe are
incompatible with either Fermi liquid behavior or
existing non-Fermi liquid proposals.  
The problematic experiments are those
on the high $T_c$ superconductors at low
or optimal doping and those
on the organic conductor (TMTSF)$_2$PF$_6$
in its low temperature, high pressure, metallic phase.

\input motiv.sec

\subsection{Theoretical Motivation}

Motivated by the apparent experimental need for
a state with coherent transport confined to
the $ab$ planes, it is a natural question
whether there is any theoretical foundation
for believing in the existence of such a state.
We believe that there is in fact a
strong foundation, which we will lay out in detail
in Sections \ref{sec:TLS}-\ref{sec:interp}; here we give a brief
sketch of the underlying physics.

We begin with the 
same starting point as RG calculations
for the relevance/irrelevance
of interchain hopping:
the recognition that,
with a truly 1D system, in turning on weak interchain hopping
one really has, in a sense, a strong coupling problem to begin with 
in that the zero hopping state is not a Fermi liquid.
If the
intrachain correlation strength is much larger than the interchain
hopping, 
it is not {\em a priori\/} appropriate to consider the electron-electron
interaction as a perturbation on an anisotropic free Fermi gas.
Rather, one should consider the {\em interliquid hopping\/} as the perturbation,
which 
opens up the possibility of finding a 2D NFL
by perturbing (in the interchain hopping) about the 1D Luttinger
liquid. Similarly,
suppose one began with a 2D NFL metal, and then coupled a macroscopic
number of these by interliquid hopping to form a 3D system.
Although increasing the dimension of a system is expected to help
stabilize a Fermi liquid, again, if the interliquid hopping 
is much smaller than the effective correlation strength in the 
%non Fermi liquid starting state, 
2D NFL starting state,
one cannot rule out the possibility
of an anisotropic  NFL state resulting.
It is exactly this path which has led 
to the proposal that the
interliquid hopping could be renormalization
group irrelevant, resulting in the stabilization
of some non-Fermi liquid groundstate. It is
generally accepted that this is the only
non-trivial way to avoid a crossover to
three dimensional Fermi liquid behavior
in the low temperature limit, however,
this belief does not result from any
demonstration that this is the case, but rather
from a simple lack of alternatives.  We believe
that the experiments we have discussed 
require another alternative {\em viz.\/}, that
the hopping in one or more
directions be intrinsically incoherent.

Recall that
the perturbative irrelevance of $t_{\perp}$ results
if the NFL states coupled by it have the property that
the low energy spectral weight of the electron
creation and annihilation operators
vanishes rapidly enough.
It is formally equivalent to the
infrared convergence of perturbation theory in
$t_{\perp}$ due to the small matrix elements
of $t_{\perp}$ 
to states with small energy denominators.
In order to give a hint of why states with
intrinsically incoherent hopping
might exist, consider 
the perturbation theory associated with
the problem of making a 2D ``stacking'' of
1D electron liquids. The simplest form of interliquid hopping operator
will remove an electron of momentum $k$ from one liquid and insert
it into an adjacent one. If the 1D liquids are free Fermi gasses,
then an electron state of momentum $k$ is a
single particle, energy eigenstate and is
{\em degenerate\/} with each of the electron states of the same
momentum in all of the other liquids. It is therefore clear that
perturbation theory in the interliquid hopping is
degenerate perturbation theory and
that, in this case,
$t_{\perp}$ is relevant and,
since the perturbation theory was degenerate,
one should 
select states which diagonalize the interliquid hopping,
and then perturb in the interactions.
These states are
built out of the creation and annihilation operators
for
Bloch states with well-defined interliquid momenta $k_{\perp}$,
and such states will be energy eigenstates.
As we will discuss, in the case of coupled Fermi liquids 
interactions do not change things
in any essential way. Therefore,
coherence is expected in both cases and the
interactions are always to be treated as a perturbation.

On the other hand, if the 1D liquids are NFL's (which is 
actually always the case) then the interliquid hopping operator 
which creates a hole of momentum $k$ in one liquid and a particle
of momentum $k$ in an adjacent liquid is no longer a zero energy
operator. This is because an electron (or hole) of momentum $k$ is
{\em not\/} an energy eigenstate. Rather, it is a (precisely specified)
superposition of energy eigenstates and does not, therefore, have a 
precise energy. Imagine that the spectral function for
the interliquid hopping were to be a flat function of
energy: in this case the perturbation theory is still
infrared divergent since the matrix elements 
between low energy states are vanishing
too slowly. However, the perturbation theory is no longer
degenerate since the states connected to the
ground state have energies which are widely different
and if anything, one's prejudice is that
Fermi's Golden Rule might be a more appropriate
starting point than Bloch states.
The non-degeneracy
of the hopping  will impede, perhaps destroy, the formation
of interliquid Bloch states. 
Roughly speaking, if there is some way to
characterize the ``width'' in energy of the eigenstates which are
in the superposition composing the in-liquid electron state of
momentum $k$, then if this width is much smaller than the interliquid
hopping rate it makes sense to 
perform degenerate perturbation theory,
form interliquid electronic Bloch
states first, and {\em then\/} consider the effect of those terms
which led to the width as a perturbation on the (anisotropic)
2D free Fermi gas. On the other hand, if the width is much larger
than the interliquid hopping rate, then 
perturbation theory in the hopping
is non-degenerate and coherent interliquid electronic
Bloch states are not a sensible starting point.  Interliquid
transport could actually be intrinsically
incoherent, {\em i.e.\/} diffusive, due to the
lack of a well-defined interliquid velocity at the Fermi surface.
In fact, the Fermi surface will not show a 2D character at all,
since that whole picture is based on the existence
of the 2D Bloch states.
{\em Intra\/}liquid transport may 
be metallic, but {\em inter\/}liquid transport
will not be.

In essence, the in-liquid interactions responsible for destroying
the Fermi liquid state give an electron of momentum $k$ an intrinsic
``lifetime'' (although, as we shall see, the term `lifetime', 
in its usual sense, is not
precisely correct: the electron spectral function is power
law, not Lorentzian) which, if shorter than the timescale for
interliquid hopping, will render all interliquid transport completely
incoherent. It is important to emphasize that this ``lifetime'' is not 
thermally, nor impurity, induced: it occurs in a {\em pure\/}
system at {\em zero temperature\/}.

As we will discuss, for the case of coupled one
dimensional chains, this physics is directly related
to the physics of the two level system, 
a model of a single two state degree
of freedom (a ``spin'') coupled to a bath of harmonic
oscillators.  The two level system model is the prototype
for studies of quantum coherence effects, and
%in the two level system problem,
there is believed to 
exist a regime
at zero temperature
in which flipping  of the 
spin 
is purely incoherent
due to the effects of the harmonic
oscillator enviroment.
The tunneling is incoherent
in the sense that it is impossible 
even in principle to
observe interference effects between histories 
of the entire system
in which the 
history of the spin
is different.
The term in the Hamiltonian which
flips the spin remains
a relevant perturbation
but the incoherent regime is
clearly
qualitatively different
from the coherent regime,
where interference effects are
in principle observable.
The existence of the
incoherent regime was the main theoretical
motivation for our
proposal that a similar regime could
exist for hopping between non-Fermi liquids
and the details of the connection between the two
problems form one of the central sections of this
paper (Section \ref{sec:TLS}).
For fermionic hopping between chains, the incoherence of
the single particle hopping would clearly
preclude Bloch states and the formation
of a two dimensional Fermi surface; it can
therefore be expected to result
in different long time, low energy properties
and constitutes a different fixed point
from any at which two dimensional coherence
obtains.  There is, however, no real reason why
an analogous proposal cannot exist for coupled
two dimensional non-Fermi liquids,
the problem that is ultimately connected
to the experiments.

%We believe that this proposal is uniquely capable of explaining
%the experiments on the organic conductor (TMTSF)$_2$PF$_6$
%discussed in Section \ref{sec:organics} and the experimental
%anomalies in $c$-axis transport in the cuprates
%discussed in Section \ref{sec:motiv} and Section
%\ref{sec:caxis}.  

\input model.sec

%% file: motiv.sec
\subsection{Experimental Motivation}
\label{sec:motiv}

The theoretical work presented in this paper was originally
motivated by the bizarre transport properties of the HTSC's.
In particular, the high $T_c$ compounds
generically display {\it qualitatively} anisotropic
transport, which, as we will argue, is not
explicable within Fermi liquid theory. At the same time,
as we will discuss,
the measured experimental properties of, for example,
Bi$_2$Sr$_2$CaCu$_2$O$_8$, rule out the possibility that
the anisotropy arises from the renormalization group 
irrelevance of the interplane hopping. We believe
there is therefore a clear and serious need for a
theoretical proposal capable of producing 
qualitatively anisotropic transport
in the face of the renormalization group relevance of
the interplane hopping. Providing such a proposal 
is the intent of this paper. 

Let us begin with a discussion of 
the evidence of 
non-Fermi liquid behavior in the cuprates.
Apart from the extraordinarily
high superconducting transition temperatures, $T_c$,
these materials exhibit a highly
anomalous metallic state at temperatures above $T_c$.
Despite attempts to do so,
we believe that no plausible scenario
has been presented to understand this metallic state within
the (Landau) Fermi liquid (FL) paradigm \cite{pines}. 
There are simply too many anomalies to account
for. The cuprates therefore present us with a much richer problem than high $T_c$
alone, {\em viz.\/} the existence of an essentially two-dimensional (2D) non-Fermi
liquid (NFL) metallic state. The problem of understanding why the $T_c$'s of the 
cuprates can be so high is intimately connected then with the problem of understanding
the physics of the anomalous normal state.

For our  purposes 
%of developing this introduction, we shall illustrate 
it suffices to illustrate
the anomalous nature of the normal state by considering the simplest transport
properties. The basic structure of all of the HTSC's is the presence of 
``stacked'' copper-oxide (CuO$_2$) planes. The standard nomenclature is to
refer to in-plane properties as ``ab-plane'', and inter-plane as ``c-axis''.
We shall begin by considering ab-plane transport. 
It has long been known that
the normal state resistivity $\rho(T)$ does not follow the Fermi liquid form.
Empirically, the normal state resistivity can be
approximately characterized by 
\begin{equation}
\rho_{ab}(T)=a + bT^p
\end{equation}
In the vicinity of ``optimal doping'', $\rho_{ab}$ exhibits the famous
``linear-$T$ dependence'', $\rho_{ab}(T)\propto T$.
In contrast, for $T~^{<}_{\sim}~0.2~\Theta_D \ll E_F$ the resistivity in a FL
is
\begin{equation}
\rho^{\rm FL}_{ab}(T)\sim \rho_{\rm imp} + c_{\rm el-el}T^2 + c_{\rm el-ph}T^5
\label{eq: FLrho}
\end{equation}
The three terms represent, respectively, the dissipation of
electrical current by scattering from impurities (temperature independent),
other electrons ($\sim T^2$) and phonons ($\sim T^5$). The Debye temperature
$\Theta_D$ of the HTSC's is estimated from specific heat measurements to be 
in the range of 300-450 K.  Although the temperature
range of the normal state is bounded below by $T_c$, the crucial observation is
that there are examples of HTSC's with 
%$T_c~^{<}_{\sim}~10K \ll 0.2~\Theta_D$
$T_c$ well below $0.2~\Theta_D$
which clearly fail to fit the FL form (\ref{eq: FLrho}).

A second anomalous ab-plane transport property involves the Hall angle,
$\theta_{\rm H}=\cot^{-1}(\sigma_{xx}/\sigma_{xy})$. As first pointed out by
Chien, Wang and Ong \cite{chien}, and by Anderson \cite{philhall}, 
there is a remarkably clean quadratic temperature 
dependence of $\cot\theta_{\rm H}$ in the normal state of the cuprates
\begin{equation}
\cot\theta_{\rm H} = A + B T^2
\end{equation}
A {\em Hall relaxation rate\/} $\tau_{\rm H}^{-1}$ can be defined via
\begin{equation}
\frac{1}{\tau_{\rm H}}=\omega_c\cot\theta_{\rm H}
\end{equation}
Assuming only weak, or no, temperature dependence of $\omega_c$
this relaxation rate is 
observed to be distinct from the transport rate $\tau_{\rm tr}^{-1}$
derived from $\rho(T)$
and the frequency dependent in-plane conductivity \cite{fnhall}.

%[footnote: Upon making fits to data, and using a self-consistent
%scattering theory
%one finds the approximate behaviors
%\begin{equation}
%\frac{\hbar}{\tau_{\rm tr}}\sim k_B T \;\;\; , \;\;\; 
%\frac{\hbar}{\tau_{\rm H}}\sim k_B T \left(\frac{T}{T_{\rm H}}\right)
%\end{equation}
%where $T_{\rm H}\sim 500-1000 K$, and factors of $\hbar$ and $k_B$ have been
%reinstated. However, recent ac Hall effect experiments (Drew {\em et al.\/},
%Phys. Rev. Lett. ???)
%suggest a much smaller
%value for $T_{\rm H}$.]

Remarkably, the $T^2$ behavior of $\tau_{\rm H}^{-1}$ appears to be very robust to 
changes in doping, {\em i.e.\/} it is not restricted to ``near optimal''
doping \cite{fn214}.
%[footnote: 214 is exceptional]. 
Indeed, it has been observed even in very under- and over-doped samples where
$\rho(T)$ is far from linear. Any successful theory of these
two relaxation rates must be able to account for this apparent empirical fact that
$\tau_{\rm H}$ is more robust than  $\tau_{\rm tr}$.

The existence of two such fundamentally different in-plane transport 
relaxation rates is incompatible with conventional scattering 
mechanisms of Landau quasiparticles, 
both quasi-elastic (impurities and phonons) and inelastic
(electron-electron). It is very difficult to see how other inelastic scattering
mechanisms (spin fluctuations for example) could be introduced within a FL
framework and be capable of reproducing these two rates
\cite{hlubina}. There are two problems
to overcome, the first to construct a physical mechanism for two relaxation rates
{\em per se\/}, the second to be able to obtain the required $\tau_{\rm tr}$
without destroying the quasiparticles. If one extracts the frequency 
dependence of $\tau_{\rm tr}$ from ab-plane conductivity, and interprets this
as the frequency dependence of the lifetime of a quasiparticle (via
$Im\Sigma(\omega)\propto\tau_{\rm tr}(\omega)$), then the quasiparticle
renormalization 
$Z_k=(1-\partial Re\Sigma_k(\omega)/\partial\omega)^{-1}_{\omega=E_k}$
is found to vanish as $k\rightarrow k_F$, {\em i.e.\/} there is no
quasiparticle at all \cite{mfl,philZ}!  This type of argument is clearly incompatible
with a FL picture.

But this is not the end of the transport story. We have so far discussed only
the ab-plane transport. However the c-axis transport is observed to be
{\it qualitatively different}, the importance of which was emphasized 
by Anderson in the very early days of high-$T_c$ \cite{phil_zou}. A more detailed
discussion is given in section \ref{sec:caxis} , but the key concepts involved
are as follows.

In La$_2$CuO$_4$, the prototypical HTSC, 
%The various
band theory calculations 
%for the parent compounds of the HTSC's
predict a single, half-filled $d_{x^2-y^2}$ band crossing the Fermi surface
\cite{band214}.
%In La$_2$CuO$_4$, the prototypical HTSC, 
The ratio of the c-axis bandwidth
$4t_{\perp}$ to the ab-plane bandwidth $4t_{\parallel}$ is calculated to be 
$t_{\perp}/t_{\parallel}\approx 1/10$. Although the band theories incorrectly
predict a metallic state for La$_2$CuO$_4$, due to their inability to
adequately account for strong correlation effects, it is perfectly correct to
use, in a starting point (``bare'') Hamiltonian,
the anisotropies predicted by these methods.
Upon doping to La$_{2-x}$Sr$_x$CuO$_4$ the anisotropy 
in the  hopping should not greatly change.
Ignoring any correlation effects beyond those incorporated into a 
FL framework, 
and within the isotropic relaxation time approximation,
the ratio of c-axis to ab-plane resistivity is
predicted to be temperature independent and of the order 
$\rho_c(T)/\rho_{ab}(T)\approx 25$. The experimental fact is that in all
superconducting samples $\rho_c(T)/\rho_{ab}(T)$ is strongly 
temperature dependent, $\rho_c(T)$ actually turning upwards
($\partial\rho_c(T)/\partial T < 0$)
at low temperatures in the underdoped samples, while
$\rho_{ab}(T)$ remains metallic
($\partial\rho_{ab}(T)/\partial T >0$ and $\rho_{ab}$ well below the
Mott-Ioffe-Regel limit).
The anisotropy can reach
a magnitude of several 
hundred as $T\rightarrow T_c^+$
\cite{uchida_dc}.
Indeed at optimal doping the projected zero temperature ratio 
%is $~^{>}_{\sim}~5000$!
can be as high as several thousand or more! 
Moreover, even as $T\rightarrow T_c^+$ the c-axis mean free
path estimated from $\rho_c(T)$ is much shorter than  the interplanar spacing,
thus precluding metallic Bloch-like conductivity. Equivalently, the c-axis
conductivity is well below the Mott-Ioffe-Regel limit. This is a highly
nontrivial result, for the interplane hopping rate is estimated from band theory
to be $t_{\perp}\sim 500 K$, yet the in-plane scattering rate is of order
$\tau_{\rm tr}^{-1}\sim T$. Thus $t_{\perp}\gg\tau_{\rm tr}^{-1}$ over much
of the experimental temperature range, and in particular in the vicinity of 
$T_c~^{<}_{\sim}~40 K$. At low temperature, and in the absence of a significant
renormalization of $t_{\perp}$, one would therefore expect to be in a region
of coherent c-axis conduction, yet the experimental evidence is strongly
contrary to this. 
%Note that while standard band narrowing 
%($t_{\perp}\rightarrow Zt_{\perp}$) effects {\em could\/} be introduced 
%to avoid the inequality $t_{\perp}\gg\tau_{\rm tr}^{-1}$, neither
%$\rho_c(T\rightarrow 0)$ nor the ratio $\rho_c(T)/\rho_{ab}(T)$ will
%be affected by such FL dynamical renormalizations, and this therefore cannot
%account for the anomalous c-axis conductivity.
Even more compelling perhaps than
the studies of the
dc conductivity are the recent experimental studies of the frequency
dependent c-axis conductivity $\sigma_c(\omega)$
\cite{cooper,uchida_ac,tajima_ac,timusk}. There is simply no 
plausible way to fit the low frequency data to a Drude, or
generalized Drude, form: the data are simply incompatible with
a zero-frequency Drude
peak with any appreciable weight and reasonable width.
The analysis of $\rho_c(T)$ for bilayer and trilayer cuprates requires
more care than for the single-layer compounds like La$_{2-x}$Sr$_x$CuO$_4$,
but the general fact of incoherence remains. 
Moreover, photoemission experiments on Bi$_2$Sr$_2$CaCu$_2$O$_8$, a 
bilayer HTSC in which the inter-bilayer coupling is very weak, point to the
existence of a {\em single\/} sharp Fermi surface
\cite{ding}. The experimental resolution
is $~^{<}_{\sim}~20~meV$. This strongly contradicts the band theory prediction
of {\em two\/} Fermi surfaces split by an energy of a few hundred $meV$.
Such a splitting would be the direct consequence of coherent single electron
hopping between the two CuO$_2$ planes within a bilayer. The absence of any 
splitting 
demonstrates that, even within a bilayer
where we expect $t_{\perp}\gg\tau_{\rm tr}^{-1}$,
interplanar hopping is incoherent. 

The empirical fact that the c-axis conduction is incoherent is therefore 
another  very strong
argument against a FL scenario, 
as pointed out long ago by Anderson
\cite{pwa_science_92}.
Indeed, it was the strange behavior of $\rho_c(T)$ which first led Anderson
to propose ``confinement'' of electrons to the CuO$_2$ planes as the driving force
behind high-$T_c$ itself \cite{phil_old}.  The idea we will advocate
in this paper is
that it is not the electrons, but the {\em coherence\/}
that is confined to the $ab$ planes.

The picture that therefore emerges is that of a macroscopic number of 2D
NFL's the conduction between which is incoherent.
Experimentally, the ``coherence'' is clearly confined to the
planes - whether the electrons are or are not is a different question.
The challenge is to understand
how the  incoherence of $c$-axis transport,
or the confinement of coherence to the $ab$ planes, 
comes about. In seeking to do so, we believe that there
are two key facts to exploit, namely: (1) electronic correlations are
strong, and (2) $t_{\perp}/t_{\parallel}$ is small, {\em i.e.\/} there is a 
significant anisotropy of nominal bandwidths. In particular,
we will take $t_{\perp} \ll t_{\parallel}$ and
$t_{\parallel}$ of order the interaction energy scale.

It is important to emphasize that anisotropy alone cannot account for the
qualitatively
anomalous c-axis conductivity in all of the HTSC's. Apart from theoretical arguments,
there is a classic experimental ``proof'' of this point.  Sr$_2$RuO$_4$,
a structural analogue of  La$_2$CuO$_4$, is predicted by band structure calculations
to be about as anisotropic as, if not more than,  La$_2$CuO$_4$ \cite{oguchi}. 
In Sr$_2$RuO$_4$, however,
 $\rho_c$ and $\rho_{ab}$ display the {\em same\/} temperature
dependence at
low temperatures, with an anisotropy of approximately 500
\cite{maeno}. Moreover, the temperature
dependence is of Fermi liquid form, $\rho=A+BT^2$. While correlation effects 
in Sr$_2$RuO$_4$ are
not negligible \cite{maeno}, as evidenced by specific heat and magnetic 
susceptibility 
measurements, they are simply not as {\em qualitatively\/} severe as those in
 La$_2$CuO$_4$, for Sr$_2$RuO$_4$ is a good metal, while La$_2$CuO$_4$ is a
Mott insulator! Thus, the experimental situation with respect to Sr$_2$RuO$_4$
strongly supports the idea that the anomalous c-axis conductivity in the
cuprates results from a conspiracy of both sufficiently high anisotropy
and sufficiently strong correlations. Paranthetically, we note that the 
superconducting transition temperature in Sr$_2$RuO$_4$ is only
$1 K$, which is 
%completely 
consistent with the interlayer tunneling scenario
for the HTSC's, a scenario which associates high $T_c$ with the absence of
coherent c-axis conduction in the normal state.

We are thus led into the following conundrum: if the correct fixed point
(ignoring the superconducting transition) governing the physics of the HTSC's
is not an (anisotropic) 3D Fermi liquid, just what is the appropriate fixed
point? In contemplating this question, we should recognize that the vast 
majority of theoretical approaches toward understanding the normal state have
been based on strictly 2D models. In doing so, there is an
implicit assumption that
c-axis transport of electrons is somehow unimportant, but the question is 
{\em why and in what sense\/} is it ``unimportant''?

One sense in which the hopping would be unimportant is
realized in the 
opposite extreme from a 3D Fermi liquid: a genuine 2D NFL. Such a
state would be possible if the interplane hopping operator was 
renormalization group (RG) irrelevant. While this is not the case if the
coupled 2D liquids are Fermi liquids, one expects quite generally 
that the scaling dimension of $t_{\perp}$ 
and its renormalization group status can be
altered by interaction effects
in a NFL.
However the observation that the HTSC's exhibit sharp Fermi surfaces is 
sufficient evidence to 
establish that, despite such renormalization, 
$t_{\perp}$ remains a relevant operator.
For example, if the singularity of $n(k)$ near $k_F$
may be parametrized by
%has the form
\begin{equation}
\label{n(k)}
\frac{\partial n(k)}{\partial k}\sim -| k-k_F|^{2\alpha-1}
\end{equation}
then at the tree level of the RG, $t_{\perp}$ is irrelevant if and
only if $2\alpha > 1$.
But in this case there would not be a sharp Fermi surface at all
because $\partial n(k)/\partial k$ would not diverge as $k\rightarrow k_F$! 
A sharp Fermi
surface therefore implies the relevance of $t_{\perp}$.
A more general, parametrization independent result is that
$t_{\perp}$ is relevant (at least at the tree level)
if the integral over energy of the
electron spectral function,
$\rho({\bf k},\omega)$, for some momentum ${\bf k}$ 
at the Fermi
surface, from zero energy 
out to  a cutoff energy $\Lambda$, does not vanish faster than
$\Lambda$.  The presence of a ``quasiparticle'' peak in the
angle resolved photoemission data is thus, in and of itself
and independent of model or interpretation, enough to
demonstrate the relevance of $t_{\perp}$ \cite{fncrossover}.  
The ``confinement'' of the electrons to the planes
in the usual renormalization group sense
is therefore ruled out.

The key theme which will run through this paper, however, is that
the {\em relevance\/} of $t_{\perp}$ in the RG sense,
and the ensuing deconfinement of the electrons,
 {\em do not 
guarantee\/} that 
single particle interliquid hopping will be {\em coherent,
i.e.\/} the deconfinement of coherence. 
Our proposal, then, is that there exists a new state
which is in some sense an intermediary
between strictly 2D and 3D coherent states. In this state there is a 
macroscopic number 
of 2D NFL's coupled by single particle hopping
which is not RG irrelevant, but the interliquid
hopping is rendered incoherent by interaction 
effects. 
We refer to this state as a state of
``confined coherence'', and schematically denote it by 
(2D)$_{\rm NFL}\otimes$(1D)$_{\rm incoh}$. 
The coherence which is clearly absent
experimentally is the coherence in the
finite temperature transport, but we believe
this to be the result of the
loss of quantum coherence even
in the pure system at zero temperature.
The defining feature of a state with
confined coherence, then, is this
incoherence at zero temperature
and in the pure material.
Our definition of the incoherence of the hopping
is essentially the fundamental definition of
quantum coherence:
the hopping is incoherent if
it is impossible, even in principle,
to observe intereference effects
between histories which involve the motion
of particles between the incoherently coupled
NFL's. The coherence persists in the individual
NFL's because
interference
effects between different histories in
which no particle hops between different
NFL's remain in principle observable.
Therefore, the coherence
can be said to be confined
to the individual NFL's.
This definition connects our proposal
for an explanation of the anomalous 
c-axis transport
in the cuprates directly to
the decohering histories approach to
the quantum to classical crossover and
the two level system problem \cite{TLS}.  
Also,
although this work was originally motivated, as described
above, by the peculiar c-axis conductivity in the cuprates, our
conclusions
will be relevant to {\em any\/} sufficiently anisotropic and
sufficiently
strongly correlated metal,
and the above definition is
the natural one for making contact
with certain mysterious  experimental results
on the
organic conductor (TMTSF)$_2$PF$_6$,
which we now discuss.

We believe that, apart from our original motivation from the cuprate data,
the need for a state with confined coherence is
also clear in the experiments on the
organic conductor (TMTSF)$_2$PF$_6$ of
Kang, {\it et al} \cite{woowon} and Danner, {\it et al.}
\cite{guynfl}.  As we have pointed out elsewhere \cite{magic},
this material exhibits an exotic low temperature metallic
phase in which magnetoresistance depends only on
the component of the magnetic field perpendicular
to its $ab$ planes.  The experiments on this
material will be discussed in detail in
Section \ref{sec:organics},
where we argue that
such an anisotropy can result
only if hopping between different $ab$ planes is
irrelevant or if it is incoherent.  The essential point
is that if the hopping is incoherent, then interference
effects between histories where particles hop
between planes cannot be observed due to the
randomization of the relative phase of the 
histories: in this case, the flux through these
paths has no effect on physical observables
(other than through Zeeman effects)
and  magnetoresistance should only be a function
of the field normal to the $ab$ planes.  Note
that this is experimentally the case
even for resistivity {\it perpendicular}
to the $ab$ planes which is totally impossible in
any semiclassical description.  The 
only possible alternative explanation would be
that the flux enclosed by paths which leave the
$ab$ plane
is unimportant because
the electrons are effectively confined and
don't leave the $ab$ planes.
However,
conduction
perpendicular to the $ab$ planes is not
insulating at low temperatures and
so the renormalization group irrelevance
of the hopping appears to be excluded as an explanation.

Further, a direct test of Danner, {\it et al.}
\cite{guynfl} of the coherence of $c$-axis
transport demonstrated that, in addition to
a phase where (TMTSF)$_2$PF$_6$ exhibits
a three dimensional Fermi surface and coherent
$c$-axis transport, the material has another 
metallic phase
where no signs of coherent transport 
out of the $ab$ planes are present and no 
evidence for a three dimensional Fermi surface exists.
The two phases are separated by only an
order one change in the magnitude of the
conductivity out of the $ab$ plane,
which is hardly what one expects if the
transverse hopping is relevant in one phase and
irrelevant in the other.
The incoherence effects were most pronounced at the lowest
temperatures studied and in the cleanest samples.
These experiments provide direct evidence that the
confinement of coherence does occur and that it 
is a truly new state of matter, rather than a
reiteration of known possibilities for the irrelevance
of $t_{\perp}$ or an
impurity or finite temperature effect.

%Note that the perturbative
%stability of Fermi liquid theory in 3D does not exclude the existence of 
%such a fixed point, nor does the instability (with respect
%to $t_{\perp}$) of a $t_{\perp}=0$ fixed
%point, since we are proposing a state in which the
%hopping is relevant, but incoherent.  

%In our picture, the incoherence
%is intrinsic to the hopping itself in the
%presence of strong interactions and does
%not result from finite temperature effects.

%% file: model.sec
%\subsection{The Model Proper}

Unfortunately, our understanding of 2D NFL states
%in the cuprates
is not well enough developed to permit precise calculations to be made.
%It is natural, therefore, 
However, it is possible
to consider the analogous problem in one less
dimension, namely, the problem of coupled 1D electron liquids. 
%Localization effects aside, 
It is
well known that interacting electrons in 1D are always 
NFL's. A very general (non-insulating)
class of these go by the name of ``Luttinger liquids''. We therefore consider the
problem of Luttinger liquids coupled by interliquid single particle
hopping as a potential paradigm for coupled NFL's. 
%The extent to which
%our results can be extended to the case of coupled 2D NFL's will be addressed
%in the conclusion.
This problem is far from new: as early as 1974, in the context of the newly
discovered quasi-1D organic conductors, Gorkov and Dzyaloshinskii discussed
how various key properties of a 1D electron ``chain'' could be destroyed
by the presence of interchain hopping \cite{gorkov}. 
More recently, many other authors have
addressed the problem, using various techniques \cite{Bunch}.
To our knowledge, however, ours is the only approach which directly 
addresses the question of interliquid {\it coherence}. This question is of
crucial importance, for many of the other approaches {\em begin\/} with
an anisotropic 2D electron gas
(or its two chain analogue),
a state with manifestly coherent interliquid hopping,
upon which interactions are treated perturbatively. In those approaches
which do not begin with the anisotropic 2D electron gas, we believe
that,
while in some cases unreasonable approximations and/or errors have been made,
in general, these works have all correctly demonstrated the relevance of
$t_{\perp}$ but have simply not addressed the question of
its coherence.
In general, past workers have argued
that the flow away from $t_{\perp}=0$ should
lead to higher dimensional coherence
and,
for infinitely many chains, to a Fermi liquid
or to some other (CDW, SDW or BCS) known higher dimensional fixed point,
mainly because of the lack of an alternative proposal.
We believe that incoherent hopping does constitute such an alternative,
that some alternative is required by the experimental situation,
and that the experiments,
particularly in the organic material
(TMTSF)$_2$PF$_6$, strongly support our proposal.

We emphasize again that we are interested
in the {\em weak interliquid hopping regime\/} 
where the bare interliquid hopping parameter is much smaller than
both the bare intraliquid hopping rate and the bare intraliquid interactions. 
This is not to say that we demand that perturbation theory in the 
interliquid hopping must be convergent, indeed quite the contrary, as shall
be discussed further below.

As we have mentioned, the key construct needed in investigating the nature
of interliquid hopping is the electron spectral function, $\rho(k,\omega)$,
defined by
\begin{eqnarray}
\rho(k,\omega)&=&\theta(\omega)
\rho^+(k,\omega)+\theta(-\omega)\rho^-(k,-\omega)
\nonumber\\
&=&\sum_n\left\{\theta(\omega)
|\langle n_{N+1}|c_k^{\dag}|0_N\rangle|^2\delta(\omega-E_n^{N+1})
+\theta(-\omega)
|\langle n_{N-1}|c_k|0_N\rangle|^2\delta(\omega-E_n^{N-1})
\right\}
\nonumber
\end{eqnarray}
where $|0_N\rangle$ is the ground state of the
$N$-particle system, and 
$|n_{N\pm 1}\rangle$ are energy eigenstates
of the $N\pm 1$-particle system.
In a FL, $\rho(k,\omega)$ is dominated by a term which sharpens up to a 
$\delta$-function as $k\rightarrow k_F$. This term, of weight $Z_k\neq 0$, is 
the quasiparticle part of $\rho(k,\omega)$. The remainder of $\rho(k,\omega)$
is featureless so that, as far as low energy properties are concerned, 
only the quasiparticle part of $\rho(k,\omega)$
matters. The physical interpretation is that an electron inserted into
(removed from) a FL with momentum $k$ not too far above (below) $k_F$
propagates coherently with a sharp energy, the quasiparticle energy.
%Clearly, 
If $\langle i,j \rangle$ label physically adjacent liquids, and $k$ in-liquid 
momenta, then
an interliquid hopping term of the form
\[
H_{\perp}=t_{\perp}\sum_{\langle
i,j\rangle,k}\left(c_{i,\sigma}^{\dag}(k)c_{j,\sigma}(k)
+{\rm h.c.\/}\right)
\]
%where $i, j$ label liquids, and $k$ in-liquid momenta,
will directly couple a quasiparticle state in one liquid with an
{\em energy degenerate\/} quasiparticle state in the physically adjacent liquids,
leading to the formation of interliquid Bloch states of precise interliquid 
momenta. An interliquid band will therefore form, entailing a coherent interliquid
velocity and hence coherent interliquid transport.

In contrast, in a Luttinger liquid (or any NFL, by definition)
there are no Landau quasiparticles. The quasiparticle weight, $Z_k$, is
zero, but in a nontrivial way. It is not simply a matter of taking
the FL spectral function and sending $Z_k$ to zero, for that would result in
a completely featureless spectral function. A ``metallic'' electron liquid
must have singularities in $\rho(k,\omega)$ in order to exhibit coherent
in-liquid transport.
The Luttinger liquid spectral function
differs from that for a FL in that its singularities are power law in nature,
even at the Fermi surface. For the physically most relevant case of spin-independent
electronic interactions, $\rho(k,\omega)$ has singularities at 
$\omega=\pm v_ck, v_sk$ determined by a single exponent $\alpha$,
and $v_c$ and $v_s$ denote velocities of propagation of charge and spin
currents. There is a singularity in $n(k)$ at $k=k_F$ which is precisely
as given in (\ref{n(k)}). The nontrivial regime is therefore $\alpha<1/2$.
Although $t_{\perp}$ is formally RG relevant in this regime, we 
emphasize that there is no {\em a priori\/} reason for the boundary
$\alpha=1/2$ separating relevant from irrelevant $t_{\perp}$ to
be the same boundary for separating coherent from incoherent interliquid
hopping. Quite generally, the issue of whether an operator is relevant or
irrelevant is simply a question of whether perturbation theory in that operator
is divergent or convergent. The problem of coherence versus incoherence is
a more subtle issue involving the {\em way\/} in which perturbation theory diverges
(and therefore always involves relevant perturbations). 
As we have mentioned, a simple model within
which to discuss quantum coherence and incoherence is that of a two level
system (TLS) coupled to a dissipative bath, and this problem is deeply
connected to our own. 
We will therefore begin by presenting a detailed discussion of the physics of 
the TLS in a way which will naturally generalize to the problem of
determining the nature of interliquid hopping between Luttinger liquids.

%To conclude, we 
%%point out 
%emphasize again 
%that 
%although this work was originally motivated, as described
%above, by the peculiar c-axis conductivity in the cuprates, our conclusions 
%will be relevant to {\em any\/} sufficiently anisotropic and sufficiently
%strongly correlated metal. 
%To the extent that our calculations have been
%done for coupled 1D liquids, they are most directly applicable to quasi-1D
%metals. 
The remainder of the paper is set out as follows. In Sec. \ref{sec:TLS}, we 
present a detailed
and germane discussion of the physics of a two level system
coupled to a dissipative bath, which provides an illustrative zero-dimensional 
analogue of the physics we wish to discuss.
Sections \ref{sec:fermions}-\ref{sec:interp} 
form the heart of the paper, in which a detailed
qualitative and quantitative discussion of the problem of weakly coupled
Luttinger liquids is presented.
In sections \ref{sec:caxis} and \ref{sec:organics}, respectively, we
present detailed arguments of why 
a state of confined coherence 
provides the only plausible way to account for 
%has been directly 
several anomalous aspects of the physics of the HTSC's and of 
%the quasi-1D 
the (highly anisotropic) organic conductor (TMTSF)$_2$PF$_6$. 
The cuprate case is perhaps of more interest to
the general reader, however the organic
case furnishes quite compelling experimental
support  for our 
%admittedly radical 
proposal.
A final conclusion is given in Sec. \ref{sec:concl}.

%% file: tls.sec
\section{Incoherence and the Two Level System}
\label{sec:TLS}

We will use the term ``incoherent'' in this paper in the same sense in
which it was used  in discussions of the two level
system, or Caldeira-Leggett, problem \cite{TLS}.
In fact,
much of what we propose is 
most clearly explained by analogies to that model
and for that reason 
we now give a self-contained, idiosyncratic review
of the model and some of its known properties. We will
use a somewhat unusual approach to the model
which is considerably less elegant than
some existing ones but which we believe to be 
transparent and, more importantly, generalizable
to the problem of fermionic hopping.

To begin with we define, following Ref. \cite{TLS},
the two level system
model with the Hamiltonian:
\begin{equation}
\label{eq:TLS_ham}
H_{\rm TLS} =  \frac{1}{2} \Delta \sigma_x + \frac{1}{2}
\epsilon \sigma_z + \sum_i  \left(\frac{1}{2} m_i \omega_i 
x_i^2  + \frac{1}{2m_i} p_i^2  \right)
+ \frac{1}{2}  \sigma_z \sum_i C_i x_i
\end{equation}
Here $C_i$ is the coupling to the $i$th oscillator,
and $m_i$, $\omega_i$, $x_i$ and $p_i$
are the mass, frequency, position 
and momentum of the $i$th oscillator,
respectively.

We restrict our discussion of the model to zero
temperature and the so called 
ohmic regime \cite{TLS} where the spectral density of 
the bath is given by:
\begin{eqnarray}
\label{eq:ohmic}
J(\omega) &= &\frac{\pi}{2} \sum_i \frac{C_i}{m_i \omega_i} 
\delta(\omega-\omega_i) \\
\nonumber
 & = & 2 \pi~ \alpha~ \omega \exp(-\omega/\omega_c)
\end{eqnarray}
$\alpha$ is a positive constant measuring the strength
of the coupling to the bath and $\omega_c$ is a cutoff
frequency. The $\alpha$ here should not be confused
with the $\alpha$
which we have previously
introduced in our discussion of the Fermi surface
in the cuprates as
defining the anomalous 
exponent of the single particle Green's function.

The two level system model describes a single quantum mechanical
degree of freedom which can be in either of two states
and which is coupled to a bath of harmonic oscillators.
We are primarily interested in the $\epsilon =0$
case and subsequent discussion refers to this case
unless otherwise stated.  
In this case, one may
think of the model as one in which a particle
tunnels with tunneling matrix element
$\Delta/2$ between two degenerate states (labelled by
$\sigma_z=\pm 1$).  The environment, represented by the bath of
oscillators,
influences the tunneling
because the bath 
is sensitive to which of the states the spin
is in.  
%In this case, one may
%think of the model as one in which a particle
%tunnels between two degenerate states 
%(labelled by $\sigma_z=\pm 1$) with 
%tunneling matrix element $\Delta$.  The environment
%influences this tunneling
%since the oscillator bath 
%is sensitive to which of the states the particle 
%is in.  
Hereafter we will refer to the discrete degree of 
freedom as a ``spin'' for convenience. This is appropriate
since we are describing the spin with Pauli matrices.

The model provides the prototypical example
of a quantum to classical  crossover, since for
$C_i =0$ the model represents the quantum mechanics
of an isolated two state system, whereas for
sufficiently strong coupling to the environment
the dynamics of the spin,
if followed without reference
to the oscillator bath, are dissipative and
no quantum coherence effects are observable \cite{TLS}.
In fact, this is how one generally expects
classical behavior to emerge for macroscopic systems:
the macroscopic degrees of freedom exchange energy with
an enormous number of unobserved microscopic degrees
of freedom and therefore 
different histories are unable to maintain
a definite relative phase long enough for 
quantum interference effects
to manifest themselves.  This is one possible solution
to the famous Schr\"odinger's cat problem of quantum mechanics,
in which apparent paradoxes arise as a result of the 
fact that
while macroscopic
objects 
%(or variables)
are never observed to be in superpositions of different states,
such superpositions are, in principle, possible according
to the standard Copenhagen interpretation of quantum mechanics.
The resolution to this problem relies on the fact that,
since 
it is impossible to measure an object in a superposition
of states with respect to the measured quantity, the real meaning of
observing an object in a superposition of states is
making an observation which is only compatible with
the object having been in a superposition of states in
the past, {\em i.e.\/} an interference experiment. 
For example, in the Young double slit experiment
the particle is not observed to be in a superposition of
having passed through each slit and, in fact, if any measurement
is made of it passing through either slit then it is not in
any such superposition. However
the interference pattern which results at the screen behind the
slits is only possible
because 
the particle was in a superposition of having passed though
both slits before being measured at the screen. 
Since the only meaningful way to detect superpositions
is through such interference experiments,
a definite phase relationship between the different
histories contributing to 
the final state is required. Hence, the interchange of
energy with the environment can totally remove the possibility
for such observations and is expected to do so for macroscopic
objects or variables (except under very special circumstances,
for example in certain  Josephson junction experiments \cite{TLS}).

What sort of quantum interference effects do we expect to
be able to observe in the TLS for sufficiently weak coupling
to the environment?  Consider a model
where the coupling to the
environment vanishes, i.e. $\alpha=0$, and the system is
prepared 
in a state where the spin is in a $\sigma_z$
eigenstate.  The exact eigenstates of the spin are 
the $\sigma_x$ eigenstates which are split by an energy
$\Delta$ so that the initial state of the system is
a superposition of these two states 
of different energy with a definite
phase between the two states in the superposition.
Since the two states have different energies, this 
phase is not time independent
(the phases evolve differently for the history
where the particle is in the symmetric state
and the history where it is in the antisymmetric
state). Furthermore, for
vanishing coupling to the environment the 
relative phase remains
well defined indefinitely. The time dependence of the phase
therefore results in observable oscillations in
the expectation value of
$\sigma_z$, in fact 
(in units where $\hbar = 1$)
$\langle \sigma_z(t) \rangle =  \cos  \Delta t$.
The oscillations are a quantum interference effect 
between histories which involve the spin spending
different amounts of time in the various
$\sigma_z$ states. In general,
we would expect such
oscillations to also occur when the spin is coupled to the environment
provided the spin is capable of
flipping without exchanging an amount of energy with
its environment sufficient for the randomization of the phase of
the history involving the spin flip.  
As we shall see, in the TLS model 
the oscillations persist for a range of couplings
to the environment,
albeit with coupling dependent
damping of the oscillations. Such damping results 
from the exchange of energy between the spin and the environment
during the course of a typical flip.

%Interference between histories can
%occur only if the spin is capable of
%flipping without exchanging an amount of energy with
%its environment sufficient for the relative
%phases of
%the various histories contributing to the final
%state to have been
%randomized.  Interference, and the resulting oscillations,
%persist for a range of couplings
%to the environment,
%albeit with coupling dependent
%damping of the oscillations resulting as more
%and more energy is exchanged with the environment
%during the course of a typical flip \cite{TLS}.  

To study these oscillations the standard
theory of the TLS focuses on the quantity
$P(t)$,
which is the probability of finding the
system in the $\sigma_z=1$ state for
$t > 0$ for a system which has been prepared by clamping the
spin into the $\sigma_z =1$ state for all $t<0$,
allowing the oscillator bath to relax to equilibrium
in this configuration, and finally releasing the spin
at $t=0$.
Note that the
calculation of $P(t)$ is equivalent to
determining $\langle\sigma_z(t)\rangle$,
since the two are simply related by 
$\langle\sigma_z(t)\rangle=2P(t)-1$ \cite{fn_P(t)}.
$P(t)$ is an appropriate quantity to study for
questions about macroscopic quantum coherence since,
if the
spin represents a generic macroscopic quantum
degree of freedom which the experimenter can observe
and control, whereas the oscillators represent
microscopic degrees of freedom which are beyond
both control and observational capacities of the
experimenter,  it is exactly the sort of
preparation used in the definition of $P(t)$ which
is possible experimentally.  The signature of quantum coherence in
$P(t)$ will be the presence of oscillations
(damped or otherwise) in contrast to the incoherent
relaxation,
$P(t)\sim \frac{1}{2}(1+e^{- \Gamma t})$,
which should result if the spin is sufficiently
strongly coupled to the environment that 
the relative phases  
associated with histories in
which tunneling between the two
$\sigma_z$ eigenstates occurs are randomized. 
A transition to
purely incoherent relaxation is indeed found to occur 
in the TLS problem, even at short times, 
%\cite{TLS}
when $\alpha > 1/2$ \cite{lowa}.
 
For further discussion of the TLS problem,
it is convenient to make a canonical transformation
on the original model 
%so that the coupling to the oscillators is removed 
by taking
\begin{equation}
H_{\rm TLS}^{\prime} = \hat{U} H_{\rm TLS} \hat{U}^{-1}
\end{equation}
where
\begin{equation}
\label{eq:xform}
\hat{U} = \exp \left(
-\frac{1}{2} \sigma_z \sum_i \frac{C_i}{m_i \omega_i^2}
\hat{p}_i \right)
\end{equation}
$\hat{p}_i$ is the momentum operator of the $i$th oscillator.
The new Hamiltonian takes the form:
\begin{equation}
\label{eq:TLS_ham_can}
H_{\rm TLS}^{\prime} = \frac{1}{2} \Delta(\sigma^+ e^{-i \Omega} + h.c.) +
H_{\rm oscillators}
\end{equation}
where $\Omega = \sum_i \frac{C_i}{m_i \omega_i^2} p_i$.
The coupling to the oscillators has been removed by the
transformation, but in the process the tunneling operator
between the
two states has been replaced by an operator which creates and
destroys excitations of the oscillator bath 
($p_i$ is expressible in terms of the creation
and annihilation operators for the excitations of
the $i$th oscillator) as well as changing
the state of the spin.
It is clear that quantum oscillations of $P(t)$
will be in danger of being destroyed should the low-frequency
oscillator density of states and/or the low-frequency couplings $C_i$
be sufficiently large.
Again we consider only the ohmic case defined by
Eq. \ref{eq:ohmic} so that the coupling is parametrized
by a single dimensionless number, $\alpha$;
the oscillations will be called into
doubt for large $\alpha$.

In this formulation, $P(t)$ can be reinterpreted as the
probability of finding $\sigma_z(t)=1$ for a system in which
$\Delta$ is suddenly switched on at time $t=0$
with the system in the $\Delta = 0$ groundstate
with $\sigma_z =1$. The previous
definition in which the spin  was clamped in  the $\sigma_z=1$
eigenstate for all negative times and the oscillators were
allowed to adapt to the clamped state is equivalent.

We begin our discussion of the physics of this model with
the two point correlation function of $\sigma^+ e^{-i \Omega}$,
which  obeys
\begin{eqnarray}
\label{eq: Omega propagator}
\langle \sigma^+ e^{-i \Omega(t)} \sigma^- e^{i \Omega(0)}\rangle & = &
\exp \left\{-\int_0^{\infty} 
\frac{1 - e^{-i\omega t}}{\omega^2}
J(\omega)\right\}\nonumber\\
& = &
\exp \left\{-2  \alpha \int_0^{\infty} 
\frac{1 - e^{-i\omega t}}{\omega} 
e^{-\omega/\omega_c} \right\}\\
\nonumber & \sim &
e^{i \pi \alpha} (\omega_c t)^{-2 \alpha}
\end{eqnarray}
 From the correlation function we can immediately construct the
spectral function of the operator $ e^{i \Omega}$
in the low energy, universal regime:
\begin{eqnarray}
\label{eq:TLSrho}
\rho_{\Omega}(\omega) & = & \sum_m |\langle m | e^{i \Omega} | GS \rangle |^2
\delta(\omega - E_m) \\
\nonumber & = & 
%\frac{\sin(2 \pi \alpha) \Gamma(1-2 \alpha)}{\pi}
\Gamma^{-1}(2 \alpha)~\theta_+(\omega)~
\omega^{-1+2 \alpha} \omega_c^{-2 \alpha}
\exp(-\omega/\omega_c)
\end{eqnarray}
where $\{m\}$ is a complete set of oscillator eigenstates with
energies $E_m$ and $|GS\rangle$ is the oscillator ground state.
The  spectral function is normalized to integrate to unity
since $\langle e^{-i \Omega(t)} e^{i \Omega(t)} \rangle = 1$.

The short time approximation to $P(t)$ can be constructed 
straightforwardly using
the spectral function above and ordinary time dependent 
perturbation theory.  We find:
\begin{equation}
\label{eq:Pt1}
P(t) = 1 -  \frac{\Delta^2}{2}  \int d\omega \rho_{\Omega}(2\omega)
\frac{\sin^2(\omega t)}{\omega^2} + \cdots
\end{equation}
Notice that when $\alpha > 1$,
$\rho_{\Omega}(\omega) \sim \omega^{-1+2 \alpha}$ results in
an infrared convergent $P(t)$; in the limit $\Delta \rightarrow
0$, $P(t) \rightarrow 1$ for all $t$. This corresponds to
the irrelevance of $\Delta$ and the localization of the
spin predicted by Chakravarty 
and Bray and Moore \cite{chakrIsing} based on
a mapping of the TLS to the inverse squared Ising model.

Conversely,
for $\alpha \rightarrow 0$, $\rho_{\Omega}(\omega)\rightarrow\delta(\omega)$
and $P(t) = 1 - \frac{\Delta^2}{4} t^2 + ...$, in agreement with the
expansion of the exact result $P(t) = ( 1 +\cos\Delta t)/2$.
For $0 < \alpha < 1$ we are in a more complicated region.
Clearly the difference between $P(t)$  and $1$ grows to
order unity for any arbitrarily small $\Delta$ throughout
this region (this simply reflects the renormalization
group relevance of $\Delta$) and one would at first sight
be tempted to conclude that throughout this
region $P(t)$ would undergo damped oscillations with
a period approximately given by $t_{\rm osc}$, where $t_{\rm osc}$
satisfies:
\begin{equation}
\label{eq:tosc}
1 = \frac{\Delta^2}{2} \int d\omega \rho_{\Omega}(2\omega)
\frac{\sin^2(\omega t_{\rm osc})}{\omega^2} 
\end{equation}
One should be cautious, however, in view of the fact that
for $\alpha > 1/2$ the spectral function
for the tunneling operator is vanishing at low frequencies
and, at $\alpha = 1/2$, it is flat and
featureless out to the cutoff scale.  A flat spectral
function is equivalent to a featureless density of
states and is exactly the condition under which 
the approximation of Fermi's Golden Rule 
should be valid, implying incoherent decay without
any recurrence effects or oscillations. 
We may scale out the time dependence in (\ref{eq:Pt1}) to obtain to
$O(\Delta^2)$ and for $\omega_c t\gg 1$
\begin{equation}
P(t) \approx 1 -  2^{2\alpha-1}\alpha\Delta^2~\omega_c^{-2\alpha}
t^{2-2\alpha}  \int_0^{\infty} dx
\frac{\sin^2x}{x^{3-2\alpha}}
\end{equation}
(for simplicity, we  have replaced the cutoff $e^{-\omega/\omega_c}$ by
a hard cutoff at $\omega_c$).
Thus,
for $\alpha > 1/2$, where the spectral function
for the tunneling operator is vanishing at low frequencies,
we see that the $O(\Delta^2)$ term in $P(t)$ grows even {\em more slowly\/}
than $t$, suggesting an even ``more incoherent'' decay of $P(t)$.
If we define $\Gamma(t)=-dP(t)/dt$, the rate at which the spin flips,
then in this regime $\Gamma(t)$ is bounded for all $t$. For the special
value $\alpha=1/2$, $\Gamma(t)=\Gamma$, a constant, and a naive
re-exponentiation
of the Golden Rule is $P(t)=(1+e^{-\Gamma t})/2$, corresponding to
$\langle\sigma_z(t)\rangle=e^{-\Gamma t}$. For $\alpha > 1/2$ it would
appear reasonable to expect exponential relaxation, too. A
self-consistent
approximation to determining the relaxation rate $\Gamma$ for small
$\Delta/\omega_c$ involves cutting off the $\omega$-integral at
$\omega\sim\Gamma$
to give $\Gamma\sim\Delta^2\omega_c^{-2\alpha}\Gamma^{2\alpha-1}$
yielding
\[
\Gamma\approx\Delta\left(\frac{\Delta}{\omega_c}\right)^{\alpha/(1-\alpha)}
\]
The right hand side is in fact nothing but $\Delta_{\rm ren}$, the
renormalized
tunneling rate which emerges from an RG analysis.

The true behavior of $P(t)$ in the region $1/2 < \alpha <1$ is actually
not
rigorously known \cite{TLS}, but there are reasons for believing that
the self-consistent argument given above is not too far from the truth. 
The true decay of $P(t)$ is probably not simply exponential relaxation, but
the
key point is that there are not any oscillations. Thus, despite the fact
that
the naive RG approach yields the same scale $\Delta_{\rm ren}$ as the
self-consistent
approach, it fails to distinguish between an essentially {\em
coherent\/}
$\Delta_{\rm ren}$ and a completely {\em incoherent\/} one.

For $0<\alpha<1/2$, $\Gamma(t)$ is unbounded and any attempt at
characterizing $P(t)$ by exponential relaxation fails, as indeed
it must as $\alpha\rightarrow 0$.
%It is believed \cite{TLS} that
%the true behavior of $P(t)$ in this region of $\alpha$ is a damped
%oscillation with oscillation frequency
%$\omega_{\rm osc}=\cos(\pi\alpha/(2-2\alpha))$ and damping
%$\Gamma=\sin(\pi\alpha/(2-2\alpha))$, plus an incoherent background.
But what is the correct interpretation of the behavior $1-P(t)\sim
t^{2-2\alpha}$?
And can we go beyond lowest order in $\Delta$? The simplest calculation
addressing these questions,
a kind of random-phase approximation, is the so-called non-interacting
blip
approximation (NIBA). The reader is referred to \cite{TLS}
for the more standard approach to this approximation, which also reveals
the
reason behind its curious name. Here, we shall present a very simple
derivation
which has the advantage of making the key approximations of the NIBA
explicit \cite{aslungal}.

Consider the canonically transformed TLS Hamiltonian
(\ref{eq:TLS_ham_can}).
The equations of motion of the spin are
\begin{eqnarray}
\partial_t
\sigma_z&=&-i\Delta\{e^{-i\Omega}\sigma^+-e^{i\Omega}\sigma^-\}
\nonumber\\
\partial_t \sigma^{\pm}&=&\mp\frac{i\Delta}{2}\sigma_z e^{\pm i\Omega}
\nonumber
\end{eqnarray}
The equations for $\sigma^{\pm}$ can be formally integrated
\begin{equation}
\sigma^{\pm}(t)=\sigma^{\pm}(0)\mp\frac{i\Delta}{2}
\int_0^t dt' \sigma_z(t')e^{\pm i\Omega(t')}
\end{equation}
which can then be substituted into the equation for $\partial_t
\sigma_z$.
Taking expectation values yields
\begin{equation}
\left\langle\frac{d\sigma_z}{dt}\right\rangle
=-\Delta^2\int_0^t dt' \left\langle\sigma_z(t')D(t-t')\right\rangle
\end{equation}
where
\[
D(t-t')=\frac{1}{2}\left\{e^{-i\Omega(t)}e^{i\Omega(t')}
+e^{i\Omega(t)}e^{-i\Omega(t')}\right\}
\]
The NIBA amounts to {\em neglecting correlations\/} in the expectation
value under
the integral, {\em i.e.\/} the replacement
\[
\left\langle\sigma_z(t')D(t-t')\right\rangle
\rightarrow\langle\sigma_z(t')\rangle\langle D(t-t')\rangle
\]
is made. Defining $D_0(t-t')\equiv\langle D(t-t')\rangle$, we arrive
at the integral equation
\begin{equation}
\frac{d\langle\sigma_z\rangle}{dt}=-\Delta^2\int_0^t dt' D_0(t-t')
\langle\sigma_z(t')\rangle
\label{eq: NIBA_int_eq}
\end{equation}

The formal solution of (\ref{eq: NIBA_int_eq}) is
\begin{equation}
\langle\sigma_z(t)\rangle=\langle\sigma_z(0)\rangle
\sum_{n=0}^{\infty}(-\Delta^2)^n
\int_0^t dt_1\int_0^{t_1} dt_2\ldots\int_0^{t_{2n-1}} dt_{2n}
D_0(t_1-t_2)D_0(t_3-t_4)\ldots D_0(t_{2n-1}-t_{2n})
\end{equation}
however the integral equation may also be solved by introducing Laplace
transforms, ${\cal L}(f(t))\equiv\hat{f}(s)$. Taking the spin to be up
at
$t=0$ gives
\begin{equation}
\hat{\sigma_z}(s)=\frac{1}{s+\Delta^2 \hat{D_0}(s)}
\label{eq: NIBA}
\end{equation}

Equation (\ref{eq: NIBA}) is the key result of the NIBA. The
determination of
$\sigma_z(t)$ is reduced to the calculation of $D_0(s)$,
which is essentially the Laplace transform of the propagator in equation
(\ref{eq: Omega propagator}), followed by the inverse Laplace transform
of $\hat{\sigma_z}(s)$. From (\ref{eq: Omega propagator}) it can be
shown that
\begin{equation}
\hat{\sigma_z}(s)=
\frac{s^{1-2\alpha}}
{\left(s^{2-2\alpha}+\Delta_{\rm eff}^{2-2\alpha}
\right)}
\end{equation}
where
\[
\Delta_{\rm eff}\equiv[\Gamma(1-2\alpha)\cos\pi\alpha]^{1/(2-2\alpha)}
\Delta\left(\frac{\Delta}{\omega_c}\right)^{\frac{\alpha}{1-\alpha}}
\]
is, up to a numerical prefactor, the renormalized hopping rate
$\Delta_{\rm ren}$ as given by the RG calculation.

We now briefly discuss the various cases
(more details are provided in \cite{TLS}), beginning with the two
simplest.

(i) $\alpha=0$ - here we have $\hat{D_0}(s)=1/s$, yielding the correct
result
for a free spin, $\langle\sigma_z(t)\rangle=\cos\Delta t$.

(ii) For $\alpha=1/2$, $\hat{\sigma_z}(s)=(s+\Delta_{\rm eff})^{-1}$,
whence $\langle\sigma_z(t)\rangle=\exp(-\Delta_{\rm eff}\,t)$.
Of course, this corresponds precisely to the situation in elementary
quantum mechanics where one would invoke Fermi's Golden Rule, there
being
a constant density of states for the system to ``decay'' into. Indeed,
the NIBA is equivalent to simply exponentiating the decay rate
$\Delta_{\rm eff}$.
Again, we emphasize that at this value of $\alpha$, $\Delta$ is RG
relevant, yet clearly its effect is purely incoherent.

(iii) In the region $0<\alpha<1/2$ \cite{lowa}, $\hat{\sigma_z}(s)$ acquires a
pair of poles on the principal Riemann sheet. A straightforward Laplace
inversion
yields the following contribution to $\langle\sigma_z(t)\rangle$
\begin{equation}
\langle\sigma_z(t)\rangle_{\rm poles}
=\frac{1}{(1-\alpha)}\cos\left\{\cos\left[
\frac{\pi}{2}\frac{\alpha}{(1-\alpha)}\right]
\Delta_{\rm eff}\,t\right\}
\exp\left\{-\sin\left[
\frac{\pi}{2}\frac{\alpha}{(1-\alpha)}\right]
\Delta_{\rm eff}\,t\right\}
\end{equation}
There is also a branch cut in $\hat{\sigma_z}(s)$ which gives a
nonoscillatory contribution to $\langle\sigma_z(t)\rangle$.
Thus, $P(t)$ in this region of $\alpha$ is the sum of a damped
oscillation, with oscillation frequency
$\omega_{\rm osc}=\cos(\pi\alpha/(2-2\alpha))\Delta_{\rm eff}$ and
damping
$\Gamma=\sin(\pi\alpha/(2-2\alpha))\Delta_{\rm eff}$, and an incoherent
background.
As $\alpha\rightarrow 0$, one approaches the free spin limit,
$\langle\sigma_z(t)\rangle=\cos\Delta t$, while as $\alpha\rightarrow
1/2$,
the oscillation frequency
$\omega_{\rm osc}\sim
\pi(1-2\alpha)\Delta_{\rm eff}\rightarrow 0$, and the limit of purely
exponential relaxation is reached.

(iv) For $1/2<\alpha<1$, the NIBA is not believed to be
valid \cite{TLS} because it predicts a non-universal
power law decay of the $\sigma_z$ two point function,
in contradiction to the connection between the TLS
problem and the Kondo problem.
We give the NIBA result for completeness,
however we believe that the true result
in this regime, at least at intermediate times,
is more likely exponential decay of
$\langle \sigma_z(t) \rangle$, as at  $\alpha = 1/2$.
In any case, within the NIBA one finds that
$\hat{\sigma_z}(s)$ has no poles on the
principal Riemann
sheet, and as a result $P(t)$ is purely incoherent \cite{TLS},
\begin{equation}
\langle\sigma_z(t)\rangle=-\frac{\sin 2\pi\alpha}{\pi}
\int_0^{\infty}dx\,
\frac{x^{2\alpha-1}\,e^{-(\Delta_{\rm eff}\,t)\,x}}
{x^2+2x^{2\alpha}\cos 2\pi\alpha+x^{4\alpha-2}}
\end{equation}

The results of the NIBA are therefore consistent with the qualitative
behavior one could intuit from the perturbative calculation of $1-P(t)$
to $O(\Delta^2)$.
% especially the perturbative calculation at $\alpha=0$,
%$\alpha=1/2$ and $\alpha>1$.
The key result is the existence of the phase $1/2<\alpha<1$ where
$\Delta$ is a relevant, but incoherent, operator.

At this point, it is useful to
%take a more physical approach to the TLS,
make some observations regarding the TLS which
%in order to
%be able to extract
help to emphasize
the general physics of incoherence.
%This will be helpful
%for generalization
%to other problems.
%To this end,
In the TLS model, the important
physical
effect of finite $\alpha$ is that there is a substantial contribution
to $P(t)$ from transitions to states with energies that are
larger than the putative renormalized oscillation frequency,
$\Delta_R \sim t_{\rm osc}^{-1}$ (see Eq. \ref{eq:tosc}) .
When the amount of weight in these transitions is larger
than the amount of weight in transitions to low energy
states, it no longer makes sense to consider the effects
of $\Delta$ to be coherent. Effectively, each change of state,
{\em i.e.\/} flipping of the spin,
is accompanied by the creation or annihilation of a sufficient
number of bosons in the environmental bath that the 
phase of that history is randomized compared to histories
with no spin flip.
Intuitively, {\em one has crossed over
from degenerate or nearly degenerate perturbation theory to
non-degenerate perturbation theory} (as opposed to the 
transition to irrelevant $\Delta$ where the long time
perturbation theory becomes convergent). To illustrate the point, first note that
we can calculate $\delta P(t)\equiv 1-P(t)$
to $O(\Delta^2)$ exactly:
\begin{eqnarray}
\delta P(t)& = &  \frac{\Delta^2}{2}  \int_0^{\infty}
d\omega\,\rho_{\Omega}(2\omega)
\frac{\sin^2(\omega t)}{\omega^2} + \cdots \nonumber \\
&=& \Delta^2\,\Gamma^{-1}(2 \alpha)\,
t^{2-2\alpha}2^{2\alpha-2}\int_0^{\infty}d\theta\,
e^{-2\theta/t}\theta^{2\alpha-3}\sin^2\theta +O(\Delta^4)\\
&=&\Delta^2
\frac{[1-(1+t^2)^{1-\alpha}\,\cos[2(\alpha-1)\arctan(t)]]}
{4(1-2\alpha)(1-\alpha)} +O(\Delta^4)\nonumber
\end{eqnarray}
For times much longer than the inverse
cutoff, $\omega_c^{-1} \equiv 1$, and for $0 < \alpha < 1$
we can use
\begin{equation}
\delta P(t)\sim\frac{\Delta^2}{4}
\frac{\cos\pi\alpha}{(1-2\alpha)(1-\alpha)}\,
t^{2-2\alpha}
\end{equation}

If we now
separate the contributions from the frequencies which
are less than $t^{-1} = \Delta_R = \Delta^{1/(1-2\alpha)}$
from those that are 
greater than $ t^{-1}$ \cite{divnote} then
(again,
for $t\gg\omega_c^{-1}\equiv 1$ and $0<\alpha<1$)
 we find that the low energy part
contributes to $P(t)$ an amount
\begin{eqnarray}
\delta P_{\rm low}(t) & = &  \frac{\Delta^2}{2}  \int_0^{1/t}
 d\omega \rho_{\Omega}(2\omega)
\frac{\sin^2(\omega t)}{\omega^2} + \cdots \nonumber \\
& \sim &
\frac{\Delta^2}{2^{2-2\alpha}} t^{2-2\alpha} \Gamma^{-1}(2 \alpha) ~ 
\frac{_1F_2(-1+\alpha;\frac{1}{2},\alpha;-1) -1}
{4 (1-\alpha) }
\end{eqnarray}
where $_1F_2$ is a generalized hypergeometric function
\cite{gradrhyz}.
This evaluates to $\frac{1}{4} \Delta^2 t^2$ at $\alpha=0$
and, for $\alpha = 1/2$, to
$\frac{1}{4}\Delta^2 t\,(2\,{\rm Si}(2)+\cos 2 -1)\approx 0.45 \Delta^2 t$
(${\rm Si}$ is the sine integral function).

The high energy part contributes
\begin{eqnarray}
\delta P_{\rm high}(t) & = &  \frac{\Delta^2}{2}  \int_{1/t}^{\infty}
 d\omega \rho_{\Omega}(2\omega)
\frac{\sin^2(\omega t)}{\omega^2} + \cdots \nonumber \\
%\nonumber & \sim &
%\Delta^2 t^{4(1-\alpha)} 2^{4(\alpha-1)} 
%\frac{ \left( 1 - 
%(1+t^2)^{1-\alpha} \cos\left(2 (\alpha-1)\arctan(t) \right) \right)}
%{ 4(1-2 \alpha)(1-\alpha)} \\
%\nonumber
%& & - \frac{\Delta^2}{2^{2-2\alpha}} t^{2-2\alpha} \Gamma^{-1}(2 \alpha) ~
%\frac{_1F_2(-1+\alpha;\frac{1}{2},\alpha;-1) -1}
%{4 (1-\alpha) }
%\end{eqnarray}
%for times much longer than the inverse
%cutoff, $\omega_c^{-1} \equiv 1$, and $0 < \alpha < 1$
%we can use:
%\begin{eqnarray}
%\nonumber
%\delta P_{\rm high}(t)  & =  &
&\sim&\frac{\Delta^2}{2^{2-2\alpha}} t^{2-2\alpha} \Gamma^{-1}(2 \alpha)
%\\ \nonumber & & 
\left(\frac{1}{4 (1-\alpha)  } +
%\frac{ \csc(\alpha \pi) \pi}{2^{2 \alpha} \Gamma(3 - 2 \alpha) }
\frac{\Gamma(2\alpha)\cos\pi\alpha}{2^{2\alpha}(1-2\alpha)(1-\alpha)}
-\frac{_1F_2(-1+\alpha;\frac{1}{2},\alpha;-1)}
{4 (1-\alpha)  } \right)
\end{eqnarray}
For $\alpha \rightarrow 0$ the high energy part vanishes like
$\Gamma^{-1}(2 \alpha)$ (the prefactor is
$\frac{1}{4} \Delta^2 t^2 \left( \frac{3}{2} - \gamma -\ln2 +
%\frac{_2F_3(1,1;2,\frac{5}{2},3;-1)}{6} \right)
\frac{1}{6}~_2F_3(1,1;2,\frac{5}{2},3;-1) \right)
\approx 0.1 \Delta^2 t^2$), while for $\alpha =1/2$
the result is 
$\frac{1}{2\pi}\Delta^2 t\,(1+\pi-_1F_2(-\frac{1}{2};\frac{1}{2},\frac{1}{2};-1)
\approx 0.34 \Delta^2 t$,
%$\frac{1 + \pi - _1F_2(-\frac{1}{2};\frac{1}{2}),
%\frac{1}{2};-1}{2 \pi} \sim .34 \Delta^2 t$,
comparable to the low energy contribution.

Clearly, the high energy contribution
is insignificant as $\alpha \rightarrow 0$ because of
the divergence of $\Gamma(2 \alpha)$ (the gamma function has
a simple pole at $0$) and,
for any arbitrary division into ``high'' and ``low'',
could always be made so by taking a
suitably small $\alpha$.
%energy pieces. 
One therefore expects to find coherence in the
limit $\alpha \rightarrow 0$. On the other hand, for finite values
of $\alpha$ the high energy part can be as important as the low energy part,
depending upon our division into high and low energy integrals.
Using the qualitatively reasonable division above, we
see that for $\alpha = 1/2$ the two contributions
are in fact comparable.  
In fact, for any division scheme involving energy scales small
compared to the oscillator cutoff,
the high energy part must dominate for
some $\alpha < 1$, since, in 
the limit where $\alpha \rightarrow 1$,
the high energy part diverges logarithmically like
%$\frac{1}{4}\Delta^2\ln(t)$ 
$\frac{\Delta^2}{2  \omega_c^{2}}
\ln(\omega_c t)$
while the low energy part is finite
and given by 
%$\frac{1}{4}\Delta^2
%(\gamma + \ln(2) - {\rm Co}(2)) \approx 0.21~\Delta^2$,
$\frac{\Delta^2}{2  \omega_c^{2}}
(\gamma + \ln(2) - {\rm Co}(2)) \approx 0.85~\frac{\Delta^2}{2 \omega_c^{2}}$,
where ${\rm Co}$ is the cosine
integral function and $\gamma$ is Euler's constant.
The high energy part can therefore
be made arbitrarily large compared
to the low energy part for any arbitrary  partition into
high and low energy pieces as we approach $\alpha = 1$.  
The dominance of the high energy part does not necessarily
imply that the quantum oscillations must cease entirely;
it could be that the oscillations would
persist but become arbitrarily heavily damped.
However, when the high energy part has become of order one,
the argument that oscillations should occur with a frequency
$\omega_{\rm osc} \sim \Delta_R$ becomes unreliable and,
in fact, as we have seen above, the conclusion of the NIBA (which
is known to be correct at $\alpha = 1/2$ from 
the exact solution \cite{TLS})
is that the oscillations 
vanish for $\alpha = 1/2$ \cite{lowa}.

The reason for the success of our perturbation theory,
which is essentially a ``short time expansion'', in
predicting a qualitative change in the tunneling
is that the expansion is valid out to precisely the time when
the spin has order one probability of flipping and is therefore
perfectly adequate to describe the nature of the states reached
by spin flip processes. In particular, it can identify whether
these states
are nearly  degenerate with the initial state
(and each other)
or of widely disparate energies,
which is the essential physical question for
coherence.
Hence, the main conclusions of this section:
the qualitative
behavior of $P(t)$, in the sense of whether or not it exhibits
oscillations,
{\em i.e.\/} quantum coherence, can actually be determined
from lowest order perturbation
theory. The special point $\alpha=1/2$, at which the Golden Rule is
naively applicable, separates a region of completely incoherent
behavior,
$1/2\leq\alpha <1$, from one of damped oscillations, $0<\alpha<1/2$.

%% file: fermions.sec
\section{The Connection to Fermionic Hopping}
\label{sec:fermions}

The existence of a third regime in the TLS problem
with behavior 
qualitatively different from that occurring for 
irrelevant tunneling or undamped tunneling is suggestive,
however, before we can claim that the lessons of the
TLS have any relevance to the
experimental peculiarities observed in, for example, the cuprates,
we must make some firmer connection between the tunneling matrix
element, $\Delta$, and the interplane, single particle hopping,
$t_{\perp}$. 
The experiments we are
interested in involve coupling planes of interacting
electrons together with a single electron hopping
operator $O_{hop} = \sum_{\vec{k},\sigma,\langle ij \rangle}
t_{\perp}(\vec{k}) c^{\dagger}_{\sigma,i}(\vec{k})
c_{\sigma,j}(\vec{k})$, where the sum over $\langle ij \rangle$
is a sum over nearest neighbor planes.
Unfortunately, as we will see, any interesting
outcome (an outcome other than three dimensional
Fermi liquid theory) will require an interesting
(i.e. non-Fermi liquid) starting point and
there at present exists no solid non-Fermi liquid
framework in two dimensions.  We therefore study
the problem of coupled one dimensional chains
rather than planes.  Once we have elucidated a connection
to the TLS in this context and understand 
what is required to produce interesting physics
in this case, we will be in a position to discuss
what might be required for similar possibilities 
to be realized for coupled planes.

It is a not entirely trivial matter to determine {\em how\/} one should go about
investigating the coherence/incoherence issue for coupled electron liquids.
The essential question is, roughly speaking, whether or not we need to use
degenerate perturbation theory in dealing with the
action of the interliquid hopping term.
Given that the excitation spectrum of a Luttinger liquid is gapless, one
might at first think that degenerate perturbation theory is unavoidable.
However, a moments reflection on the physics of the TLS will convince the
reader that such is not necessarily the case.

We begin with the coupling of a pair of one dimensional,
interacting electronic systems with a transverse hopping.
We will study this problem using
bosonization techniques \cite{bosonization,duncan}.
General, gapless, one dimensional interacting electronic systems and 
higher dimensional Fermi liquids can both be studied
via this approach so our results can be made at least
that general.

The Hamiltonians of the isolated systems are of the
general Luttinger liquid form:
\begin{eqnarray}
H & = & 
\frac{1}{4 \pi} \int dx~ \left(
v_{\rho} K_{\rho} (\partial \Theta_{\rho})^2 +
v_{\rho} K_{\rho}^{-1} (\partial \Phi_{\rho})^2 +
v_{\sigma} K_{\sigma}  (\partial \Theta_{\sigma})^2 +
v_{\sigma} K_{\sigma}  (\partial \Phi_{\sigma})^2
\right)
\label{eq:ham_hubb} \\
\nonumber
 & = &  
\frac{1}{4 \pi} \int dx~ \left(
v_{\rho,N}  (\partial \Theta_{\rho})^2 +
v_{\rho,J}  (\partial \Phi_{\rho})^2 +
v_{\sigma,N}   (\partial \Theta_{\sigma})^2 +
v_{\sigma,J}   (\partial \Phi_{\sigma})^2
\right)
\end{eqnarray}
where $K_{\rho}$ is interaction dependent and less than
one for  a repulsive interaction, while $K_{\sigma}$ is
set to one hereafter as a consequence of considering
only interactions which preserve the $SU(2)$ spin invariance.
The bosonized form for the
electron operator is given by
\begin{equation}
\Psi^{\dagger}_{\uparrow}(x) \sim
\sqrt{ \frac{\partial \Phi_{\uparrow}(x)}{ \pi}}
\sum_{m ~odd} A_m ~\exp\left(i [m \Phi_{\uparrow}(x) +
 \Theta_{\uparrow}(x)]\right)
\label{eq:elop}
\end{equation}
so that the  inter-liquid hopping
is given by:
\begin{eqnarray}
\label{eq:hopping}
\Psi^{\dagger,(1)}_{\uparrow}(x) \Psi^{(2)}_{\uparrow}(x)  & \sim &
\sqrt{ \frac{\partial \Phi^{(1)}_{\uparrow}(x)}{ \pi}}
\sum_{m ~odd} A_m ~\exp\left(i [m \Phi^{(1)}_{\uparrow}(x) +
 \Theta^{(1)}_{\uparrow}(x)]\right) \\
\nonumber & \times  & 
\sqrt{ \frac{\partial \Phi^{(2)}_{\uparrow}(x)}{ \pi}}
\sum_{m ~odd} A_m^{\star} ~\exp\left(i [m \Phi^{(2)}_{\uparrow}(x) -
 \Theta^{(2)}_{\uparrow}(x)]\right)
\end{eqnarray} 
where 
\begin{equation}
\Theta_{\uparrow} = 2^{-\frac{1}{2}} (\Theta_{\rho} + \Theta_{\sigma})
\end{equation}
\begin{equation}
\Theta_{\downarrow} = 2^{-\frac{1}{2}} (\Theta_{\rho} - \Theta_{\sigma})
\end{equation}
and
\begin{equation}
\Theta_{\rho}(x) = \Theta_{\rho}^0 + N_{\rho} x/L -i \sum _{q \neq 0} 
|\frac{2 \pi}{q L}|^{\frac{1}{2}} K_{\rho}^{-\frac{1}{2}} ~{\rm sgn}(q)
e^{i q x} \left( b^{\dagger}_{\rho}(q) + b_{\rho}(-q) \right)
\end{equation}
\begin{equation}
\Phi_{\rho}(x) = \Phi_{\rho}^0 + J_{\rho} x/L -i \sum _{q \neq 0} 
|\frac{2 \pi}{q L}|^{\frac{1}{2}} K_{\rho}^{\frac{1}{2}} ~{\rm sgn}(q)
e^{i q x} \left( b^{\dagger}_{\rho}(q) - b_{\rho}(-q) \right)
\end{equation}
where the $b_{\rho}$ operators create and annihilate the bosonic,
charge density eigenexcitations.
Similar expressions obviously apply for $\Theta_{\sigma}$ and $\Phi_{\sigma}$.
The expression for
hopping of down spin electrons is easily obtained by changing
the sign of $\Theta_{\sigma}$ and $\Phi_{\sigma}$
in Eq. \ref{eq:hopping}, while that for
hops in the other
direction can be obtained by interchanging the chain labels in
\ref{eq:hopping}.

In the above expressions the operators $\Theta^0_{\uparrow}$
and $\Phi^0_{\uparrow}$
are canonically conjugate to the the conserved quantum numbers
$J_{\uparrow}$ and $N_{\uparrow}$ and are not expressible in 
terms of the bosons.  The role of these operators was stressed
by Haldane in his solution of the Luttinger model \cite{duncan},
however they are generally ignored since they do not enter into
single particle correlation functions. They will be crucial for
our discussion since it is the quantum numbers 
$N$ and $J$ that are analagous to $\sigma_z$ in the TLS problem.
This is readily apparent when the canonically transformed
form for the tunneling matrix element,
$\frac{\Delta}{2} \sigma^{+} e^{-i \Omega} + ~h.c.$
(see Eq. \ref{eq:TLS_ham_can}), is compared to the bosonized form
for the interchain hopping in Eq. \ref{eq:hopping}.  Both
contain operators which act to raise and lower otherwise 
conserved quantum numbers ($\sigma_z$ for the TLS and 
$N_{\uparrow,1}-N_{\uparrow,2}$, $J_{\uparrow,1}-J_{\uparrow,2}$,
etc. for the fermion hopping).  In addition to the raising
and lowering operators, both contain exponentials in bosonic
creation and annihilation operators which are responsible for
the interesting dynamics and determine the correlation functions
of the operators.  In fact, if the fermionic hopping occurred at only
a single point in space, that problem could be mapped onto
the TLS problem.  We do not believe that such a mapping exists
for the physical model of a uniform interchain hopping.
However, the formal connection
between the models is quite strong and suggests that the interesting
incoherent regime for the TLS might well have a fermionic analogue.

Let us now construct the analogue to $P(t)$ for the fermionic problem.
As we argued above the natural analogy  maps the Tomonaga bosons
to the harmonic oscillator bath and the conserved quantum  numbers
$N$ and $J$ to $\sigma_z$.  In this case a clear analogy exists
to  $P(t)$,as defined for the canonically
transformed TLS Hamiltonian of  
Eq. \ref{eq:TLS_ham_can}.  Instead of taking a system adapted to 
$\Delta =0$ with $\sigma_z = 1$ and then turning on $\Delta$ suddenly,
we adapt a system with some non-zero values for
$N_{\uparrow,1}-N_{\uparrow,2}$,
$J_{\uparrow,1}-J_{\uparrow,2}$, etc. to $t_{\perp}= 0$ and then turn on
$t_{\perp}$ suddenly.  Instead of studying the resulting oscillations
(or lack thereof)
in $\sigma_z(t)$ we study them in $N_{\uparrow,1}(t) - N_{\uparrow,2}(t)$,
etc.  For simplicity we will hereafter consider only the case
where  the initial condition has
$N_{\uparrow,1} - N_{\uparrow,2}=J_{\uparrow,1} - J_{\uparrow,2}=
N_{\downarrow,1} - N_{\downarrow,2}=J_{\downarrow,1} - J_{\downarrow,2}
=  \delta N(t=0) \neq 0 $, i.e. equal numbers of up and down spin electrons
are added at the right Fermi point of one chain. 
We will follow the dynamics of $\langle \delta N(t \neq 0) \rangle$.

This is a somewhat unfamiliar approach to studying $t_{\perp}$ so it is worth
examining the results for the simple case of free fermions.  In that case,
the problem is exactly soluble. The requirement that the two chains 
be prepared in
states adapted to $t_{\perp}= 0 $ and
in which no Tomonaga bosons are excited but in which
$\delta N(t=0) \neq 0 $ is easily satisfied by simply taking
$n_{1,\sigma}(k) =\Theta ( k_F - k+\frac{2 \pi}{L} \delta N(t=0))\Theta (k_F + k)$
while
$n_{2,\sigma}(k) =\Theta ( k_F - k )\Theta (k_F + k)$.
Since the free fermion problem is a single particle one
every $k$ is independent and independent oscillations
occur for the $\delta N(t=0)$ states for which
$n_{1,\sigma}(k) - n_{2,\sigma}(k) \neq 0$.
The exact result for $\langle \delta N(t) \rangle$ is 
$\delta N(t=0) \cos(2 t_{\perp} t)$.  This is exactly analogous to the 
$\alpha=0$ case and clearly corresponds to the interchain hopping
being coherent.  Given this coherence,
it is reasonable to expect the correct 
eigenstates of the system 
to be built out of single particle states
involving superpositions of 
the particles on different chains; 
this should occur even though
these states
are connected only by $t_{\perp}$. 
In fact the exact groundstate in this case is built up
with exactly this sort of 
the creation operators for single particle
eigenstates, specifically the symmetric and antisymmetric
superpositions of the single particle eigenstates
of the individual chains. 
%The ground state thus involves states with different
%possible numbers of particles on each chain
%{\it and} the phases of these states are all 
%well defined in relation to each other.
Because of this, we have 
\begin{eqnarray}
\langle c^{\dagger}_1(k) c_2(k) & =  &
\frac{1}{2} \langle n_S(k) - n_A(k) +
c^{\dagger}_S(k) c_A(k) - c^{\dagger}_A(k) c_S(k) \rangle
\\ \nonumber & = &  O(1) \\
\nonumber & \gg & O(t_{\perp}/E_F)
\end{eqnarray}
From this it is clear
that the spectral function for the interliquid
hopping acting on the
ground state also
acquires a delta function
term with finite weight.
The coherence is thus manifest in the
nature of the groundstate.

The Fermi liquid case
is very similar to the free fermion case, but
there are a number of added complexities.
To deal with these the natural language
to use is that of spectral functions, as
it was for the case of  the TLS.  In a Fermi liquid,
the typical one electron spectral function 
consists, for momenta near the 
Fermi surface, of a $\delta$-function at energy $\epsilon(k)$
with weight $Z$
and a broad incoherent background of weight $(1-Z)$ which vanishes
near $\omega=0$ at least as fast as $\omega^2$.  The typical
spectral function for an operator $c^{\dagger}_1(k)c_2(k)$ then
consists of a $\delta$-function at $\omega=0$ \cite{fkkworry}
with weight $Z^2 \Theta ( k_F^1 - k ) \Theta (k - k_F^2) $.
The background contains a piece with weight $(1-Z)^2$ vanishing
at small $\omega$ like $\omega^5$ and a piece with weight
$ 2Z(1-Z)$  which is zero for all 
$\omega < \epsilon(k)- E_F^1 + E_F^2$
and vanishes linearly with 
$\omega - \epsilon(k)- E_F^1 + E_F^2$.
In a quasiparticle picture, $t_{\perp}$ would consist primarily
 of an
operator which hopped single quasiparticles $k$ diagonally
with matrix element $t_{\perp} Z$. 
There would also be terms that removed
one quasiparticle from one chain and inserted two quasiparticles
and a quasihole in the other while conserving momentum and so on.
The first operator results in the $\delta$-function 
in the spectral function, while the second results in the
term which is zero for all 
$\omega < \epsilon(k) - E_F^1 + E_F^2$
and vanishes quadratically with 
$\omega - \epsilon(k)- E_F^1 + E_F^2$ and so on.
When summed over $k$ the second term vanishes
linearly in $\omega$ since the coefficient at a given
$k$ at small $\omega$ is proportional
to $\left( \omega - \epsilon(k) - E_F^1 + E_F^2 \right)^{-2}$.
This is fast enough that,
for small $t_{\perp}$, only the delta function term is
important because of the vanishing of the other
contributions to the spectral
function at low frequency.  To see this,
examine the slowest vanishing contribution, the
term quadratic in $\omega - \epsilon(k) - E_F^1 + E_F^2$,  
which, 
summed over $k$, leads to a 
contribution to the spectral function
of $\sum_k c_1^{\dagger}(k) c_2(k)$
vanishing like $\omega$ for 
$\omega \gg v_F \delta N(t=0)$.  The contribution of
this term to the probability to have made a transition is
\begin{eqnarray}
P_{\rm next leading}(t) & \propto & \int d\omega \omega 
\frac{\sin^2(\omega t)}{\omega^2 } \\
 & \propto & \int d\omega  \omega^{-1} \sin^2(\omega t)
\nonumber
\end{eqnarray}
which is only logarithmically divergent
in the infrared at long times, signalling that the
operator is only marginally relevant. As one would expect,
this can be made arbitrarily smaller than the contribution
from the relevant operator. This occurs because  
the spectral function for the marginal one
contains two inverse powers of $E_F$ and thus
at long times contributes only
\begin{equation}
P_{\rm marginal} \sim L t_{\perp}^2 \ln[\delta N(t=0)] E_F^{-2}
\end{equation}
which is much smaller than the contribution from
the relevant operator:
\begin{equation}
P_{\rm relevant} \sim \delta N(t=0)\, t_{\perp}^2  t^2
\end{equation}

For $\omega < v_F \delta N(t=0)$ the leading correction
to the single quasiparticle part of the spectral function
is finite but this leads again
only to a correction
that is 
\begin{equation}
P_{{\rm low} \omega} \sim  L t_{\perp}^2 t^2 (v_F \delta N(t=0))^2 E_F^{-2}
\end{equation}
at short times and
\begin{equation}
P_{{\rm low} \omega} \sim  L t_{\perp}^2 t v_F \delta N(t=0) E_F^{-2}
\end{equation}
at long times.
Both are much smaller than the
quasiparticle contribution.

In general,
an operator whose spectral function vanishes 
faster than linearly will lead to an infrared
convergent contribution
to $P(t)$, corresponding directly to the renormalization
group definition of irrelevance.  
The higher order (in terms of quasiparticles)
contributions to $\rho$ are thus irrelevant.

We can therefore neglect the
irrelevant and marginal
contributions to the spectral function of
the single particle hopping operator (though these can
certainly generate finite renormalizations they will not
change the physics qualitatively) provided that
$t_{\perp}$ is small enough.  
The contribution to $P(t)$ coming from the delta function
in the single particle hopping spectral function then
leads to the conclusion that a Fermi liquid behaves,
at lowest order, in  the
same way as non-interacting particles, except that
the hopping matrix element $t_{\perp}$ is renormalized
by a factor of $Z$.  Since, in this case,
the important hops occur only in one direction,
the leading behavior of $\langle \delta N(t) \rangle$
is the same as $P(t)$.  
Further, since only the quasiparticle contribution
is important, 
we expect that the behavior at higher order
will be simple: $\langle \delta N(t) \rangle$
will oscillate with frequency $Z t_{\perp}$.
With modest assumptions about ergodicity
it is also clear that,
at higher order, $P(t)$ must vanish in the thermodynamic limit 
after times of order $L^{-1}$ since the 
system will never return to its original microstate,
but it is possible that observable oscillations
might be present in $P(t)^{\frac{1}{L}}$, signalling
coherence.  If present, these oscillations will
also be at frequency $Z t_{\perp}$. 

These results are entirely consistent with the
usual picture for a Fermi liquid where the quasiparticles
dominate the physics.  In fact in going  to higher 
orders the only difference that appears
in the Fermi liquid calculation is the presence of
interactions among the quasiparticles.  It is also
easy to
see that if the Landau interaction parameters remain
finite, then the effect of the interactions 
on the response is proportional
to the number of quasiparticles involved in the
response, so that in the limit of small
$t_{\perp}$ and small $\delta N(t=0)$, the effects of 
the interactions  on $\langle \delta N(t) \rangle$
are negligible and the free particle picture with 
its coherent oscillations at 
a renormalized $t_{\perp}$ obtains.
Again, the coherence of the response to $t_{\perp}$
is consistent with the fact that the groundstate
of the coupled Fermi liquids is built up out
of quasiparticle creation operators which act on
both chains simultaneously, either symmetrically or
antisymmetrically, but generating  
phase coherent superpositions.
These phases can only be 
meaningful if we can observe interference effects
between histories in which 
$t_{\perp}$ acts.

Notice that in this sense
free electrons and  Fermi liquids exhibit
a heretofore unremarked on
{\it macroscopic quantum coherence}: the total number 
difference between
two chains (or planes) of free electrons is a macroscopic
variable which would undergo oscillations, 
rather than incoherent
relaxation, if a finite interchain hopping were suddenly turned
on. Viewed in this light it is not surprising that there should exist
states in which this macroscopic variable loses its coherence.
Rather, it is surprising (though undoubtedly correct for all normal 
metals) that macroscopic quantum behavior should 
occur in generic materials.
It is interesting that this macroscopic quantum coherence 
has not previously occasioned some concern in
the theory of interacting electronic systems. For example,
the above arguments guaranteeing 
the presence of coherent oscillations  in Fermi liquids
relied
in several places  on 
the quasiparticle structure of the Fermi liquid,
which fails totally for interacting
fermions in one spatial dimension.  We therefore
believe that the postulate
of previous works, on arrays of chains of
interacting fermions coupled by a single particle hopping
\cite{Bunch}, that
the relevance of $t_{\perp}$ 
signals a crossover to a three dimensionally coherent
Fermi liquid is just that: a postulate.  In fact, we
will see that the
extension
to coupled Luttinger liquids  of the tools
we have used for the TLS problem and
coupled Fermi liquids
does not support the conclusion that a relevant $t_{\perp}$
is always a coherent $t_{\perp}$.

To begin our analysis of coupled Luttinger liquids,
we require the spectral function of the
single particle hopping operator between otherwise
isolated Luttinger liquids.  This is easily obtained 
from the spectral function of the single particle Green's
function.  The universal features of this function
are readily accessible \cite{spectral,steve_net} and in 
\ref{sec:appendix}  we give a simple
method for calculating the hopping spectral function
from the single particle spectral function. 
The results
of the method when applied to the single electron
creation operator are in agreement with those previously
obtained \cite{spectral} and an example is shown
in Figure
\ref{fig:spectral1}.  At the level
of a linearized dispersion relation, the annihilation operator
for momentum $k$ \cite{fermipoint}
has the same spectral function (when the Fermi energy 
contribution to the energies is taken out) as the 
creation operator for momentum $2 k_F -k$.  The hopping operator's
spectral function can be obtained by convolving the
spectral function of the individual creation and annihilation
operators.  

For $\delta N(t=0) = 0$, the spectral function
for $\sum_k c^{\dagger}_1(k) c_2(k)$ is given by
$L ~\alpha \omega^{4 \alpha} \Lambda^{-(1+4 \alpha)}$
where $2 \alpha = \frac{1}{4}(K_{\rho} + K_{\rho}^{-1} - 2)$ 
is the anomalous exponent of the 
single particle Green's function for the case with
spin and $2 \alpha = \frac{1}{2}(K_{\rho} + K_{\rho}^{-1} - 2)$
for the spinless case.
Since the spectral function
vanishes as $\omega \rightarrow 0$ the response to 
$t_{\perp}$ is always incoherent for $\delta N(t=0) = 0$.
This should not be surprising since for $\delta N(t=0) = 0$
there is no possibility of coherent oscillations in 
$\langle \delta N(t) \rangle$.  It is for this reason
that for free particles there
would be no response for $\delta N(t=0) = 0$
since their response is
entirely coherent.  Fermi liquids would have a response
but there would be no long time singular behavior 
due to relevant operators \cite{relevance_caveat},
the relevant part of $t_{\perp}$, the single
quasiparticle hopping, having been  completely blocked for
$\delta N(t=0) = 0$.  

Notice that for the Luttinger liquid,
the long time incoherent response {\it is singular}
provided that $4 \alpha < 1$,
despite the fact that coherent hopping is 
totally blocked.  This suggests that the
incoherent hopping which was marginal for a Fermi 
liquid is relevant here and that flows away from
the $t_{\perp} = 0$ fixed point may be dominated by this
relevant operator, rather than the relevant operator
corresponding to coherent interliquid hopping.
If this is the case, then the renormalization group flows
should
end elsewhere than Fermi liquid theory.
In any case, it shows that
the failure of the quasiparticle picture has a profound
effect here.  This is due not only to the anomalous exponent
of the Luttinger liquid, but also the destruction of the
Fermi surface.  No such effect would be present for
a model with a single particle Green's function
of the form $G(k,\omega) \sim (\omega - v k)^{-1+2 \alpha}$.

To see the long time singularity,
in addition to $\langle \delta N(t) \rangle$
which at lowest order involves the difference of the
spectral function for $\sum_k c_1^{\dagger}(k) c_2(k)$
and $\sum_k c_2^{\dagger}(k) c_1(k)$, and therefore
vanishes for $\delta N(t=0) = 0$ (as it should):
\begin{equation}
\label{eq:N}
\langle \delta N(t) \rangle
 = \delta N(t=0) - 4 t_{\perp}^2
\int d\omega \frac{\sin^2(\omega t/2)}{\omega^2}
\left( \rho_{1\rightarrow2}(\omega) - \rho_{2\rightarrow1}(\omega)
\right) + \dots
\end{equation}
it is useful to consider the quantity
$P(t)$, defined as the probability to remain in the
initial state: 
\begin{eqnarray}
\label{eq:samestate}
P(t) & = & |\langle O | \exp(i \int_0^t dt' H'(t') ) | 0 \rangle|  \\
 & \sim & 1 - 4 t_{\perp}^2
\int d\omega \frac{\sin^2(\omega t/2)}{\omega^2}
\left(   \rho_{1\rightarrow2}(\omega) +
  \rho_{2\rightarrow1}(\omega)
\right) + \dots
\nonumber
\end{eqnarray}
Note that oscillations in $\delta N$ are the natural signature
of coherence and no oscillatory behavior in $P(t)$ is expected
in general, 
%\cite{Mila},
however it is useful to talk about both
since there may be important effects in $P(t)$
that are obscured in $\delta N(t)$ by the subtraction.
The potential for an important effect found at lowest order
in $P(t)$ and not visible in the
lowest order version of $\langle \delta N(t) \rangle$
 is not present in
a Fermi liquid since the only relevant terms 
in that case come from
hopping in a single direction and therefore
are contained in  $\langle \delta N(t) \rangle$.

Let us now procede with a short
time expansion analogous to that which we used for the TLS.
This should be valid for determining the presence or absence
of coherence for exactly the same reason as it was for that
problem: the presence or absence of coherence is equivalent
to the near degeneracy or non-degeneracy of the states
connected to the initial state by $t_{\perp}$. The short
time expansion is capable of revealing such features
since it is valid out to precisely the timescale 
where  the  initial state has been left behind.

For spinless fermions and
finite $\delta N(t=0) $ ($k_F^1 > k_F^2$)
the initial spectral
function for $\sum_k c^{\dagger}_1(k) c_2(k)$ is
given by (see Sec. 
\ref{sec:luttliq} and \ref{sec:appendix}) :
\begin{eqnarray}
\rho_{2\rightarrow1} (\omega)  & = & 
\Gamma^{-1}(2\alpha) \Gamma^{-1}(2+2\alpha) (2v_S)^{-(1+4\alpha)}
\Theta\left(\omega - (E_F^1 -E_F^2) -v_S(k_F^1-k_F^2) \right) \\
\nonumber &  &
\left(\omega - (E_F^1 -E_F^2) -v_S(k_F^1-k_F^2) \right)^{1+2\alpha} 
\left(\omega - (E_F^1 -E_F^2) +v_S(k_F^1-k_F^2) \right)^{-1+2\alpha} 
\end{eqnarray}

\noindent
likewise

\begin{eqnarray}
\rho_{1\rightarrow2} (\omega)  & = & 
\Gamma^{-1}(2\alpha) \Gamma^{-1}(2+2\alpha) (2v_S)^{-(1+4\alpha)}
\Theta\left(\omega - (E_F^1 +E_F^2)  -v_S(k_F^1-k_F^2) \right) \\
\nonumber &  &
\left(\omega - (E_F^2 -E_F^1) -v_S(k_F^1-k_F^2) \right)^{-1+2\alpha} 
\left(\omega - (E_F^2 -E_F^1) +v_S(k_F^1-k_F^2) \right)^{1+2\alpha} 
\end{eqnarray}

The short time behavior of $\delta N(t)$ would naively look
highly coherent for small $\delta N(t=0)$
due to the subtraction of the two
spectral functions and is not that different
from what occurs in a Fermi liquid (at least at small
$\alpha$).   On the other hand, if we look only
at hops in one direction (even the favored one) then
the incoherent high energy tail present in the
spectral function for $P(t)$ enters and things look much less
coherent.   $P(t)$ is radically different, even at
small $\alpha$ from the Fermi liquids case.
In fact for small $\delta N(t=0)$,
the incoherent part completely dominates
$P(t)$.  
One way to disentangle these effects 
is to consider the spectral functions
broken down into the contributions coming from
individual momenta.
First, examine the spectral function
for $c^{\dagger}_1(k) c_2(k)$, which may be obtained
by convolving the spectral functions for
$c^{\dagger}_1(k)$ and $c_2(k)$.  
%We consider first the
%spinless case for simplicity since the effects of spin
%charge separation on the calculation for such small
%$\delta N$ are negligible. However, we will give a
%treatment of the spectral functions valid 
%for general positive $\delta N$.
The spectral function
for $c^{\dagger}_1(k)$ has support for 
$\omega > E_F^1+ v_S |k - k_F^1|$,
where $v_S$ is the sound velocity of the Luttinger liquid.
The spectral function  behaves at large
$\omega$ like 
\begin{equation}
\rho_{\dagger,1}(\omega ~ large) \sim \omega^{-1+2\alpha}
\end{equation}
and behaves for $\omega \rightarrow E_F^1+ v_S |k - k_F^1|$
like 
\begin{equation}
\rho_{\dagger,1}(\omega ~ small) \sim
\left(\omega - (E_F^1+ v_S |k - k_F^1|) \right)^{\alpha - H(k-k_F^1)}
\end{equation}
where $H(x) = 1$ if $x \ge 0$ and $O$ if $x \le 0$. 
The integrated weight is $1-n_1(k)$.  
The spectral function for
$c_2(k)$ has support for $\omega > -E_F^2+ v_S |k - k_F^2|$,
also behaves at large
$\omega$ like 
\begin{equation}
\rho_2(\omega ~ large) \sim \omega^{-1+2\alpha}
\end{equation}
and behaves for $\omega \rightarrow ~-E_F^2+ v_S |k - k_F^2|$
like 
\begin{equation}
\rho_2(\omega ~ small) \sim
\left(\omega - (v_S |k - k_F^2| -E_F^2)\right)^{\alpha - H(k_F^2-k)}
\end{equation}
The integrated weight is $n_2(k)$.  

The convolution has support for
$\omega > E_F^1 - E_F^2 + v_S |k - k_F^1| + v_S |k - k_F^2|$
which means that except for $k_F^2 \le k \le k_F^1$ the
threshold is $E_F^1 - E_F^2 + 2 v_S |k - k_F^{avg}|$
where $k_F^{avg} = (k_F^1 +  k_F^2)/2$.
The behavior as 
$\omega \rightarrow E_F^1 - E_F^2 + 2 v_S |k - k_F^{avg}|$
is 
\begin{equation}
\rho_{\dagger,1,2}(\omega ~ small) \sim
\left(\omega - E_F^1 - E_F^2 + 2 v_S |k - k_F^{avg}|\right)^{4 \alpha}
\end{equation}

For the case, where $k_F^2 \le k \le k_F^1$, 
the threshold is smallest; there the
$k$ independent threshold is given by
\begin{eqnarray}
\label{eq:wmin}
\omega_{min}(k) & = &
E_F^1 - E_F^2 + v_S (k - k_F^2) - v_S (k - k_F^1) \\
\nonumber
 & = & E_F^1 - E_F^2 +v_S(k_F^1 - k_F^2) \\
\nonumber
 & > & 0 
\end{eqnarray}
The behavior of the spectral function as 
$\omega \rightarrow E_F^1 - E_F^2 + v_S(k_F^1 - k_F^2)$
is 
\begin{equation}
\rho_{\dagger,1,2}(\omega ~ small) \sim
\left(\omega - E_F^1 - E_F^2 + v_S(k_F^1 - k_F^2)\right)^{1 +4 \alpha}
\end{equation}
The positivity of the minimum energy
results  and the large exponent with which
the spectral function vanishes result
from the fact that hops in this direction
are `wrong way' hops, that is they increase
rather than decrease the initial $\delta N$.

The behavior for large $\omega$
is given by
\begin{equation}
\rho_{\dagger,1,2}(\omega ~ large) \sim
\omega^{-1+4\alpha}
\end{equation}
The integrated weight is $(1-n_1(k)) n_2(k)$.  

The spectral function for $c^{\dagger}_2(k) c_1(k)$
is similar.  The spectral function
for $c^{\dagger}_2(k)$ has support for
$\omega > E_F^2+ v_S |k - k_F^2|$ behaves at large
$\omega$ like
\begin{equation}
\rho_{\dagger,2}(\omega ~ large) \sim
\omega^{-1+2\alpha}
\end{equation}
and behaves for $\omega \rightarrow E_F^2+ v_S |k - k_F^2|$
like
\begin{equation}
\rho_{\dagger,2}(\omega ~ small) \sim
\left(\omega - (E_F^2+ v_S |k - k_F^2|) \right)^{\alpha - H(k-k_F^2)}
\end{equation}
The integrated weight is
$1-n_2(k)$. 
The spectral function for
$c_1(k)$ has support for $\omega > -E_F^1+ v_S |k - k_F^1|$,
also behaves at large
$\omega$ like
\begin{equation}
\rho_1(\omega ~ large) \sim
\omega ^{-1+2\alpha}
\end{equation}
and behaves for $\omega \rightarrow ~-E_F^1+ v_S |k - k_F^1|$
like
\begin{equation}
\rho_1(\omega ~ small) \sim
\left(\omega - (v_S |k - k_F^1| -E_F^1)\right)^{\alpha - H(k_F^1-k)}
\end{equation}
The integrated weight is
$n_1(k)$. 

The convolution has support for
$\omega > E_F^2 - E_F^1 + v_S |k - k_F^1| + v_S |k - k_F^2|$
which means that except for $k_F^2 \le k \le k_F^1$ the
threshold is $E_F^2 - E_F^1 + 2 v_S |k - k_F^{avg}|$.
The behavior as
$\omega \rightarrow E_F^2 - E_F^1 + 2 v_S |k - k_F^{avg}|$
is
\begin{equation}
\rho_{\dagger,2,1}(\omega ~ small) \sim
\left(\omega - E_F^2-E_F^1+2 v_S |k - k_F^{avg}|\right)^{4\alpha}
\end{equation}

For the case, $k_F^2 \le k \le k_F^1$
threshold is $E_F^2 - E_F^1 +  v_S (k_F^1 -k_F^2)$,
which vanishes for weak interactions
and is always smaller than 
the threshold for hops in the
other direction.
The behavior as
$\omega \rightarrow E_F^2 - E_F^1 +  v_S (k_F^1 - k_F^2)$
is
\begin{equation}
\label{eq:shiftsing}
\rho(\omega ~ small) \sim
\left(\omega-(E_F^2 - E_F^1 + v_S(k_F^1-k_F^2)) \right)^{-1+2\alpha}
\end{equation}
There is a power law divergence.

The behavior for large $\omega$
is given by
\begin{equation}
\rho_{\dagger,2,1}(\omega ~ large) \sim
\omega^{-1+4\alpha}
\end{equation}
The total weight in the spectral function is given by
$n_1(k)(1-n_2(k))$.

In the region  where
$k_F^2 \le k \le k_F^1$
a Fermi liquid spectral function would
be a delta function at zero frequency, but here
there is a power law singularity at a non-zero,
negative frequency
since $E_F^1 - E_F^2 = (k_F^1 -k_F^2) \frac{v_J + v_N}{2}$
is always larger   than than 
$v_S (k_F^1 - k_F^2) = \sqrt{v_J v_N} (k_F^1 - k_F^2)$ \cite{vJfootnote}.
The essential points are that the singularity is in general
not at zero energy and is a power law rather than a delta
function.

Notice that when $4 \alpha >1$, none of the spectral functions
for the individual momenta in either direction
is decreasing for large $\omega$.  The high energy behavior
is thus incoherent
which implies
that the response in
the $t_{\perp} \rightarrow 0$ limit is
always incoherent at every $k$ at short times.   
The low energy behavior for
$\delta N(t=0) = 0$ is also incoherent. 
This implies that the proposal of
incoherence for $\alpha ^{>}_{=} 1/4$ is self-consistent,
since incoherence leads to a vanishing 
oscillation frequency, $\omega_{osc}$,
and therefore the natural $k_F^1-k_F^2$
to consider is $\omega_{osc}/v_F = 0$, a case
where all the spectral functions are in fact incoherent.
Notice that for
$\alpha < 1/4$ this consistency is not present since
even for $k_F^1-k_F^2 = 0$, some of the spectral
functions associated with momenta near to the Fermi
surface are coherent \cite{lowa}.
The self-consistency is therefore not trivial
and we believe that there must be incoherence for
all $\alpha > 1/4$.

This is despite the fact that the low energy form of the
spectral function for finite $\delta N(t=0)$ can have
singular behavior (see Eq. \ref{eq:shiftsing}).
The reason for this is that
the question of coherence is not 
one that involves the singularities of the spectral 
function, but rather
comparing the low and high
energy contributions to 
$\int d\omega \rho(2\omega) \frac{\sin^2(\omega t)}{\omega^2}$,
as defined previously.
From that point of view 
it is clear that a low energy singularity  
does not necessarily imply coherence if the high energy
behavior is incoherent and the singularity is integrable.
Instead, one needs to compare the energy scale at which
the spectral functions behavior crosses over between coherent
and incoherent to the energy
scale which defines the division into high and low energy parts.
The appropriate energy scale for that division is approximately
given by the inverse of the time to leave the initial
state $\sim t_{\perp}^R \sim t_{\perp}^{1/(1-2 \alpha)}$.
Based on the results for the
TLS problem, we expect $\omega_{osc} \sim t_{\perp}^R$
for small $\alpha$ but for $\alpha$ finite,
it is possible that
$\omega_{osc} \ll t_{\perp}^{R}$
or even $\omega_{osc} = 0$.

It is clear that if $v_F(k_F^1- k_F^2) << t_{\perp}^R$,
the spectral function differs from the $v_F(k_F^1- k_F^2)=0$
case only by corrections at energies low compared to the
scale for the high-energy low energy division.
Then neither the high nor low energy integrals is greatly
changed
and the incoherence for $\delta N(t=0)$ certainly remains.
Since the natural choice of $v_F(k_F^1- k_F^2)$
is $\omega_{osc}$ this should apply
for $\omega_{osc} < t_{\perp}^R$, however we also consider the two
other possibilities.

In the case  $v_F(k_F^1- k_F^2) >> t_{\perp}^R$,
and
strong interactions, the spectral
function for hops in the correct direction (the direction
which results in a decrease in the initial $\delta N$)
has a singularity at such
a negative frequency (much larger than $t_{\perp}^R$)
that coherence is again ruled out
(see Eq. \ref{eq:shiftsing}, also there are more
important effects when spin is included that preclude
coherence in this regime). Therefore, the only threatening  
case is that where $v_F(k_F^1- k_F^2) \sim t_{\perp}^R$.
In this case, the crossover of the spectral function
to its high energy $\omega^{-1+4\alpha}$ behavior 
occurs at roughly the same frequency as the division into
high and low energy pieces and the high energy contribution
to transitions out of the initial state is not greatly 
affected.  The effect on the low energy part is difficult
to calculate as it depends on the division into
high and low energy and on the exact time scale at which one
is making the comparison.  In general, the velocity inequality
produces an $O(1)$ decrease in the low energy part, whereas
the integral divergence in the
spectral function at low energies produces an
$O(1)$ increase in the low energy part and the effect on 
coherence is ambiguous.   In view of this,
the self-consistency
of incoherence
for $\alpha >1/4$ and the fact that for
$\alpha \rightarrow 1/4$ we
expect $\omega_{osc} \ll t_{\perp}^R$,
it seems unlikely that there exists
a second self-consistent solution which 
is coherent for $\alpha > 1/4$, although we cannot rule this out.
We can rule out the possibility that the incoherent phase 
is eliminated by being  pushed all
the way to $\alpha = 1/2$, where irrelevance sets in.
This cannot occur because, as $\alpha$ approaches
$1/2$, the 
high energy piece is diverging relative to the low 
energy, and, as we have seen, the initial $\delta N(t=0)$
makes only an $O(1)$ change in these quantities.

In conclusion, we believe that the behavior at long times
will be incoherent for all $\alpha>1/4$
and, in that case, the ground state should
not involve a phase coherent superposition of 
quasiparticle states, and, more generally,
interference effects between histories involving
$t_{\perp}$ should be unobservable.
The $\alpha < 1/4$ case is less favorable for
coherence, but there are other more complicated
reasons for believing that incoherence may extend into
that region, as we will now discuss.

%% file: smalla.sec
\subsection{Dynamics for   $\alpha  < \frac{1}{4}$ }
\label{sec:smalla}
Since, at $\alpha = 1/4$,
the spectral function for single particle hopping
at any $k$ in any direction is less coherent
than the spectral function of the TLS at
the onset of incoherence \cite{lowa},  coherence in the
dynamics of the fermion model would have to result
from an interaction of electrons hopped at
different $k$'s.  While we can not explicitly rule
this out without a full solution of the strongly
interacting problem at strong coupling and finite
$t_{\perp}$, the idea seems implausible since such
interactions are irrelevant in Fermi liquid theory
and, in so far as they generate finite renormalizations
there, they are hostile to coherence.  
Certainly, the states connected {\it to the
appropriate initial state for small} $\delta N(t=0)$
by the single
particle hopping operator are effectively nondegenerate,
and all that is required for incoherence is that
this nondegeneracy is typical. From this point of
view the possibility that interactions somehow restore
coherence at higher orders seems highly unlikely
and we know of no examples of this behavior
in soluble models.
Given this,
we take the incoherence of the dynamics for
$\alpha > 1/4$ for granted and pass now to
a discussion of
the more complicated case of smaller
$\alpha$.  Readers not interested
in the more technical arguments for
incoherent dynamics for $\alpha <\frac{1}{4}$
should skip to section \ref{sec:interp}.

%While we believe that the incoherent fixed point is
%certainly realized for $\alpha > 1/4$ as 
%t_{\perp} \rightarrow 0$,
%incoherence in the case
%where $4 \alpha < 1$ is more subtle and the
%possibility of such behavior is what we are 
%interested in establishing in this section. 

Consider the case where there is only
one extra particle which may oscillate coherently back and forth,
i.e. $\delta N (t=0) = 1$.
In this case,
for the Fermi liquid, the arguments for general 
$\delta N$ show that one expects various irrelevant
effects plus coherent oscillations of period
approximately equal to $(Z t_{\perp})^{-1}$, however
the Luttinger liquid case is quite different.
Let us consider the lowest order contributions to 
$P(t)$ and $\langle \delta N(t) \rangle$.
$P(t)$ is the simpler to treat since it 
is finite for $\delta N = 0$.  So long as
$\delta N(t=0) \ll t_{\perp}/v_F$, the 
$\delta N(t=0) = 0 $ results can be used
for $P(t)$ so that the lowest order result is
\begin{eqnarray}
P(t) & =  & 1- 4 t_{\perp}^2
\int d\omega \frac{\sin^2(\omega t)/2}{\omega^2}
L ~\alpha \omega^{4 \alpha} \Lambda^{-(1+4 \alpha)} \\
 & \sim & 1 - \frac{\alpha}{1-4 \alpha} 2^{1+4\alpha}
t_{\perp}^2 L t^{1-4 \alpha} \nonumber
\end{eqnarray}
For $\delta N(t=0) = 1 $, 
there is only
one extra particle which may oscillate coherently back and forth.
The total weight in the spectral function
that enters into computing 
%$P(t)$ coming from a given k is 
%$n_1(k) + n_2(k) - n_1(k) ~n_2(k)$
%while the total weight in the
%spectral function for
$\langle \delta N(t) \rangle$
%for finite $\delta N(t=0)$
is 
$n_1(k)-n_2(k)$, and therefore, since 
the extra added electron is spread out over a range of 
$k$'s, the spectral function for
$\langle \delta N(t) \rangle$ then has support over many momenta.
In frequency things
are more localized.
Since the spectral function  contributing
to $\langle \delta N(t) \rangle$ involves the difference
of the spectral function of
$c^{\dagger}_2 c_1$ and $c^{\dagger}_1 c_2$,
it is sharply peaked at the lowest possible
energies at a given $k$, decaying for
$\omega \gg v_F \frac{2 \pi}{L} \delta N$ like
$\omega^{-2+4 \alpha}$.
As a result, 
the spectral function for
$\langle \delta N(t) \rangle$
can be approximated by
a delta function 
for small $\delta N$ and $\alpha < 1/4$.
There is a complication due to the fact that the weight
must be reduced to reflect the fact that
the amount of weight in the low energy
part is not
\begin{eqnarray} 
n_1(k)-n_2(k) & \sim  &
\frac{2\pi}{L} \left( -\frac{\partial n}{\partial k} \right)
\\ \nonumber & \sim &
|k-k_F^{avg}|^{-1+2\alpha}
\end{eqnarray}
This 
arises because the $\omega^{-2+4 \alpha}$
behavior is valid only for energies much
smaller than the cutoff
(but much larger than
$v_F \delta N$, of course). 
There is weight pushed out past the cutoff on the single particle
Green's function by convolving two of them
so that in fact
\begin{eqnarray} 
\left( n_1(k)-n_2(k) \right)_{\rm low ~energy} & \sim  &
O(\frac{\alpha}{L}) \left( -\frac{\partial }{\partial k} 
|k-k_F^{avg}|^{2\alpha} \int_{|k-k_F^{avg}|}^{2|k-k_F^{avg}|} d\omega
(\omega-|k-k_F^{avg}|)^{-1+2\alpha} \right)
\\ \nonumber & \sim &
O(\frac{\alpha}{L}) |k-k_F^{avg}|^{-1+4\alpha}
\end{eqnarray}
We then find:
\begin{eqnarray}
\langle \delta N(t) \rangle & \sim & 
\delta N(t=0) \left( 1 -
t_{\perp}^2  O(\frac{\alpha}{L}) 
\int d\omega \sum_k  \frac{\sin^2(\omega t)}{\omega^2}
\delta(\omega - 2 v_S |k - k_F^{avg}|) 
|k-k_F^{avg}|^{-1 + 4 \alpha}  \right)\\
\nonumber
& \sim &
\delta N(t=0) \left(1 -  
 t_{\perp}^2  O(\alpha)
\int_0^{\infty} dx 
\frac{\sin^2( v_S x t))}{v_S^2 x^2} x^{-1 + 4 \alpha} \right) \\
\nonumber
& \sim & \delta N(t=0) \left( 1- O(1)
t_{\perp}^2  t^{2 - 4 \alpha} \right)
\end{eqnarray}
The time dependence of
$\langle \delta N(t) \rangle$
is thus intrinsically coherent at lowest
order
for $\alpha < 1/4$.
This is in agreement with the
finding that some of the momentum channels
contributing to $P(t)$ had coherent
spectral functions for $\alpha < 1/4$.

We then have clear indications of coherence
in $\delta N$ for $\alpha < 1/4$, however,  the 
number of transitions
contributing to $P(t)$ outnumber by a factor of 
$L/\delta N$  those contributing to
$\langle \delta N(t) \rangle$. This factor is
thermodynamically large 
for $\delta N =1$ and still very large
for 
$\delta N \sim L t_{\perp}/v_F$.  The question then 
becomes whether the coherence in $\delta N(t)$ can be
believed in the presence of $P(t)$'s behavior.

If we adopt the maximally collective point
of view and ask only if the system has left
its initial state 
before the extra particle has moved between the two chains,
then it is clear that, in the thermodynamic limit
($L \rightarrow \infty)$
for finite $\delta N(t=0)$,
the system always changes state first.
The incoherent $P(t)$ might then be expected
to dominate the
coherent $\delta N(t)$.  This seems extreme, however, since
it is clear that there must be some interaction
between the hops associated with $P(t)$ and
those associated with $\delta N(t)$ for the
coherence of  $\delta N(t)$ to be spoiled and
the maximally collective assumption amounts
to assuming that every electron participating
in an  $P(t)$-type hop will interact with 
every electron which would have
participated in a $\delta N(t)$-type hop. 
A more reasonable approximation
might be  to take the other extremal point
of view, that those hops contributing to 
$P(t)$ that occur at a given wavevector, $k$, 
affect only the contribution to $\delta N(t)$
associated with that same wavevector and the same 
direction.
This amounts to allowing the electrons 
to interact only ``statistically'', and should be
a minimal assumption.  Clearly, if there has just been a
transition induced by $c^{\dagger}_1(k) c_2(k)$
to a high energy state, no transition induced
by $c^{\dagger}_1(k) c_2(k)$ to a low energy
state (or any other state for that matter) is
possible until both the hole and particle
have scattered.  This is similar again to the
TLS problem where the spin can not flip
in the same direction twice and thus
high energy flips completely destroy the
coherence once their number becomes of order
one per low energy flip.
In our case, 
for the incoherent hops
to plausibly destroy the coherence,  
we need that
probability to have undergone an $P(t)$-type 
hop (at every wavevector $k$ for those $k$'s
that contribute to $\delta N(t)$) be of order
unity or larger for times of order
the inverse of the putative
oscillation time 
(recall that the expected oscillation time is
of order $t_{\perp}^{1/(4\alpha-1)}$
or much longer).  For $\delta N \ll L$,
we can use the spectral functions for
$\delta N =0$ in computing $P(t)$. For the most important
momenta (those near the Fermi surface)
we find contributions to $P(t)$
\begin{eqnarray}
\delta P(t) & \sim & 4 \alpha t_{\perp}^2 \int d\omega \sin^2(\omega t/2)
\omega^{-1+4\alpha} \Lambda^{-4\alpha} \\
\nonumber & \sim & \frac{1}{4} t_{\perp}^2 t^{2-4 \alpha} \Lambda^{-4\alpha}
\\ \nonumber & \sim & 1
\end{eqnarray}
We see that for one extra particle, there is no
compulsion to believe in coherence for any $\alpha$!
However, it is not clear either that 
the case of one extra particle settles the question or
that our counting all $P(t)$ type processes as
incoherent is entirely fair.
If we instead take the incoherent processes
to be those involving 
energies larger than
$t_{\perp}^R$ \cite{energyfootnote},
we find an effect of
order $O(\frac{\alpha}{1-4 \alpha})$ \cite{14alphacaveat}.
The effect is essentially
identical to  what occurs in the
TLS if one took the TLS spectral
function for
($\alpha_{TLS} = 2 \alpha_{fermion}$).
This destroys coherence for
$\alpha > 1/4$, as we argued it
should earlier when we compared the
spectral function of 
$c^{\dagger}_1(k) c_2(k)$ 
to that of the TLS.
This seems more reasonable since it is in agreement
with our findings for the spectral functions
for individual momenta, and, as we shall see,
is also in agreement with our results for
larger $\delta N$.

Let us now consider the effect on $P(t)$
of the natural choice
for $\langle \delta N(t=0)$:
$\langle \delta N(t=0) \rangle \sim t_{\perp}^{R}/v_S$.
For this case, the spectral function
for $\delta N(t)$ essentially contains
all of the weight in the correct
direction which is also low energy,
i.e. lower than $t_{\perp}^R$, paralleling the
definition we introduced in the TLS discussion.
In this case our crude division of processes into
high and low energy at the energy
scale $t_{\perp}^{R}/v_S$
is equivalent to  the division into
$\delta N(t)$ and $P(t)-\delta N(t)+\delta N(t=0)$.
This further motivates the study of
$\delta N(t=0)$ of this size.
Now, the spectral function for $P(T)$
is essentially modified because of the
substantial $\delta N(t=0)$. In fact,
is easy to see that there is essentially
no weight in $P(t)$ for energies below
$v_F\delta N(t=0)$ so that the above 
contribution from $P(t)$ type
hops is modified to
\begin{eqnarray}
\delta P(t) & \sim & 4 \alpha t_{\perp}^2 \int_{\lambda}^{\infty} d\omega \frac{\sin^2(\omega t/2)} 
\omega^{-1+4\alpha} \Lambda^{-4\alpha} \\
\nonumber & \sim & \frac{\alpha}{1-2\alpha} t_{\perp}^2  \lambda^{-2+4\alpha} \Lambda^{-4\alpha}
\\ \nonumber 
\delta P(t = \lambda^{-1}) & \sim &
\frac{\alpha}{1-2\alpha}
\end{eqnarray}
where $\lambda = t_{\perp}^R$.
This is then in complete agreement with our earlier discussions.

We see that, by itself, the minimal statistical interaction
is equivalent to looking at the 
individual momentum channels and 
reproduces our
earlier criterion 
for incoherent dynamics. 
Let us now consider 
the non-statistical
interaction of the hopped
electrons with each other.
The first important point to
consider is that,
in a Luttinger liquid,
the four point function does not factorize
into products of two point functions,
signaling that inserting two electrons
has a very different effect than inserting one.
This is due to the strong interactions between the inserted
electrons {\it even as they approach the Fermi surface}.
If the two added electrons are very much farther
from each other in space time
than they are from the added holes, 
then the four point function will  
approach the product of two point functions;
otherwise, the two added particles should be regarded
as interacting strongly with each other. In
our calculation, the creation and annihilation
operators are generally separated by times 
of order $1/t_{\perp}^R$, so that if an additional hop has
occurred closer than $v_S/t_{\perp}^R$, that hopped electron
will interact strongly and we can expect an $O(1)$
scattering effect and the loss of coherence in that way.
Note that this effect is completely absent in Fermi
liquids since the four point function always factorizes
in the low energy limit where quasiparticle
interactions are irrelevant.
The density of hopped electrons from the bare incoherent
effects at time $1/t_{\perp}^R$ is roughly
given by
\begin{equation}
\label{eq:hops}
\rho_{hops} \sim P(t = 1/t_{\perp}^R)/L \sim 
\frac{\alpha}{1- 4 \alpha} t_{\perp}^R
\end{equation}
so that in general the chance that an electron
attempting to hop coherently
will be scattered before the amplitude can be built up to
order one is $O(\frac{\alpha}{1- 4 \alpha})$ \cite{14alphacaveat},
of the
same order of importance as the statistical interaction,
so that we have here an entirely separate source of
incoherence from that in the TLS problem.
Further, that second source  is
as strong as the first.  It is this additional effect
which of course makes our problem a
truly many body one and keeps it
from being rigorously tractable via any of the known
approaches to the TLS.
We believe that it is safe to conclude that
the critical $\alpha$ will be pushed substantially
lower by the combination of the two effects.

There are additional reasons 
unrelated to $P(t)$ for believing in a lower
critical
$\alpha$ which appear when macroscopic particle
number inequalities are considered.
As we make $\delta N(t=0)$  larger,
effects related to the plethora of
velocities in Luttinger liquids
set in. 
Consider again the case:
$\langle \delta N(t=0) \rangle ^{>}_\sim t_{\perp}^{R}/v_S$.
In this range of $\delta N(t=0)$,
several factors relating to velocities in
Luttinger liquids neglected in the 
previous discussion  become important.
Recall that, in a Luttinger liquid, 
the velocities for current and charge excitations
are in general unequal ($v_N \neq v_S \neq v_J$),
implying that the shift of the Fermi
energies need not exactly cancel the 
energy of the bosons associated with
hops occurring at momenta between the two $k_F$'s in the
correct direction.
We found in Eq. \ref{eq:shiftsing} that:
\begin{equation}
\rho(\omega ~ small) \sim
\left(\omega-(E_F^2 - E_F^1 + v_S(k_F^1-k_F^2)) \right)^{-1+2\alpha}
\end{equation}
There is no analog of this
in a Fermi liquid where $\epsilon(k)$
is by definition the same in the two identical liquids
and, as we suggested earlier,
we expect potentially important effects on coherence.
In particular, coherence will be substantially disrupted
if we make 
$\delta N(t=0) \gg t_{\perp}^{R}(v_S/|v_N-v_J|^2)$,
since this shifts the singularity in the spectral function
away from zero energy by an amount much larger than
$t_{\perp}^{R}$.
The shift of the spectral weight due to the
inequality of $v_N$ and $v_J$ away from
zero energy
is an order $O(\alpha)$ effect since 
$v_N - v_J = O(\sqrt{\alpha})$ while the
energy mismatch is of order 
$O((v_N - v_J)^2)$, and can therefore
be considerable at finite $\alpha$,
substantially reducing the critical $\alpha$
required for incoherent dynamics.

An even more substantial velocity inequality occurs
between the spin and charge excitation velocities
for Luttinger liquids with spin.
This spin-charge separation also diminishes
coherence as we make $\delta N(t=0)$
larger since the width of the spectral 
function associated with the electron 
far from the Fermi surface grows linearly with
distance. This drastically changes the spectral
function for hopping in the case of finite
$\delta N(t=0)$, completely eliminating the 
divergence
in $\rho_{2\rightarrow1}$,
replacing it with a threshold singularity 
at energy $ v_{\sigma}(k_F^1 -k_F^2) - (E_F^2 - E_F^1) $
of the form
\begin{equation}
\rho_{2->1}(k) = \left( \omega - v_{\sigma}(k_F^1 -k_F^2) - E_F^2 - E_F^1
\right)^{2\alpha}
\end{equation}
for hops at momenta, $k$, between $k_F^1$
and $k_F^2$. In addition to the elimination of the divergence,
its placement is now at an energy which is very different
from zero, particularly if $v_{\sigma}$ is small.  

There are additional singularities associated with
other energies, but no divergences and the physical effect is really
to spread all of the weight from the divergence in the spinless
case out over a region of size $(v_{\sigma} -v_{\rho})(k_F^1-k_F^2)$,
drastically reducing the possibility for coherence.
The effect resembles introducing an effective lifetime proportional to
$\delta N(t=0)$ and severely restricts the possibilities for
coherence for initial conditions with large $\delta N(t=0)$.

The effects  due to the spin-charge
separation should be
large for the large $U$ Hubbard model
where
$v_{\sigma} \rightarrow \frac{t_{\parallel}^2}{U} \sim 0$,
while $v_{\rho}$ remains finite.
For small $U$ where $v_{\rho} - v_{\sigma} = O(U)$
the effect is $O(U)$ or, since
$ \alpha = O(U^2)$, $O(\sqrt{\alpha})$.
We again expect that the $\alpha$ required for
incoherent dynamics will be substantially reduced by
the inclusion of these velocity inequality effects,
in effect allowing us to consider only
the case  $\delta N(t=0) ^{<}_\sim t_{\perp}^{R}/v_S$.
For that case,
there is a significant
statistical  and non-statistical interaction of
the high energy hops contributing
to $P(t)$ on the hops contributing
to $\delta N(t)$ through the same momentum channel.
In fact, as we saw, both effects are little changed
from the single extra particle case and remain
$O(\alpha/(1-4\alpha)$, so that the
velocity inequality effects are 
truly additional to our earlier sources of
incoherence.

Notice that,
for the case of coupled Hubbard
models as $U \rightarrow 0$, the 
velocity inequalities become
arbitrarily small, as does $\alpha$,
and for any partition into high and 
low energy processes, the low energy will
dominate.  The scattering from
incoherently hopped electrons, as 
estimated above, will
vanish as well. Together these
things would imply a critical $\alpha$
below which hopping would be coherent
even in the $t_{\perp}\rightarrow 0$
limit, however,
the arguments in favor of this are
not entirely conclusive.
We did find suggestions of incoherence
for one additional particle for any $\alpha$
due to interparticle interactions.
Also,
in the limit as $t_{\perp}\rightarrow 0$
the incoherent hops have arbitrarily long
time to scatter the coherent hops so
that if this scattering is enhanced over our
estimate by any divergent amount in the
long time limit then incoherence should
result as $t_{\perp}\rightarrow 0$.
Our original work \cite{prl}
claimed a logarithmic correction
so that the critical $U$ for
the coherence-incoherence transition
was zero for $t_{\perp}\rightarrow 0$
and the $t_{\perp}$ required for
coherence went like $\exp(const ~U^{-5})$.
We have not been able to make the arguments
put forward there 
compelling, so we regard the question
of the existence of a critical
interaction strength
as $t_{\perp}\rightarrow 0$ as open,
but favor a finite
interaction strength as most likely
necessary for incoherence.
It seems clear that, since $U$ is a
marginal operator while $t_{\perp}$
is relevant, any phase boundary passing through
$U=0$, $t_{\perp} = 0$ would
have to have the form $t_{\perp} \sim \exp(const ~U^{-x})$.
This would involve a logarithmic enhancement of the
interaction at long times of the coherent
hops by the incoherent ones  and
the question is therefore rather subtle.

In any case, we have given strong arguments 
for the existence of incoherence for  $\alpha > 1/4$ and
demonstrated
three factors which should result in incoherence
for $\alpha$ substantially lower values of
$\alpha$.

%% file: wen.sec
\subsection{Connection to Green's Function Approaches}
\label{sec:G}
Hints of the incoherence we are proposing
can be seen in more
conventional calculations. 
The interpretation of these calculations is complicated since
they are not really intended for
studying coherence effects, but in light of
what we have learned there are a number of
suggestive results. Consider as an
example the calculation of the  
Green's function for a Luttinger liquid of
spinless fermions
with vertex corrections 
neglected, as in Wen's work \cite{Bunch}
(there have been similar calculations by a number of
others \cite{Bunch} and what we have to say
applies, in general, to all of them - we focus on one for the sake
of clarity).  In that
work, Wen studied how the Green's function
\begin{equation}
G_0(k,\omega) = (v^2 k^2 - \omega^2  )^{ \alpha} (\omega-v k )^{-1}
\end{equation}
%(here $k$ is momentum measure from the Fermi surface, in
%this case the right Fermi point)
is modified, if vertex corrections are neglected and infinitely many
chains are coupled.
Here $k$ is momentum along the one-dimensional chains
measured from the right Fermi point.

It is easy to see that 
the neglect of vertex corrections amounts to incorporating 
$t_{\perp}(k_{\perp})$ as an energy
independent self-energy and
one expects:
\begin{equation}
G^{-1}(k,k_{\perp},\omega) = G_0^{-1}(k,\omega) - t_{\perp}(k_{\perp})
\end{equation}
Wen actually proposed the form for $k$ near the right Fermi point:
\begin{equation}
G(k,k_{\perp},\omega) = 
\frac{(v^2 k^2 - \omega^2)^{\alpha}}
{(\omega-vk)-t_{\perp}(k_{\perp})(v^2 k^2 -\omega^2)^{\alpha}}
\end{equation}
where we have set the dimensionful
high energy cut-off to $1$. 
This is what one expects,
from the incorporation of $t_{\perp}$ as a self-energy.
The expression is, however,  incomplete in that no
discussion of analytic properties of
the unperturbed or the 
perturbed Green's function is given.
All the singularities in $G_0$ are for real $\omega$
and must be moved off the real axis.  When this 
is done with care we will see that rather interesting
physics
results.

Putting aside for the moment the question of analytic properties,
Wen found that a zero of the denominator 
for $\omega = 0$ occurs when
\begin{equation}
vk^{1-2\alpha} + t_{\perp}(k_{\perp}) = 0
\end{equation}
which has no solution for $\alpha > 1/2$
for small $t_{\perp}$ and $k$.  As Wen correctly
pointed out, this signals the renormalization group
irrelevance of $t_{\perp}$ in this regime.

For $\alpha < 1/2$, he suggested that the solutions
may be interpreted as specifying the position of a
two dimensional Fermi surface.
If the hopping is relevant, 
then the conclusion of Wen and various others \cite{Bunch} 
was that
a crossover to
behavior with coherence in all
directions should occur. The  transverse
bandwidth would be of order the scale
at which the splitting was comparable in magnitude to 
the bare energy, $v k $, leading to
$t_{\perp}^R \sim \left( t_{\perp} \right)^{\frac{1}{1-2 \alpha}}$
\cite{wenfootnote}.  We agree with this interpretation
for sufficiently small $\alpha$ for spinless Luttinger liquids
(although even here a note of caution should be sounded),
however, if we examine the behavior close
to the Fermi surface as $\alpha \rightarrow 1/4$ there is
rather interesting behavior. 

First, recall that singularities of the Green's function,
particularly poles, have sensible physical interpretations
only 
in the second and fourth quadrants of the complex $\omega$
plane.  Since, for $k \neq 0$, the Green's function has
two branch cut singularities, one for each sign of $\omega$,
these originate just off the real axis in the second and fourth quadrants.  
Further, $G_0$ must be real
for $-vk < \omega < vk$ since in that region no on-shell decay 
of an injected fermion is possible.
This implies that 
the phase of $G_0$ for $\omega > v k$
should be given by $-\alpha \pi$, and
by $-\pi - \alpha \pi$ for $\omega < -vk$.
Recalling this, let us look at the effect on the
Green's function at the right Fermi point, $k=0$, when $t_{\perp}$ is turned on.
Consider $t_{\perp}(k_{\perp})>0$,
then the pole for $\alpha = 0$ is at $t_{\perp}(k_{\perp})$.
If we now turn on $\alpha$, the pole must shift into the
fourth quadrant since the pole equation is
\begin{equation}
0 = G_0^{-1}(k,\omega) - t_{\perp}(k_{\perp})
\end{equation}
and $G_0^{-1}(k,\omega)$ has a phase which is $\alpha \pi$ on the real
axis for $\omega > 0$ and
decreases as we move down into that quadrant.
Moving off the axis into the
fourth quadrant an angle $\Theta$ changes the phase
to  $\alpha \pi - (1-2\alpha) \Theta$ and it is again possible
to have a pole if 
\begin{equation}
\Theta = \pi \frac{\alpha}{1-2\alpha}
\end{equation}
For small, $\alpha$ this pole can be sensibly interpreted
as a weakly damped quasiparticle pole with the 
corresponding quasiparicle state unoccupied, as it should be. 
However, for $\alpha \rightarrow 1/4$, $\Theta \rightarrow \pi/2$ 
and the ``quasiparticle energy''
of the pole, given by the real part of $\omega$,
becomes smaller
and smaller.  For $\alpha = 1/4$, $\Theta = \pi/2$
and the real part of the quasiparticle energy is not
changed at all. The quasiparticle Fermi
surface, determined by the
zero crossings of the real part of
the pole frequency, is also not moved by $t_{\perp}$. 

For $\alpha >1/4$, the pole
moves out of the fourth quadrant and there is no 
sensible physical interpretation of the pole.
As a quasiparticle pole, it would be signalling 
an unoccupied, negative energy
quasiparticle state.  Note that, while this is suggestive
of an instability, the spectral function in this approximation
remains positive definite for $\alpha > 1/4$ and we know
of no argument which rigorously demonstrates an instability setting
in for  $\alpha > 1/4$.
We do, however, regard it as significant that 
the pole which occurs for $\alpha = 1/4$ has a
purely imaginary frequency, entirely in keeping with the idea
that $t_{\perp}$ is  acting incoherently at this value
of $\alpha$.

For a negative $t_{\perp}$, an exactly parallel scenario
involving the second, instead of the fourth quadrant,
would have resulted. In both cases, for $\alpha > 1/4$,
there is no physically sensible interpretation of the 
pole obtained by the incorporation of
$t_{\perp}$ as a self-energy, and the results are extremely
suggestive of incoherence.

%\vspace*{-0.3cm}
%\begin{figure}
%\centerline{ \epsfxsize = 4.5in
%\epsffile{wenFS.eps}}
%\vspace*{0.3cm}
%\caption{ Behavior of the poles of the Green's function 
%for $k=k_F$ in
%the approximation discussed in the text.
%There is no physically sensible solution for
%$\alpha > 1/4$ since the poles move into
%physically inaccessible regions as
%$\alpha \rightarrow 1/4$.
%}
%\label{fig:wenFs}
%\end{figure}

The effect is very closely analogous to the behavior of the
Laplace transform of $P(t)$ found in \cite{TLS}
at the onset of incoherence.   A similar
analogy between the locations of the poles of
the single particle Green's function approximated in
this way and the Laplace transform  
of $P(t)$ in the TLS problem was noted in
\cite{Sudip}.
%, although there a different criterion
%for incoherence was reached due to the somewhat
%peculiar analytic properties of the Green's function 
%considered there.

If we now move away from the Fermi surface,
are there physically sensible poles in $G$?
First, note that the arguments for the poles we have considered
for
$k=0$ will apply essentially without 
modification to all $k \ll t_{\perp}^R$
so, for these poles, we must consider $k$'s that are at least
$O( t_{\perp}^R)$.
Our approach is to begin for $\alpha < 1/4$
and track the poles as we increase $k$.
Then we will see if an increase in $\alpha$
past $1/4$ still allows sensible poles at finite $k$.
We consider both signs
of $t_{\perp}$ but only positive $k$ since
negative $k$ is essentially identical 
for the opposite sign of $t_{\perp}$.

Consider first the case $t_{\perp}>0$.
As we move some distance away from the Fermi surface,
the singularity at $-v k$ becomes more distant and
its effect on the phase less important.
It now becomes possible to circle the singularity at
$vk$ without moving appreciably with respect to
the singularity at $-vk$.
% and in fact,
%for $vk \gg t^R_{\perp}$
%the pole can do exactly that as we 
%increase $\alpha$ at fixed $k \neq 0$.
Therefore the phase of $G_0^{-1}(k,\omega)$
varies like
\begin{equation}
{\rm phase} = \alpha \pi - (1-\alpha) \Theta
\end{equation}
where $\Theta$ is
measured downward
from the real $\omega > vk $ half-line.
%If we look at fixed $\alpha$ and increase
%$k$ the pole moves counterclockwise
%around $vk$ and we might expect that 
%by increasing $k$ and $\alpha$
%together, 
Because of this change, we can 
reach larger $\alpha$ without the pole
being forced  into unphysical regions.
In fact, it is possible to attain a phase of
$(2 \alpha - 1) \pi$ for
$k \gg t^R_{\perp}$ by going around
the singularity at $vk$.
Under these circumstances there is an
allowed pole for $\alpha < 1/2$
at an angle 
given by:
\begin{equation}
\Theta = \frac{\alpha}{1-\alpha} \pi
\end{equation}

For $\alpha > 1/3$,
the behavior of this pole is rather bizarre
since then $\Theta > \pi/2$ is required.
In that case, the addition of  a positive real self energy shifts
the singularity to a complex energy with a real part less than $vk$.
If we were to add in spin-charge separation and then
turn on $\alpha$ then 
the behavior of the pole is even more bizarre.
In that case, the phase of the Green's function
changes from $-\pi/2$ to $\pi/2 - 2 \alpha \pi$ as
$\omega$ circles the head of the branch cut
at $v_{\rho} k$.
For $\alpha > 1/6$, the pole lies
at an energy whose real part is shifted  lower
than $v_{\rho} k$, and for $\alpha > 1/4$, it is
not clear that there is a pole at all.  The phase
angle 
is less than zero  after circling
$\omega = v_{\rho} k$, although it reaches 
the positive value (for $\alpha < 1/2$)
$\pi(1-2 \alpha)$ as we move
to $\omega = v_{\sigma} k$.
Since the phase part of the pole equation
requires 
an $\omega$ which is a finite distance
away from  
$v_{\rho} k$, the Green's function is in general
finite where the phase is correct
and it is not clear that the modulus aspect
of the pole equation is satisfiable.  In fact,
for large $k$ it is not, while for sufficiently
small $k$, the Green's function can
be made large enough and eventually the pole
equation will be satisfiable.  In this case,
there 
is no physically sensible solution for large enough $k$ and $\alpha > 1/4$,
as there was none for
small $k$ and $\alpha > 1/4$. 
These peculiarities probably arise because of
the uncontrolled nature of the diagrammatic
calculation. Also,
without an entirely correct solution near $k=0$,
the analytic properties
of $G$ far from $k=0$ cannot be determined
reliably since they  depend on the
accessible decay channels 
near $k=0$. At the level of approximation
we are working at, all we can say is that
there are definite indications of incoherence
for $\alpha > 1/4$ and that such indications
of coherence as there are for
$\alpha > 1/4$ are very dubious,
particularly in spin charge separated 
models.

Now, let us follow the pole for $t_{\perp}<0$
as we increase $k$.  For $\alpha< 1/4$ this pole lay
in the second quadrant and the  phase
requirement upon increasing $k$
will push it first further off the axis
into this quadrant 
and then towards
the first quadrant,
eventually it will
cross over to the first quadrant and
a physical interpretation in terms of a quasiparticle
pole is impossible.
We can calculate at what $k$ this occurs
and find 
\begin{equation}
k_{\rm critical} = \frac{1}{v} t_{\perp}^{1/(1-2\alpha)} \cos(2\pi\alpha)
\end{equation}
(recall that we are working in units where the high
energy cut-off is $1$).
At this point the pole lies
at its last allowed $\omega$:
\begin{equation}
\omega_{\rm critical} = i\, t_{\perp}^{1/(1-2\alpha)} \sin(2\pi\alpha)
\end{equation}
We see that there are allowed $k$'s only for
$\alpha < 1/4$ and that $k_{\rm critical}$ is monotonically
decreasing with $\alpha$, so that the pole is always
unphysical for $\alpha > 1/4$.
Notice also that the correct behavior for
$\alpha = 0$ is obtained for this pole
which then moves along the real axis and crosses 
through the origin for $t_{\perp} = v k$.

In addition to this pole,
a new pole always appears for negative $t_{\perp}(k_{\perp})$
for any finite $k$.
For small enough $k$, the pole sits at
\begin{equation}
\omega \sim  -vk + \left( \frac{2 v k}{t_{\perp}} \right)^{1/\alpha} (2vk)^{-1}
\end{equation}
where the equation is valid as long as the distance of
the pole from $-vk$ is much less than $vk$.
The 
appearance of this pole is not simultaneous with the disappearance of
the other pole, except when $\alpha = 1/4$.
The new pole 
moves to some $\omega$ very close
to but less than $vk$ as we increase $k$, always remaining on
the real axis. 

This
pole must
occur because the Green's function is
always real on this region of the real $\omega$
axis in this approximation and so a pole
is guaranteed for $\alpha>1/2$ and several authors regard
this as important ({\em e.g.\/}, Boies {\em et al.\/}\cite{Bunch}).  
The appearance of this undamped pole is, however, an artifact. 
The Green's function is purely real in this region
because there
is nothing for a fermion at this momentum and
energy to decay into in the unperturbed model.  However,
in the perturbed model, at this level of approximation,
all of the other electron states have been pulled
down to lower energies also and there are many 
accessible decay
channels. These are neglected by the neglect of
vertex corrections in this approximation.  The interpretation
of this pole as physical is therefore highly dubious.
This is especially true in light of 
the pole's origin.  It not only appears discontinuously
as we increase $k$ from zero,
but if we take the limit $\alpha \rightarrow 0$
at some finite $k$ for which $2vk < t_{\perp}$,
the weight in this pole vanishes rapidly and, moreover,
the pole does not approach the position of the pole for the
$\alpha = 0$ case (it approaches $-vk$ instead).  
Conversely, the pole which is connected
continuously to the $k=0$ pole approaches the correct position.

In  addition to this strange behavior for small $k$,
the pole on the real axis has very strange properties 
in the $t_{\perp}^R \ll v_{\rho} k$ limit
for spin charge separated models.
For $\alpha > 1/4$, it
should no longer exist
since the unperturbed Green's function
does not diverge as $v_sk$ is approached.
For $\alpha < 1/4$,
the pole lies just below $v_{\sigma}k$,
at an energy
essentially unrelated to the charge velocity of
the model and the original hopping integral
in the chain direction.  Because of this,
in the large $U$ limit of the Hubbard model,
where the spin velocity vanishes,
the pole is completely dispersionless
along the chains. The ``quasiparticles'' defined by this
pole would have
zero bandwidth
in the direction of large hopping, but propagate
coherently with finite bandwidth
in the direction of small hopping!

Even in the spinless case,
one should also be cautious in interpreting this
pole because, as we have seen, it 
cannot be obtained, for
any finite $\alpha$, by a continuous deformation
from poles occurring for small $k$  and
$\alpha = 0$.
The other poles we have discussed, which do connect up
to that limit in a sensible manner, at least
in the calculable spinless case,
become unphysical  
immediately after developing a purely imaginary 
frequency.
This occurs for sufficiently large $k$ even for $\alpha < 1/4$.
Only for $\alpha = 0$ does the imaginary frequency not arise,
since, in this case, the all important branch cuts are
absent.

In summary, we see that in this approximation:
(1) there no physical poles
for $k=0$ for $\alpha > 1/4$, (2) the last physically
interpretable poles occur with purely imaginary frequencies
(3)  the undamped
pole that appears  for finite $k$ has  a number of
unphysical features,
(4) for ${\rm sgn}(t_{\perp}) \neq {\rm sgn}(k)$
there is no physically meaningful pole for $\alpha > 1/4$,
(5) for $\alpha < 1/4$ and
${\rm sgn}(t_{\perp}) \neq {\rm sgn}(k)$, the last
physically sensible pole again lies at a purely
imaginary frequency,
(6) the physical pole for
${\rm sgn}(t_{\perp}) = {\rm sgn}(k)$
and $\alpha > 1/3$ ($1/6$ for the spin-charge
separated case) lies to the wrong side of $v_{\rho} k$,
and finally (7) this pole is not even present for
large $k$ in spin-charge separated models with
$\alpha > 1/4$.
Taken together, these observations clearly
demonstrate that there is no support 
for coherence beyond $\alpha = 1/4$ in the
Green's function approach. The resummation
leading to the Green's function examined is 
an uncontrolled approximation and should be interpreted
with caution, but in so far as it is believable at all,
it seems to support the proposal of incoherent
interliquid hopping for $\alpha$'s well below the
$1/2$ required for renormalization group irrelevance.

%This is true even if one accepts as a starting point
%a perturbative study of $t_{\perp}$, 
%which is after all a relevant operator.
%A perturbative, self-energy approach 
%works in the case of free fermions
%because the $t_{\perp} = $ finite
%and $t_{\perp} = 0$ fixed points
%are so similar and because incorporating $t_{\perp}$ 
%as a self-energy resums the entire
%perturbation theory to all orders,
%since there are no interactions.  The latter
%is not true for coupled Fermi liquids
%and, in fact, incorrect analytic properties result
%for that case, however, the former still holds and
%the results can be interpreted to give more
%or less correct physics. For coupled Luttinger liquids,
%there in no obvious justification for the approach
%especially at finite $\alpha$.

It is worth noting that
many of the changes in the singularities of $G$ as we move away from
the Fermi surface parallel what occurred 
with the spectral functions in our calculation
of $\langle \delta N(t) \rangle$
when we moved away from
$\delta N(t=0) = 0$, although the  
results for the spectral functions are in general
more sensible.
Recall that a singularity appeared
for $\delta N(t=0) \neq 0$
that naively suggested coherence,
at least for intermediate $\delta N$.  The
dynamical calculation made it clear that
the increase in coherence was not as dramatic
as naive exponent counting would suggest, since
coherence depended on the properties of the spectral
function integrated out to an energy $\sim t_{\perp}^R$.
Incoherence remained a self-consistent
possibility for all $\alpha > 1/4$ (at least).  
In both the Green's function calculation
and the $\langle \delta N(t) \rangle$ calculation,
the introduction of
spin-charge separation effects is clearly
important for considering
momenta away from $k_F$ or,
equivalently, large $\delta N(t=0) $,
and it is clearly
detrimental to coherence,
removing many of the 
naively coherent behaviors for $k \neq 0$
and $\delta N(t=0) \neq 0$, respectively.

In summary, guided by  our results for the 
behavior of $\delta N$,
we have been able to identify
many strange features of the  Green's 
function approach to the coupled chains problem.
What at first sight appear to be peculiarities 
in that approach are, in fact,
fully consistent with our results
for incoherence from our
earlier calculation.
In both cases, coherence for
$\alpha > 1/4$ requires 
a leap of faith that the calculations
do not support.
%The results are not conclusive:
Note that the Green's function calculation
neglects vertex corrections and,
likewise, 
our dynamical calculation works 
to lowest order and therefore also neglects
the effects of interactions among the 
hopping fermions.  However, we know of
no reason to believe that these effects
can somehow magically restore coherence
to the interchain hopping. In fact, as we argued
before they should substantially decrease it
since fermions which hop coherently
may interact  with those hopping incoherently.
This may lead to a two particle instability
but that is by no means clear since it is not the
case that the single particle hopping is 
renormalizing to zero for $\alpha < 1/2$ (see Sec.
\ref{sec:otheroperators}).

Which approach to studying incoherence is preferable
might therefore appear to be a matter of taste.  In general,
we believe that the dynamical calculation
based upon $\delta N$ 
is more appropriate for settling questions
of coherence.
The primary reason for this is that the dynamical
calculation is specifically aimed at the question
of coherence, being closely analogous 
to the calculation done for the TLS problem.
This carries with it the advantage that what
is being calculated is the outcome of a
Gedanken experiment and is in principle physically
observable, whereas the single particle Green's
function is not.  To see why this is advantageous
consider the fact that the single particle
Green's function {\it can never be correctly obtained
directly from zero temperature perturbation theory when
the perturbation produces a Fermi surface shift}
\cite{kohn_luttinger}.
The physical reason for this is that Fermi
surface shifts, such as those predicted in
the coherence regime, are not perturbatively
accessible since they involve the rearrangement of
thermodynamically many fermions. Moreover, for
perturbations which conserve $n^1_k+n^2_k$, as
$t_{\perp}$ does, there can be no change 
at {\it any} order in perturbation theory in $n^1_k+n^2_k$.
A split Fermi surface clearly requires such a change.
On the other
hand, the dynamical question we have asked naturally
involves initial states from the $t_{\perp} = 0$
problem, which we understand, and is in principle answerable
from a direct calculation.  We need not
assume in advance what form $n^1_k+n^2_k$
will have since our choice of initial
conditions explicitly allows for coherence (but does
not force it, unlike a perturbative calculation {\it beginning}
from a split Fermi surface starting point \cite{Bunch}).
In contrast, as we have seen above, the lowest order approximation
($\Sigma=t_{\perp}$) to the self energy is generally an uncontrolled
approximation, leading to a Green's function with properties generally
difficult to interpret, and often quite bizarre.

It is worth noting that,
for the case of infinitely many coupled chains
the behavior of $n(k_x,k_y)$, which is not
identically conserved, has been
studied at lowest order in perturbation theory
by Castellani {\it et al.} \cite{Bunch}.
They find a shift of $n(k_x,k_y)$ which at
lowest order is proportional to 
$\cos(k_y) |k^F_x - k_x|^{-1+4\alpha}$
and interpret this as signaling the instability
of the Luttinger liquid.  Since they work only
to lowest order, the effects of vertex corrections,
or equivalently the effects of interactions
between hopped fermions,
are neglected. Taking this into
account, their results
are
fully consistent with ours, since they
find a qualitative change in the behavior 
exactly for $\alpha = 1/4$.  Their instability
is equivalent to the renormalization group
relevance of $t_{\perp}$, with which we agree,
and the change in behavior at $\alpha = 1/4$
would be perfectly compatible with flows
leading off to an incoherent but finite $t_{\perp}$
fixed point, rather than a coherent and finite
$t_{\perp}$ fixed point.

To conclude, there is strong evidence for incoherence in
Green's function calculations,
consistent with our dynamical calculation; 
however, the evidence is more difficult to
interpret in the former, and the dynamical calculation is more suited
to the study of coherence (or its absence).

%% file: luttliq.sec
\section{Coupled Luttinger Liquids - Exact Results}
\label{sec:luttliq}

In this section, we present exact analytic results for the interliquid
hopping rate using Luttinger liquid spectral functions. 
We are interested in the problem of $N$ coupled Luttinger liquids, 
$N\rightarrow\infty$. At $O(t_{\perp}^2)$, however, our results are
equivalent to those for $N=2$, and we therefore consider
the problem of two Luttinger liquids coupled
by a spatially uniform, single particle hopping.
The Hamiltonian is
\begin{equation}
H= H_{{\rm LL}}^{(1)}+H_{{\rm LL}}^{(2)}+t_{\perp}\sum_{x}
\{c_{\sigma}^{(1)\dag}(x)c_{\sigma}^{(2)}(x) +{\rm h.c.}\}
\end{equation}
As discussed in the previous section,
the connection to the TLS-type physics is made by first bosonizing the
Luttinger liquids, which then play the role of two baths of spin and charge
bosons. Under the bosonization the interliquid hopping operators become
exponentials of spin and charge boson creation and annihilation operators.
They resemble the operators $e^{\pm i\Omega}$ of the TLS Hamiltonian,
$H_{\rm TLS}^{\prime}$. Further, the $t_{\perp}$ operator acts to raise
the particle number of one chain by 1, and lower the other by 1, analogous
to the action of the spin flip operators in the TLS.
Moreover, the $t_{\perp}$ operator has a power law two-point function.
Thus, it is very similar to the tunneling operator $(\sigma^+e^{-i\Omega}
+\sigma^-e^{i\Omega})$ in the ohmic regime of the TLS, and 
$H_{{\rm LL}}^{(1)}+H_{{\rm LL}}^{(2)}$ plays a role similar to the
oscillator bath in the TLS.

Despite the striking similarity, however, there is no precise mapping
of $H$ to $H_{\rm TLS}$ for the simple reason that in the TLS problem
there is just a {\em single\/} tunneling particle, 
whose tunneling is associated with a single exponential in
boson creation and annihilation operators.
For fermions, the hopping can occur anywhere along
the chains and
so the operator which changes the relative particle number
of two chains is coupled to an integral over space of exponentials
in boson creation and annihilation operators.
In a TLS the tunneling particle
is {\em distinct\/} from the oscillator bath, while in the coupled LL
problem there are $N$ particles which can hop from liquid to liquid and,
moreover, these particles are themselves the {\em source\/} of the
dissipative bath.

Nevertheless, as discussed in the previous section,
it is clear that $\delta N \equiv N_2-N_1$,
the particle number difference between the two
liquids, is the most natural variable analogous to $\sigma_z$ in
the TLS, apart from the obvious difference that $\delta N$
is not two-valued. In our previous work, we introduced
the function $P(t)$ (not to be confused with the $P(t)$ in
the TLS problem) defined by
\begin{equation}
P(t)\equiv|\langle O_1O_2| e^{iH_0 t}e^{-iHt}| O_1O_2\rangle|^2
\end{equation}
Here $| O_1O_2\rangle$ denotes the product of the ground states
$| O_1\rangle$, $| O_2\rangle$ of each Luttinger liquid in the
absence of $t_{\perp}$. For $t<0$, $H= H_{{\rm LL}}^{(1)}+H_{{\rm LL}}^{(2)}$,
and at $t=0$ the interliquid hopping is turned on. The particle number
difference $\Delta N \equiv \delta N(t=0)$ 
entails a Fermi momentum difference $\Delta k$
and a chemical potential difference $\Delta\mu$.
At $O(t_{\perp}^2)$, it is of no consequence, from the computational
point of view, whether one calculates $P(t)$ or $\delta N(t)$, and
we should emphasize that in all our work, reference to $P(t)$ is
always to be taken to be at the $O(t_{\perp}^2)$ level. If one
is interested in attempting calculations 
(in particular, numerical calculations, {\em e.g.\/} \cite{mila})
beyond lowest order in $t_{\perp}$, it is clear that $\delta N(t)$
is a more appropriate function to calculate than $P(t)$. 
The latter has a far too
stringent condition on the need for the system to return to the
initial state: at $O(t_{\perp}^2)$ this restriction is innocuous,
but it clearly will not be at higher order \cite{fn:P(t)}.

The prescription, then, is to take the ground state 
$|0\rangle\equiv | O_1O_2\rangle$ in the absence of interliquid hopping
and to propagate that ground state for a time $t$. The 
nature of the
departure of the propagated state from the initial state will determine
the nature of the interliquid hopping processes.

Again, let us begin by supposing that, instead of being Luttinger liquids, 
the 1D liquids
were free Fermi gasses. Then the Hamiltonian becomes a direct product
$H=\otimes_k H_k$, where
\[
H=\left(\begin{array}{cc} E_k & t_{\perp} \\
	t_{\perp} & E_k
	\end{array}\right)
\]
so that 
\[
P(t)=\cos^{2\Delta N}(t_{\perp}t)
\]
Perturbation theory picks up the $O(t_{\perp}^2)$ term correctly,
\[
P(t)\sim 1-\Delta N t_{\perp}^2 t^2
\]
This is precisely the type of behavior for which Golden Rule 
or extended Golden Rule, {\em i.e.\/} incoherent, type methods
fail and for a very clear reason: quantum coherence is established 
separately for each $k$.
% in a very trivial way.

When interactions between electrons within a given liquid are included,
$H$ can no longer be written in this direct product form. We might
suspect, however, that for true (Landau) Fermi liquids, where the
Landau quasiparticle concept is valid, an approximate decomposition
into a direct product of quasiparticle Hamiltonians would be possible.
On the other hand, the situation for coupled Luttinger liquids, where the
quasiparticle concept completely breaks down, is not at all obvious.

The calculation of $P(t)$ using space-time Green's functions was
presented in our earlier work \cite{prl}. Here we shall 
use spectral function
methods.
This is more illuminating than the space-time Green's
function method in that the ``shape'' of the spectral
function which determines $P(t)$ 
%can be simply examined to determine the nature
provides a qualitative idea
of the interliquid hopping processes. Moreover, the spectral function method
allows us to more easily calculate key correlation functions,
and hence $P(t)$, {\em exactly\/}.

To $O(t_{\perp}^2)$ we have
\begin{equation}
1-P(t)=2t_{\perp}^2L{\rm Re}\int_0^t dt_1 \int_0^{t_1} dt_2
\int dx\left\{\langle c^{(1)}(x,t_1)c^{(1)\dag}(0,t_2)\rangle
\langle c^{(2)\dag}(x,t_1)c^{(2)}(0,t_2)\rangle +(1\leftrightarrow 2)
\right\}
\end{equation}
where the superscripts on the electron operators label the chain in
which the operator acts. The {\em interliquid hopping rate\/}
$\Gamma(t)\equiv -dP(t)/dt$
%defined by $\Gamma_{12}(t)+\Gamma_{21}(t)\equiv -dP(t)/dt$ 
can be written in a spectral function
form as
\begin{equation}
\Gamma(t)=2t_{\perp}^2L\int\frac{d\omega}{2\pi}
\frac{\sin\omega t}{\omega}\{A_{12}(\omega)+A_{21}(\omega)\}
\label{eq: gamma(t)}
\end{equation}
where
\begin{equation}
A_{ij}(\omega)=\int\frac{d\omega'}{2\pi}
\int\frac{dk}{2\pi}
{\cal J}_1^{(i)}(k,\omega'){\cal J}_2^{(j)}(k,\omega'-\omega)
\label{eq: A's}
\end{equation}
and ${\cal J}_{1,2}(k,\omega)$ are the Fourier transforms of
\begin{eqnarray*}
{\cal J}_1(k,t)&\equiv&\langle c(k,t)c^{\dag}(k,0)\rangle\\
{\cal J}_2(k,t)&\equiv&\langle c^{\dag}(k,0)c(k,t)\rangle
\end{eqnarray*}
In this paper we shall only consider the zero temperature limit, in which case
\begin{equation}
{\cal J}_{1,2}(k,\omega')=
\theta_{\pm}(\omega'-\mu)\rho(k,\omega'-\mu)
\end{equation}
where $\rho(k,\omega)$ is the electron spectral function as conventionally
defined.

Physically, $A_{12}(\omega)$ is the effective spectral function governing hops
in which an electron hops {\em to\/} liquid 1, {\em from\/} liquid 2,
and $A_{21}(\omega)$ the opposite. There is then also a natural definition
of interliquid hopping rates $\Gamma_{12}(t)$ and $\Gamma_{21}(t)$
satisfying $\Gamma_{12}(t)+\Gamma_{21}(t)=\Gamma(t)$.

At this point it should be emphasized that while, for concreteness, we restrict
our discussion to the well-defined case of coupled 1D Luttinger liquids, the
derivation of (\ref{eq: gamma(t)}) is also quite generally applicable to the 
problem of coupled 2D liquids. All that one needs to know are the effective
interliquid hopping spectral functions $A_{12}(\omega)$ and $A_{21}(\omega)$.
The definition of these given in (\ref{eq: A's}) may be easily extended to 2D
by replacing $k$ by ${\bf k}$ in the $k$-integrals and in the definition of
${\cal J}_1$ and ${\cal J}_2$.

Before presenting the calculation of $P(t)$ for coupled spin-1/2 Luttinger
liquids, let us first observe how the coherence of interliquid hopping
manifests itself in the case of coupled (Landau) Fermi liquids.
%\cite{fnFL1D}.

\subsection{Free Fermi Gasses, and Fermi Liquids}

To begin with, we note that 
for free Fermi gasses, $A_{12}(\omega)\propto\Delta\mu\delta(\omega)$ and
$A_{21}(\omega)=0$. Thus $\Gamma(t)\propto\Delta\mu~t$, a clear signal of
coherent hopping and hence of a fundamental rearrangement of the ground
state.

In a true Fermi liquid 
%one finds, using Fermi liquid spectral functions,
the (retarded) Green's function is
\[
G_R^{-1}(k,\omega)=Z^{-1}(\omega-E_k)+i\gamma\omega^2
\]
where $Z\sim(1+2\gamma\Lambda/\pi)^{-1}$ is the quasiparticle 
renormalization factor, $\gamma$ a (positive) parameter
characterizing the strength of the electron-electron interactions,
and $\Lambda$ is an ultraviolet cutoff for these interactions.
The spectral function is then given by 
$\rho(k,\omega)=-2\;{\rm Im}G_R(k,\omega)$, {\em i.e.\/}
\[
\rho(k,\omega)=\frac{2Z^2\gamma\omega^2}
{(\omega-E_k)^2+Z^2\gamma^2\omega^4}
\]
from which we obtain (see the Appendix for details)
\begin{eqnarray}
A_{12}(\omega)&\sim&\frac{1}{v_F}
\{Z^2\Delta\mu\delta(\omega)+\frac{1}{3\pi} Z^3 \gamma \omega\}
\,\theta_+(\omega+\Delta\mu) \\
A_{21}(\omega)&\sim&\frac{1}{3\pi v_F}
\,Z^3 \gamma\,\frac{(\omega-\Delta\mu)^3}{\omega^2}
\,\theta_+(\omega-\Delta\mu)
\end{eqnarray}
Note that $Z\rightarrow 1$ if and only if $\gamma\rightarrow 0$,
recovering the free Fermi gas result.
From the point of view of coherence, only $\Gamma_{12}(t)$ is
of interest. We see that
$\Gamma_{12}(t)$ is a sum of a term $\propto Z^2\Delta\mu t$
representing fundamentally coherent processes, and a term 
$\propto\gamma Z^3 t^{-1}$ which is on the border of incoherent 
and irrelevant ({\em i.e.\/}, it is marginal).
By choosing a sufficiently small $t_{\perp}$ one can
find a time $t$ such that $(1-P(t))/N\ll 1$ ({\em i.e.\/} we are not
outside of the reasonable range of our $O(t_{\perp}^2)$ expansion),
yet the ratio of the coherent contribution to the incoherent contribution
is arbitrarily large. This is true {\it regardless of how small $Z$ is\/}.
Thus, a perturbative calculation in $t_{\perp}$ does not reveal any
likelihood of a loss of coherence of interliquid tunneling, and there is
therefore no impediment to the formation of an interliquid band of
width $\sim Zt_{\perp}$. This result is entirely consistent with
what we would expect from a calculation based upon (Landau) quasiparticles,
and is but another illustration of the power of the quasiparticle concept.

\subsection{Luttinger Liquids}

We now turn to the problem proper, that of Luttinger liquids coupled
by interliquid, single-particle hopping. In order to calculate $A_{12}(\omega)$
and $A_{21}(\omega)$ we need the spectral functions ${\cal J}_1(\omega)$ and
${\cal J}_2(\omega)$ for a Luttinger liquid.

In \cite{prl} we used the space-time Luttinger
liquid Green's functions, $G(x,t)$, to calculate $P(t)$, since
these are more directly calculated within the bosonization framework
than are the corresponding $G(k,\omega)$. The electron spectral function
$\rho(k,\omega)$
{\em can\/} be calculated by direct Fourier transform of $G(x,t)$
\cite{spectral}, but
it can actually be determined in a much simpler way
which we shall now describe.
The idea is to write the
electron space-time Green's function as a product of ``fracton''
Green's functions, where the fracton operators are exponentials of spin
and charge boson operators. 
In contrast to the electron spectral function,
the fracton spectral functions are sharp
$\delta$-functions. By ``fusing'' the fracton spectral functions together
via convolution, one obtains simple integral expressions for the electron
spectral function. From our point of view, the great utility of this method
over the space-time approach
is that it provides a method of explicitly obtaining the effective interliquid hopping
spectral functions $A_{12}(\omega)$ and $A_{21}(\omega)$. These spectral functions
have direct physical significance and can be used to obtain exact expressions
for $P(t)$.

The general asymptotic form of the space-time electronic Green's function in
a Luttinger liquid is
\begin{equation}
\langle\psi_{R,\uparrow}(x,t)\psi^{\dag}_{R,\uparrow}(0,0)\rangle
=\frac{e^{i(k_Fx-\mu t)}}{2\pi a}
\frac{(i\pi a/v_c\beta)^{\alpha+1/2}}
{\left(\sinh\left[\frac{\pi(x+ia-v_ct)}{v_c\beta}\right]\right)^{\alpha+1/2}}
\frac{(i\pi a/v_c\beta)^{\alpha}}
{\left(\sinh\left[\frac{\pi(x+ia+v_ct)}{v_c\beta}\right]\right)^{\alpha}}
\frac{(i\pi a/v_s\beta)^{1/2}}
{\left(\sinh\left[\frac{\pi(x+ia-v_st)}{v_s\beta}\right]\right)^{1/2}}
\label{eq: Greens function}
\end{equation}
where $\beta$ is inverse temperature, $\alpha$ is the so-called ``anomalous''
(or Luttinger liquid) exponent, 
$v_c$, $v_s$ are the charge- and spin-velocities, and $a$ is a short distance
cutoff. 
Recall that $2\alpha$ is the
exponent which characterizes the singularity in $n(k)$ near $k_F$ (see
(\ref{n(k)})) and is not to be confused with the $\alpha$
used in the discussion of the TLS problem.

The expression (\ref{eq: Greens function})
is for right moving electrons (hence the subscript `R');
but this is trivially changed to produce the left-moving part and hence the
full Green' function. For our purposes, however, we need only consider the
right-moving part.

For non-interacting electrons, $v_s=v_c=v_F$ and $\alpha=0$. In this case
we recover the free fermion propagator from (\ref{eq: Greens function}).
In the Hubbard model, $\alpha\sim U^2$, $v_c-v_s \sim U$ for small $U$, and
$0< \alpha<1/16$ for all $U$. Larger exponents can occur, for
example in extended Hubbard models.

The key observation to make at this point
is that the Green's function is a product of
fracton correlation functions,
\begin{equation}
\langle\psi_{R,\uparrow}(x,t)\psi^{\dag}_{R,\uparrow}(0,0)\rangle
=\frac{e^{i(k_Fx-\mu t)}}{2\pi a}
\left[G_{c+}^{(\alpha+1/2)}(x,t)\right]
\left[G_{c-}^{(\alpha)}(x,t)\right]
\left[G_{s}^{(1/2)}(x,t)\right]
\label{eq: fracton_product}
\end{equation}
We shall call $G^{(p)}(x,t)$ the Green's function of a fracton of 
{\em weight p\/} (further discussion may be found in the Appendix).
The spectral function associated with $G^{(p)}(x,t)$ is therefore\begin{equation}
{\cal J}_1^{(p)}(k,\omega)=\left(\frac{i\pi a}{v\beta}\right)^p
\;2\pi\delta(\omega-vk)\int_{-\infty}^{\infty}
\frac{dz\;e^{-ikz}}{\left[\sinh\left(\frac{\pi(z+ia)}{v\beta}\right)\right]^p}
\end{equation}
where $v$ is the velocity of the relevant charge or spin bosonic excitation. 
In the zero temperature limit this simplifies to
\begin{equation}
{\cal J}_1^{(p)}(k,\omega)
\stackrel{\beta\rightarrow\infty}{=}
 \frac{4\pi^2}{\Gamma(p)}a^p\;k^{p-1}\theta_+(\omega)\delta(\omega-vk)
\end{equation}
%Complete details of the evaluation are given in the Appendix.

It is then a simple matter to obtain the {\em electron\/} spectral function
by convolving the appropriate fracton spectral functions as determined by
(\ref{eq: fracton_product}). We find
\begin{eqnarray}
{\cal J}_1&(k,\omega)&\propto\int dk_1 dk_2 dk_3
\int d\omega_1 d\omega_2 d\omega_3
\delta(\omega-\mu-\sum_i\omega_i)\delta(k-k_F-\sum_i k_i)\nonumber\\
&&\delta(\omega_1-v_ck_1)\theta_+(\omega_1)(\omega_1/v_c)^{\alpha-1/2}
\delta(\omega_2-v_sk_2)\theta_+(\omega_2)(\omega_2/v_s)^{-1/2}
\delta(\omega_3+v_ck_3)\theta_+(\omega_3)(\omega_3/v_c)^{\alpha-1}
\nonumber\\
&\propto&
\int_0^{\infty} d\omega_1 d\omega_2 d\omega_3
\delta(\omega-\mu-\sum_i\omega_i)
\delta\left(k-k_F-\frac{(\omega_1-\omega_3)}{v_c}-\frac{\omega_2}{v_s}\right)
\nonumber\\
&&(\omega_1/v_c)^{\alpha-1/2}
(\omega_2/v_s)^{-1/2}
(\omega_3/v_c)^{\alpha-1}
\label{eqn: electron spectral function}
\end{eqnarray}
The various singularities near $\omega=\pm v_ck,~v_sk$ can be readily 
determined from this expression and are illustrated in figure \ref{fig:spectral1}.

We again emphasize that the fracton spectral functions are sharp, due
to the fact that a fracton is an exponential of one of the free bosonic
fields which have well defined energy-momentum dispersion relations.
It is therefore quite natural
in this language to think of the electron as a 
{\em composite\/} particle made up of 
{\em spin- and charge-fracton quasiparticles\/}.

For reasons of pedagogy, it is convenient to first consider the case of 
{\em chiral\/} Luttinger liquid models which are forward-scattering-only (FSO), 
{\em i.e.\/} 
there
is no coupling between left- and right-moving electrons.
Then we shall consider the generic Luttinger liquid case.

\subsubsection{Chiral Luttinger liquid}

This model exhibits spin-charge separation, but no anomalous exponent
\cite{fabrizio_parola}.
The eigenexcitations are spin and charge bosons with
velocities $v_s$ and $v_c$, respectively, and $v_c-v_s\equiv\Delta v >0$.
The electron Green's function may be obtained from 
(\ref{eq: Greens function}) by simply setting $\alpha\rightarrow 0$.
There is no ``holonic backflow'', and so the electron spectral function may
be written as the convolution of the spin fracton spectral function and a 
single charge fracton spectral function
\begin{eqnarray}
{\cal J}_{1,2}(k,\omega)=\frac{1}{(2\pi)^2}&&\int dk_1 \int dk_2
\int d\omega_1 \int d\omega_2 
{\cal J}_{1,2}^{(c)}(k_1,\omega_1){\cal J}_{1,2}^{(s)}(k_2,\omega_2)
\nonumber\\
&&\delta(\omega-\mu-\omega_1-\omega_2)\delta(k-k_F-k_1-k_2)
\end{eqnarray}
where
\begin{eqnarray}
{\cal J}_1^{(c)}(k,\omega)
=4\pi^{3/2}a^{1/2}\theta_+(\omega)\delta(\omega-v_c k)
\left(\frac{\omega}{v_c}\right)^{-1/2} \\
{\cal J}_1^{(s)}(k,\omega)
=4\pi^{3/2}a^{1/2}\theta_+(\omega)\delta(\omega-v_s k)
\left(\frac{\omega}{v_s}\right)^{-1/2} \\
{\cal J}_2^{(c)}(k,\omega)
=4\pi^{3/2}a^{1/2}\theta_-(\omega)\delta(\omega-v_c k)
\left(\frac{-\omega}{v_c}\right)^{-1/2} \\
{\cal J}_2^{(s)}(k,\omega)
=4\pi^{3/2}a^{1/2}\theta_-(\omega)\delta(\omega-v_s k)
\left(\frac{-\omega}{v_s}\right)^{-1/2} 
\end{eqnarray}
We have changed notation slightly, the superscripts `c' and `s' referring
to `charge' and `spin'.
Note that, as in the case of a free Fermi gas, there is no response to
$t_{\perp}$ if $\Delta\mu=0$, {i.e.\/} both $A_{12}(\omega)$ and
$A_{12}(\omega)$ vanish.
This is the result of there being a step-function
singularity at the Fermi surface, introducing the well-known problem of
the inability to incorporate ``anomalous diagrams'' in
zero temperature perturbation theory \cite{kohn_luttinger}. 

For $\Delta\mu > 0$ we find (see Appendix) the rather simple expression
\begin{eqnarray}
A_{12}(\omega)&=&\frac{1}{\Delta v}\theta_+((v_c-v)\Delta k-\omega)
\theta_+(\omega-(v_s-v)\Delta k) \nonumber \\
A_{21}(\omega)&=& 0 \label{eq: chiral eff spectral}
\end{eqnarray}
where we have introduced the velocity $v$ corresponding
to the chemical potential shift, $\Delta\mu\equiv v\Delta k$.
That is, $A_{12}(\omega)$ has constant non-vanishing weight in the interval
$\omega\in[(v_s-v)\Delta k,(v_c-v)\Delta k]$. Note that, in the
limit $\Delta v\rightarrow 0$, this step function of width $\Delta v$, height
$\propto 1/\Delta v$, goes over to a $\delta$-function, as it should, for
the limit $\Delta v\rightarrow 0$ is the limit of free electrons.
For $\Delta v\neq 0$, $A_{12}(\omega)$ is peaked around $\omega=0$,
but has a width $\tau^{-1}_{\Delta k}\sim\Delta k\Delta v$.

It is then a simple matter to calculate
$\Gamma(t)$ for a chiral Luttinger liquid. We have, using
(\ref{eq: gamma(t)}),
\begin{eqnarray}
\Gamma(t)&=&\frac{1}{\pi\Delta v}t_{\perp}^2 L 
\int_{(v_s-v)\Delta k}^{(v_c-v)\Delta k} d\omega\frac{\sin\omega t}{\omega}
\nonumber \\
&=&\frac{1}{\pi\Delta v}t_{\perp}^2 L \left\{
{\rm Si}((v_c-v)\Delta k t)-{\rm Si}((v_s-v)\Delta k t)\right\}
\end{eqnarray}
where ${\rm Si}(x)$ is the Sine-integral function. 
The behavior of ${\rm Si}(x)$ for small $|x|$ is ${\rm Si}(x)\sim x$ so that
provided $(v_c-v)\Delta k t\ll 1$ and $(v-v_s)\Delta k t \ll 1$, we have
\begin{eqnarray}
\Gamma(t)&\sim&\frac{1}{\pi\Delta v}t_{\perp}^2 L \Delta v\Delta k t
\nonumber \\
&=&\Delta N t_{\perp}^2 t
\end{eqnarray}
Thus, for all times $t~^{<}_{\sim}\tau_{\Delta k}$, $\Gamma(t)$ behaves just as
it does for coupled free Fermi gasses. Moreover, $\tau_{\Delta k}$ can be 
made arbitrarily long by choice of sufficiently small $\Delta k$,
while remaining in the perturbative regime, $N^{-1}\int_0^t\Gamma(t')dt'\ll 1$.
The physics here is that $\tau_{\Delta k}$ plays the role of an intrinsic 
lifetime due to spin-charge separation, but this lifetime diverges as $k_F$
is approached ({\em i.e.\/} as $\Delta k\rightarrow 0$). This is the only
lifetime in the model.

We conclude, therefore, that for coupled chiral Luttinger liquids, 
as for coupled Fermi liquids, 
there is no obvious impediment to
interliquid coherence for arbitrarily small $t_{\perp}$.

%the essential physics can be obtained by inspection. In the limit
%of $\Delta v\rightarrow 0$, $A_{12}(\omega)\rightarrow\delta(\omega)$.

\subsubsection{Spinless Luttinger liquid}

After the chiral case, the next simplest Luttinger liquid to consider is that
formed by interacting spinless fermions. The space-time Green's function
has the form (\ref{eq: Greens function})
 with $v_c=v_s$ ({\em i.e.\/} $\Delta v=0$). In place of
(\ref{eqn: electron spectral function}) we have
\begin{equation}
{\cal J}_1(k,\omega)\propto
%\int dk_1 dk_2
\int_{0}^{\infty} d\omega_1 d\omega_2
\delta(\omega-\mu-\omega_1-\omega_2)
\delta\left(k-k_F-\frac{(\omega_1-\omega_2)}{v_c}\right)
%&&
%\theta_+(\omega_1)
(\omega_1/v_c)^{\alpha}
%\delta(\omega_1-v_ck_1)
%\theta_+(\omega_2)
(\omega_2/v_c)^{\alpha-1}
%\delta(\omega_2+v_ck_2)
\end{equation}
for the spinless fermion spectral function. 
The hole spectral function is 
\begin{equation}
{\cal J}_2(k,\omega)\propto
%\int dk_1 dk_2
\int_{0}^{\infty} d\omega_1' d\omega_2'
\delta(\omega-\mu+\omega_1'+\omega_2')
\delta\left(k-k_F+\frac{(\omega_1'-\omega_2')}{v_c}\right)
%&&
%\theta_+(\omega_1)
(\omega_1'/v_c)^{\alpha}
%\delta(\omega_1-v_ck_1)
%\theta_+(\omega_2)
(\omega_2'/v_c)^{\alpha-1}
%\delta(\omega_2+v_ck_2)
\end{equation}

The effective interliquid 
hopping spectral functions $A_{12}(\omega)$ and $A_{21}(\omega)$
are then obtained from (\ref{eq: A's}). We find
\begin{eqnarray}
A_{12}(\omega)&=&
\frac{1}{\Gamma(2\alpha)\Gamma(2+2\alpha)}
\left(\frac{a}{2v_c}\right)^{4\alpha} 
\frac{1}{2v_c}
\theta_+(\omega-(v_c-v)\Delta k)
(\omega+(v_c+v)\Delta k)^{2\alpha+1}(\omega-(v_c-v)\Delta k)^{2\alpha-1}
 \\
A_{21}(\omega)&=&
\frac{1}{\Gamma(2\alpha)\Gamma(2+2\alpha)}
\left(\frac{a}{2v_c}\right)^{4\alpha} 
\frac{1}{2v_c}
\theta_+(\omega-(v_c+v)\Delta k)
(\omega-(v_c+v)\Delta k)^{2\alpha+1}(\omega+(v_c-v)\Delta k)^{2\alpha-1}
\end{eqnarray}
In the noninteracting case ($\alpha=0$ and $v_c=v$) $A_{21}(\omega)$
vanishes and so interliquid hopping occurs only from liquid 2 to liquid 1.
It is straightforward to see that for $\alpha\neq 0$ we need only examine
$A_{12}(\omega)$, and hence $\Gamma_{12}(t)$, to address the coherence issue.
Representative plots of $A_{12}(\omega)$ for various $\alpha$ are shown in
Fig. \ref{fig: inter_spectral_spinless}.
%For a noninteracting gas, $A_{12}(\omega)$ is proportional to $\delta(\omega)$,
%signalling coherent interliquid hopping. We 

In calculating $\Gamma_{12}(t)$ it is simplest first of all to consider
its time derivative, since this is just a Fourier transform of the spectral
function $A_{12}(\omega)$, and therefore easier to deal with analytically:
\begin{equation}
\frac{d\Gamma_{12}(t)}{dt}=
\frac{t_{\perp}^2 L}{\pi}\int_{-\infty}^{\infty}d\omega\;\cos\omega t\;
A_{12}(\omega)
\end{equation}
Inserting the above expression for $A_{12}(\omega)$ gives (omitting prefactors)
\begin{equation}
\frac{d\Gamma_{12}(t)}{dt}\propto
t^{-(1+4\alpha)}{\rm Re}\left\{
e^{i(v_c-v)\Delta kt}\int_0^{\infty}d\theta\;\theta^{2\alpha-1}
(\theta+x)^{2\alpha+1} e^{i\theta}\right\}
\end{equation}
where, for convenience, we have introduced the variable $x=2v_c\Delta k t$.

The integral may be done via contour integration, and the final result 
 is
\begin{eqnarray}
\frac{d\Gamma_{12}(t)}{dt}
&=&\frac{t_{\perp}^2 L}{\pi}
\frac{1}{\Gamma(2\alpha)\Gamma(2+2\alpha)}
\left(\frac{a}{2v_c}\right)^{4\alpha} 
\frac{1}{2v_c}\frac{1}{\Gamma(1-2\alpha)}t^{-(1+4\alpha)} \nonumber\\
&&{\rm Re}\{
e^{i(v_c-v)\Delta kt}
[ie^{i2\pi\alpha}\Gamma(1-2\alpha)\Gamma(1+4\alpha)~_1F_1(-1-2\alpha,-4\alpha;-ix)
\nonumber \\
&&
+\frac{1}{2}\frac{(1+2\alpha)}{(1+4\alpha)}
\Gamma(2\alpha)\Gamma(1-4\alpha)\;
x^{1+4\alpha}~_1F_1(2\alpha,2+4\alpha;-ix)
]\}
\label{eq: spinless exact}
\end{eqnarray}
where $~_1F_1$ is the confluent hypergeometric function.

Equation (\ref{eq: spinless exact}) is an {\em exact\/} result for
the (time derivative of the) interliquid hopping rate, to lowest
order in $t_{\perp}$, obtained by the use of fracton spectral functions.
It allows us to extract the important physics. To begin with, even
when the particle number difference $\Delta N$ vanishes ({\em i.e.\/}
$\Delta k=0$) there are {\em incoherent\/} interliquid transitions, for
\begin{equation}
\frac{d\Gamma_{12}^{\Delta k=0}(t)}{dt}=
-\frac{t_{\perp}^2 L}{\pi}
\frac{\Gamma(1+4\alpha)}{\Gamma(2\alpha)\Gamma(2+2\alpha)}
\sin(2\pi\alpha)\left(\frac{a}{2v_c}\right)^{4\alpha}\frac{1}{2v_c}
t^{-(1+4\alpha)}
\end{equation}
thus
\begin{equation}
\Gamma_{12}^{\Delta k=0}(t)=
\frac{t_{\perp}^2 L}{\pi}
\frac{\Gamma(1+4\alpha)}{\Gamma(2\alpha)\Gamma(2+2\alpha)}
\frac{\sin(2\pi\alpha)}{4\alpha}
\left(\frac{a}{2v_c}\right)^{4\alpha}\frac{1}{2v_c}
t^{-4\alpha}
\end{equation}
Such relevant, incoherent interliquid hopping does 
not occur for coupled
Fermi liquids, nor for chiral Luttinger liquids. 
In a Fermi liquid,
coherent interliquid hopping manifests itself within perturbation theory via
a $t_{\perp}^2\Delta N\,t$ contribution to $\Gamma_{12}(t)$, so we need
to examine the $\Delta k$ dependence of (\ref{eq: spinless exact}).

Upon using the expansion 
\[
~_1F_1(a,b;z)=1+\frac{a}{b}\frac{z}{1!}+\frac{a(a+1)}{b(b+1)}\frac{z^2}{2!}
+\ldots
\]
we deduce from (\ref{eq: spinless exact}) that as $\alpha\rightarrow 0$ only 
the $O(x)$ and $O(x^{1+4\alpha})$ terms survive: all other terms are
proportional to $\alpha$. Indeed, one can readily check that the contribution
of these two terms correctly recovers the free fermion result in the limit
$\alpha\rightarrow 0$ (and $v_c-v\rightarrow 0$).
Moreover, it makes little physical sense to suppose that terms of 
$O(x^2)$ or higher ({\em i.e.\/} terms of $O(n^2)$ or higher, where 
$n=\Delta N/L$ is the density difference between the liquids) are important
in determining the coherence or incoherence of {\em single particle\/} hopping.
We therefore retain only the $O(x^0)$, $O(x)$ and $O(x^{1+4\alpha})$ terms.
Again, the latter two continuously develop into the free fermion limit of
coherent interliquid single particle hopping  as $\alpha\rightarrow 0$.

Retaining only these terms gives
\begin{eqnarray}
\frac{d\Gamma_{12}(t)}{dt}
&=&\frac{t_{\perp}^2 L}{\pi}
\frac{1}{\Gamma(2\alpha)\Gamma(2+2\alpha)}
\left(\frac{a}{2v_c}\right)^{4\alpha} 
\frac{1}{2v_c}\,\cos[(v_c-v)\Delta k t]\,t^{-(1+4\alpha)} \nonumber\\
&&\{(1+2\alpha)
\left\{-\sin(2\pi\alpha)\frac{\Gamma(1+4\alpha)}{(1+2\alpha)}
+\cos(2\pi\alpha)\Gamma(4\alpha)\,x
+\frac{\Gamma(2\alpha)\Gamma(1-4\alpha)}{2(1+4\alpha)\Gamma(1-2\alpha)}\,
x^{1+4\alpha}\right\} \nonumber \\
&&-\tan[(v_c-v)\Delta k t]\,\cos(2\pi\alpha)\Gamma(1+4\alpha)\}
\label{eqn: spinless gamma dot}
\end{eqnarray}
For $2\pi\alpha\ll 1$ we have
\begin{equation}
\frac{d\Gamma_{12}(t)}{dt}\approx\frac{t_{\perp}^2 L}{\pi}
\frac{1}{2v_c}\,\cos[(v_c-v)\Delta k t]\,t^{-(1+4\alpha)}
\left\{-4\pi\alpha^2+v_c\Delta k\,t
+\frac{1}{2}(2v_c\Delta k\,t)^{1+4\alpha}\right\}
\end{equation} 
%whence
%\begin{equation}
%\Gamma_{12}(t)\approx\frac{t_{\perp}^2 L}{\pi}
%\frac{1}{2v_c}\,\cos[(v_c-v)\Delta k t]
%\left\{\pi\alpha\,t^{-4\alpha}
%+\frac{1}{(1-4\alpha)}v_c\Delta k\,t^{1-4\alpha}
%+\frac{1}{2}(2v_c\Delta k)^{1+4\alpha}\,t\right\}
%\end{equation}

Let us now discuss the physics of each of the terms in 
(\ref{eqn: spinless gamma dot}).
%We note that a term in
%$d\Gamma_{12}(t)/dt$ decaying faster than $t^{-1}$ describes incoherent
%interliquid hops.
As already remarked, the
$\Delta k$-independent term represents fundamentally incoherent
interliquid hops. 
By analogy with the situation in the TLS \cite{lowa},
if the term linear in $\Delta k$ decays slower than
$t^{-1}$ it should be interpreted
as a potentially coherent term. For $\alpha > 1/2$,
the term linear in $\Delta k$
is irrelevant in the sense that its contribution to $P(t)$ is long-time
convergent.
This is consistent with the renormalization group characterization of the
relevance/irrelevance of the interliquid hopping operator.

The $O(\Delta k^{1+4\alpha})$ term requires careful thought to interpret
when $\alpha\neq 0$. The key thing to note is that the oscillatory prefactor
$\cos[(v_c-v)\Delta k t]$ will force $\Gamma_{12}(t)$ to be essentially
time-independent for times $t~^>_{\sim}[(v_c-v)\Delta k]^{-1}$. This prefactor
is analogous to the prefactor which occurs in the tunneling rate of a TLS
when the two levels are not energy degenerate. It indicates that, unlike
for hopping between Fermi liquids, hopping between Luttinger liquids is never
an energy degenerate process. This is another important factor working against
coherent interliquid hopping.

We now utilize this oscillatory prefactor to interpret the 
$O(\Delta k^{1+4\alpha})$ term. In order for this term to be of
coherent type, one must remain at times short enough to avoid the
cutoff effect of the $\cos[(v_c-v)\Delta k t]$ prefactor. The maximum
possible $\Delta k$ for a given time $t$ is 
$\Delta k^{\rm max}\sim [(v_c-v) t]^{-1}$. The $O(\Delta k^{1+4\alpha})$ term
in $d\Gamma/dt$ is therefore bounded by 
$\sim \Delta k\,t^{-4\alpha}/(v_c-v)^{4\alpha}$ which has the same form
as the term linear in $\Delta k$. It therefore suffices to consider only
the latter term.

We now observe that for $\alpha > 1/4$ the 
$O(\Delta k)$ term in $d\Gamma/dt$ behaves as $t^{-4\alpha}$,
which is weaker than $t^{-1}$.
In fact, for $\alpha=1/4$ this term is identically zero,
reflecting the fact that at this particular value of $\alpha$,
the linear-in-$\Delta k$ contribution to $\Gamma(t)$ is actually
time-independent.
Therefore, for  $1/4 \leq\alpha < 1/2$ \cite{lowa} the $O(t_{\perp}^2)$
calculation indicates that interliquid single particle hopping
is completely incoherent, {\em despite the relevance of 
$t_{\perp}$ in the RG sense\/}. This is completely analogous
to the regime in the TLS model where inter-state tunneling 
is relevant but non-oscillatory. 
Indeed, ignoring the $\cos[(v_c-v)\Delta k t]$ prefactor for the moment,
if the term linear in $\Delta k$ was the {\em only\/} contribution
to $\Gamma_{12}$ the result would be formally equivalent, at $O(t_{\perp}^2)$,
to that of $\Delta N$ TLS's in parallel. The modification from the FL
result $\Gamma(t)\sim\Delta N\,t_{\perp}^2 t$ to a behavior
$\Gamma(t)\sim\Delta N\,t_{\perp}^2 \Lambda ^{4\alpha} t^{1-4\alpha}$
($\Lambda\sim a/v_c$ being an UV cutoff)
would be completely analogous to the behavior of a TLS upon turning on
coupling to the ohmic bath. In particular, in the nomenclature
of the TLS problem
the interliquid transport of electrons would be localized when 
$\alpha>1/2$, purely incoherent when $1/4<\alpha<1/2$, and coherent
if $0\leq\alpha<1/4$. At the very least then, there would be a nontrivial 
incoherent interliquid hopping phase for $1/4 < \alpha < 1/2$.
However, unlike in the (degenerate) TLS problem,
for the problem of coupled Luttinger liquids
there are additional
factors enhancing incoherence over and above the time exponent of
the $O(\Delta k)$ term.
The question then arises as to the nature of interliquid transport in
the regime $0<\alpha<1/4$.

The factors enhancing incoherence are as follows. First of all, there are
the incoherent processes contributing to the $\Delta k$-independent term.
This term represents a ``channel'' in which the electrons hop in a purely
incoherent way from liquid to liquid.

Secondly, as already remarked upon, there is a ``dephasing'' prefactor
$\cos[(v_c-v)\Delta k t]$ which is analogous to a biassing term in a TLS.
This will quite generally enhance incoherence.

Thus, for $0 < \alpha <1/4$ the interliquid hopping rate is essentially
a sum of incoherent and coherent parts, 
$\Gamma_{12}(t)=\Gamma_{12}^{\rm incoh}(t) + \Gamma_{12}^{\rm coh}(t)$.
Alternatively, one can consider the integrated transition probabilities
up to time $t$ and write $P_{12}(t)=P_{12}^{\rm incoh}(t)+P_{12}^{\rm coh}(t)$.
The key point to recognize is that the coherent term remains so only
for times $t~^<_{\sim}[(v_c-v)\Delta k]^{-1}$. As such, $P_{12}^{\rm coh}(t)$
is bounded above in magnitude by 
$\sim t_{\perp}^2 v_c \Lambda ^{4\alpha} t^{1-4\alpha}/(v_c-v)$ so that,
approximately,
%roughly speaking, 
\[
\frac{P_{12}^{\rm incoh}(t)}{P_{12}^{\rm coh}(t)}
~^{>}_{\sim}\,\alpha\frac{(v_c-v)}{v_c}
\]
which is {\em independent\/} of $t_{\perp}$.
In particular, the purely incoherent channel cannot eliminated in the 
$t_{\perp}\rightarrow 0$ limit, as it would have been were the dephasing
factor absent. Thus, as we decrease $\alpha$ from $\alpha=1/4$ we expect
incoherence to be stabilized down to some critical value $\alpha_c<1/4$
by a combination of the purely incoherent term,
which will always dominate if $\Delta k$ is too small, and the 
dephasing prefactor of the TLS-like coherent term linear in $\Delta k$
which kills coherence of this term if $\Delta k$ is too large. 
The reader should contrast this with the situation in a chiral Luttinger
liquid discussed earlier. In that case, the absence of a purely incoherent 
interliquid hopping term allows $\Delta k$ to be chosen sufficiently small
for a given $t_{\perp}$ that dephasing factors can be suppressed to an arbitrary
degree.

Finally, there is a further effect which will push $\alpha_c$ even
lower, namely the influence of interliquid hops upon one another via
intraliquid interactions. Interhop correlations (effectively absent
for hopping of electrons between Fermi liquids) are not included in our
$O(t_{\perp}^2)$ calculation.  It is physically clear that such correlations
will only hinder coherence and an estimate of their effect has already been given in
the previous section. In particular, if the effect of incoherent hops in
disrupting potentially coherent hops is such that the the amplitude
for the latter is premultiplied by any 
function of time asymptoting to zero
for $t \gg (t_{\perp}^R)^{-1}$, it can be shown that in the limit 
$t_{\perp}\rightarrow 0$ the incoherent phase extends over all
$0 < \alpha <1/2$. In that case, one would have $\alpha_c=0$
%which would be a spectacular statement to the effect that 
and the operations
of turning on in-liquid interactions and of turning on interliquid hopping
would not commute.

%%%%%%% end of spinless %%%%%%%%%

\subsubsection{Generic (Spinny) Luttinger liquid}

We now turn to the generic case where there is spin-charge separation, 
$v_c-v_s=\Delta v > 0$, {\em and\/} an anomalous exponent, $\alpha$.
This case is relevant, for example, to the 1D Hubbard model, and to
most physical models which are not ``chiral''. 

For ease of notation we shall omit constants of proportionality in most
expressions, restoring them only at the end. To begin with, we have the electron
spectral functions
\begin{eqnarray}
{\cal J}_1(k,\omega)&\propto&
\int_0^{\infty} d\omega_1 d\omega_2 d\omega_3
\delta(\omega-\mu-\sum_i\omega_i)
\delta\left(k-k_F-\frac{(\omega_1-\omega_3)}{v_c}-\frac{\omega_2}{v_s}\right)
\nonumber\\
&&(\omega_1/v_c)^{\alpha-1/2}
(\omega_2/v_s)^{-1/2}
(\omega_3/v_c)^{\alpha-1}
\nonumber
\end{eqnarray}
and
\begin{eqnarray}
{\cal J}_2(k,\omega)&\propto&
\int_0^{\infty} d\omega_1' d\omega_2' d\omega_3'
\delta(\omega-\mu+\sum_i\omega_i')
\delta\left(k-k_F+\frac{(\omega_1'-\omega_3')}{v_c}+\frac{\omega_2'}{v_s}\right)
\nonumber\\
&&(\omega_1'/v_c)^{\alpha-1/2}
(\omega_2'/v_s)^{-1/2}
(\omega_3'/v_c)^{\alpha-1}
\nonumber
\end{eqnarray}
The effective spectral function for hopping is then given by (\ref{eq: A's})
\begin{eqnarray}
&A_{12}(\omega)&\propto\int\frac{d\omega'}{2\pi}\int\frac{dk}{2\pi}
\int_0^{\infty} d\omega_1 d\omega_2 d\omega_3
\delta(\omega'-\sum_i\omega_i)
\delta\left(k-k_F-\frac{(\omega_1-\omega_3)}{v_c}-\frac{\omega_2}{v_s}\right)
\omega_1^{\alpha-1/2}
\omega_2^{-1/2}
\omega_3^{\alpha-1}
\nonumber\\
&&\int_0^{\infty}d\omega_1' d\omega_2' d\omega_3'
\delta(\omega-\omega'-\Delta\mu+\sum_i\omega_i')
\delta\left(k-k_F+\frac{(\omega_1'-\omega_3')}{v_c}+\frac{\omega_2'}{v_s}\right)
(\omega_1')^{\alpha-1/2}
(\omega_2')^{-1/2}
(\omega_3')^{\alpha-1} \nonumber
\end{eqnarray}
Some straightforward, if laborious, algebra (see Appendix) reduces this to
\begin{eqnarray}
A_{12}(\omega)&\propto& I(a,b) \nonumber \\
&=&\int_0^{\infty}dx\int_0^{\infty}dy x^{-1/2}y^{-1/2}
(a-(x+y))^{2\alpha}(b+(x+y))^{2\alpha-1}\theta_+(a-(x+y))\theta_+(b+(x+y))
\nonumber
\end{eqnarray}
where
\begin{eqnarray}
a&=&\frac{v_s}{\bar{v}}(\omega+(v+v_c)\Delta k)\nonumber\\
b&=&\frac{v_s}{\Delta v}(\omega+(v-v_c)\Delta k)\nonumber
\end{eqnarray}
and we have introduced $\bar{v}\equiv v_c+v_s$.

Further simplification of the double integral is possible, and we finally
arrive at the following exact result, illustrated in
Fig. \ref{fig: inter_spectral_spinny}
\begin{eqnarray}
&=& 0,\;\;\; \omega < (v_s-v)\Delta k \nonumber \\
A_{12}(\omega)&=&\frac{1}{\Gamma(1+4\alpha)}
\left(
\frac{a^2}{\bar{v}\Delta v}\right)^{2\alpha}
\left(\frac{1}{\Delta v}\right)
(\omega+(v-v_s)\Delta k)^{4\alpha},
\;\;\;(v_s-v)\Delta k < \omega < (v_c-v)\Delta k 
\label{eq: LL eff spectral}\\
&=&\frac{1}{(1+2\alpha)}\frac{1}{\Gamma(2\alpha)\Gamma(1+2\alpha)}
\frac{1}{\bar{v}}
\left(\frac{a}{2v_c}\right)^{4\alpha} 
%\frac{1}{\bar{v}(v_c)^{4\alpha}}
(\omega+(v_c+v)\Delta k)^{2\alpha+1}(\omega-(v_c-v)\Delta k)^{2\alpha-1}
\nonumber \\
&&_2F_1\left(1, 1-2\alpha;2+2\alpha;-\left(\frac{\Delta v}{\bar{v}}\right)
\left[\frac{\omega+(v_c+v)\Delta k}{\omega-(v_c-v)\Delta k}\right]\right),
\;\;\;\omega > (v_c-v)\Delta k \nonumber
\end{eqnarray}
%A plot of $A_{12}(\omega)$ is presented in figure ???.

For notational purposes, we shall divide $A_{12}(\omega)$ into its natural
``high'' and ``low'' frequency parts according to
\begin{equation}
A_{12}(\omega)=\theta_+(\omega-(v_s-v)\Delta k)\theta_+((v_c-v)\Delta k-\omega)
A_{12}^{\rm low}(\omega)+
\theta_+(\omega-(v_c-v)\Delta k)A_{12}^{\rm high}(\omega)
\end{equation}
Note that this definition is based on the structure of $A$
and not on the nature of the hopping and is not equivalent
to the division into energies ``high'' and ``low'' compared
to the renormalized hopping, $t_{\perp}^R$
which we consider elsewhere in this paper.

Note that both the chiral and spinless
limits are correctly reproduced by this
spectral function. The chiral limit is $\alpha\rightarrow 0$ in which case
$A_{12}^{\rm high}(\omega)\propto
\Gamma^{-1}(2\alpha)\sim 2\alpha\rightarrow 0$,
and $A_{12}^{\rm low}(\omega)\rightarrow (\Delta v)^{-1}$.
For the spinless limit, we take $\Delta v \rightarrow 0$, so 
$~_2F_1(a,b,c,-z(\Delta v/\bar{v}))\rightarrow ~_2F_1(a,b,c,1)=1$
recovering the correct expression for $A_{12}^{\rm high}(\omega)$,
and we can set $A_{12}^{\rm low}(\omega)\rightarrow 0$
because its integrated weight 
$\sim (\Delta v)^{2\alpha}\stackrel{\Delta v\rightarrow 0}{\rightarrow} 0$.

Some qualitative 
remarks are in order. The first is to note that $A_{12}(\omega)$
has non-zero weight over an energy range from just below $\omega=0$ all
the way up to the ultraviolet cutoff.
As for the spinless case, 
$A_{12}(\omega)\propto\omega^{4\alpha}$
at high frequencies, $\omega\gg(v_c-v)\Delta k$.
Again, for $\Delta k=0$ ({\em i.e.\/} $\Delta N=0$) this is all there is
to $A_{12}(\omega)$. A spectral function $\propto\omega^{4\alpha}$ implies
a $P(t)\propto t^{1-4\alpha}$ which describes manifestly incoherent 
interliquid hopping. 
We again emphasize that such incoherent processes are not present
for coupled Fermi liquids, nor for chiral Luttinger liquids.

Upon comparing Figs. \ref{fig: inter_spectral_spinless} and 
\ref{fig: inter_spectral_spinny} we see that one  feature  of 
$A_{12}(\omega)$ differentiating
the spinny from the spinless Luttinger liquid is the absence of any
low frequency singularity. This follows by 
using the integral form for $~_2F_1$:
\begin{eqnarray}
~_2F_1(1,1-2\alpha;2+2\alpha;-z)
&=&\frac{\Gamma(2+2\alpha)}{\Gamma(1+2\alpha)}
\int_0^1 dt\;(1-t)^{2\alpha}(1+tz)^{2\alpha-1} \nonumber\\
&\propto& z^{2\alpha-1}
\end{eqnarray}
in the limit $z\gg 1$. Thus in the vicinity of $\omega-(v_c-v)\Delta k=0$,
the contribution 
$\sim(\omega-(v_c-v)\Delta k)^{1-2\alpha}$ from the $~_2F_1$ part of 
$A_{12}^{\rm high}(\omega)$ cancels the $\sim(\omega-(v_c-v)\Delta k)^{2\alpha-1}$
piece, and there is no singularity.
%(note, however, that in the spinless case,
%$\Delta v=0$, and there {\em is\/} a power law singularity at
%$\omega=(v_c-v)\Delta k$. We shall discuss the spinless case later).

Recall that Fermi's Golden Rule, which describes incoherent decay from an 
initially prepared state, cannot be applied if the effective spectral
function of final states is singular, and/or has too narrow support. The
nonsingular nature of $A_{12}(\omega)$ and the fact that it has support
over the full energy range, gives us good reason to believe that incoherent
methods like the Golden Rule will work. 
We reiterate that this was not the case for coupled Fermi
liquid or chiral Luttinger liquids.

We now turn to evaluating the interliquid hopping rate. 
Again, it is easiest to consider 
$d\Gamma(t)/dt=-d^2P(t)/dt^2$,
for then the integrals simplify and we obtain the following exact 
result:
\begin{eqnarray}
&&\frac{d\Gamma_{12}(t)}{dt}=\frac{d\Gamma_{12}^{\rm low}(t)}{dt}
+\frac{d\Gamma_{12}^{\rm high}(t)}{dt} \nonumber \\
&&\frac{d\Gamma_{12}^{\rm low}(t)}{dt}
=\frac{t_{\perp}^2L}{\pi}
\frac{1}{\Gamma(1+4\alpha)}
\left(\frac{a^2}{\bar{v}\Delta v}\right)^{2\alpha}
\frac{1}{\Delta v}\,
%\nonumber \\
t^{-(1+4\alpha)}\,{\rm Re}\left\{e^{i(v_s-v)\Delta k t}
\int_0^{\Delta v\Delta k t}d\theta\;\theta^{4\alpha} e^{i\theta}
\right\} \\
%&&\left\{\cos[(v_s-v)\Delta k t] t^{-(1+4\alpha)}\int_0^{\Delta v\Delta k t}
%d\theta\;\theta^{4\alpha} \cos\theta
%-\sin[(v_s-v)\Delta k t] t^{-(1+4\alpha)}\int_0^{\Delta v\Delta k t}
%d\theta\;\theta^{4\alpha} \sin\theta 
%\right\}\\
&&\frac{d\Gamma_{12}^{\rm high}(t)}{dt}
=\frac{t_{\perp}^2L}{\pi}
\frac{1}{(1+2\alpha)}\frac{1}{\Gamma(2\alpha)\Gamma(1+2\alpha)}
\left(\frac{a}{2v_c}\right)^{4\alpha} 
\frac{1}{\bar{v}}\,t^{-(1+4\alpha)}\nonumber \\
&&{\rm Re}\left\{e^{i(v_c-v)\Delta k t}
\int_0^{\infty}d\theta\;\theta^{2\alpha-1}(\theta+2v_c\Delta k t)^{2\alpha+1}
e^{i\theta}
~_2F_1\left(1,1-2\alpha;2+2\alpha;\left(\frac{-\Delta v}{\bar{v}}\right)
\left(\frac{\theta+2v_c\Delta k t}{\theta}\right)\right)
\right\} \nonumber 
%&&\cos[(v_c-v)\Delta k t] t^{-(1+4\alpha)}\int_0^{\infty}
%d\theta\;\frac{\cos\theta}{\theta^{1-2\alpha}}(\theta+2v_c\Delta k t)^{2\alpha+1}
%~_2F_1\left(1,1-2\alpha;2+2\alpha;\left(\frac{-\Delta v}{\bar{v}}\right)
%\left(\frac{\theta+2v_c\Delta k t}{\theta}\right)\right) \nonumber \\
%&&-\sin[(v_c-v)\Delta k t] t^{-(1+4\alpha)}\int_0^{\infty}
%d\theta\;\frac{\sin\theta}{\theta^{1-2\alpha}}(\theta+2v_c\Delta k t)^{2\alpha+1}
%~_2F_1\left(1,1-2\alpha;2+2\alpha;\left(\frac{-\Delta v}{\bar{v}}\right)
%\left(\frac{\theta+2v_c\Delta k t}{\theta}\right)\right) \nonumber
\end{eqnarray}

We will now show that, as expected on physical grounds, the effect of
spin-charge separation is to enhance incoherence further beyond the
effects discussed in the spinless case. We begin with 
$d\Gamma_{12}^{\rm high}/dt$ which requires evaluation of the integral
\[
\int_0^{\infty}d\theta\;\theta^{2\alpha-1}(\theta+x)^{2\alpha+1}
e^{i\theta}
~_2F_1\left(1,1-2\alpha;2+2\alpha;\left(\frac{-\Delta v}{\bar{v}}\right)
\left(\frac{\theta+x}{\theta}\right)\right)
\]
Unfortunately, we do not know of an exact result for this integral when
$\Delta v\neq 0$. Moreover, its nonanalytic nature makes an asymptotic
analysis nontrivial.

The important effect of spin-charge separation, as has already been pointed
out, is that it removes the $\theta\rightarrow 0$ singularity in the integrand.
For $\Delta v=0$ the $O(x)$ contribution from this integral is proportional to
$\alpha^{-1}$. In contrast, in the almost chiral limit 
($\Delta v\neq 0$, $\alpha\ll 1$) the integral can be approximated by 
setting $\alpha=0$, {\em i.e.\/} evaluating
\[
\int_0^{\infty}d\theta\;\theta^{-1}(\theta+x)
e^{i\theta}
~_2F_1\left(1,1-2\alpha;2+2\alpha;\left(\frac{-\Delta v}{\bar{v}}\right)
\left(\frac{\theta+x}{\theta}\right)\right)
\]
This integral can be evaluated exactly using 
$~_2F_1(1,1;2,-z)=\log(1+z)/z$. The result is
\[
i\left(\frac{\bar{v}}{\Delta v}\right)
\left\{
\gamma
%\log\left(1+\frac{\Delta v}{\bar{v}}\right)
+\log\left(\frac{-i\Delta v x}{\bar{v}}\right)
+\exp\left(\frac{-i\Delta v x}{\bar{v}+\Delta v}\right)
E_1\left(\frac{-i\Delta v x}{\bar{v}+\Delta v}\right)
\right\}
\]
where $\gamma\approx 0.577$ is Euler's constant, and $E_1$ is the exponential
integral function. The small $x$ behavior is
\[
-\frac{\bar{v}\,x}{\bar{v}+\Delta v}\log\left(\frac{\Delta v\,x}{\bar{v}+\Delta v}\right)
\]
(which is singular in the $\Delta v\rightarrow 0$ limit, as required,
but otherwise finite).

In the spinless case, the incoherent ($\Delta k$-independent) contribution
to $d\Gamma/dt$ is $O(\alpha)$ for small $\alpha$, while the coherent
$O(\Delta k)$ term is $O(\alpha^0)$. The free Fermi gas limit is correctly
recovered as $\alpha\rightarrow 0$. In contrast, using the results above,
for an almost chiral
Luttinger liquid both the incoherent and coherent contributions
to $d\Gamma^{\rm high}/dt$
are $O(\alpha)$. In fact, the dominant coherent contribution comes from
the low frequency part of the spectral function. We have
\begin{eqnarray*}
\frac{d\Gamma_{12}^{\rm low}(t)}{dt}
&\propto&
t^{-(1+4\alpha)}\,{\rm Re}\left\{e^{i(v_s-v)\Delta k t}
\int_0^{\Delta v\Delta k t}d\theta\;\theta^{4\alpha} e^{i\theta}
\right\} \\
&=&t^{-(1+4\alpha)}\,\left\{\cos[(v_s-v)\Delta k\,t]
\int_0^{\Delta v\Delta k t}d\theta\;\theta^{4\alpha}\cos\theta
+O((\Delta k)^2)\right\}
\end{eqnarray*}
(note that this vanishes for $\Delta k=0$). The integral is bounded
above by $\sin(\Delta k\,\Delta v\,t)$ which provides a short time
expansion. At long times ($\Delta k\,\Delta v\,t\gg 1$)
\[
\int_0^{\Delta v\Delta k t}d\theta\;\theta^{4\alpha}\cos\theta
\rightarrow -\sin(2\pi\alpha)\Gamma(1+4\alpha)
\]
and the contribution crosses over to a purely incoherent
time dependence.

The almost chiral limit provides an illustration of the enhancement of
incoherence due to spin-charge separation. In this limit the coherent
contribution to $d\Gamma/dt$ is dominated by the low frequency contribution
and for small $\alpha$ is bounded by
\begin{equation}
\frac{t_{\perp}^2L}{\pi}\frac{1}{\Delta v}\,t^{-(1+4\alpha)}\,
\cos[(v_c-v)\Delta k t]\sin[(v_c-v_s)\Delta k t]
\label{almost chiral}
\end{equation}
In comparison, the coherent contribution in the spinless case at small
$\alpha$ is
\[
\frac{t_{\perp}^2L}{\pi}\Delta k\,t^{-4\alpha}\,\cos[(v_c-v)\Delta k t]
\]
Equation (\ref{almost chiral}) is equivalent to this at short times,
$t~^<_{\sim}[(v_c-v_s)\Delta k)]^{-1}$
(using $v_c-v=v-v_s$ at small $\alpha$),
but at finite time the spin-charge
separation provides a further decohering factor beyond those present
in the spinless limit. As such, spin-charge separation can 
lead to a smaller $\alpha_c$ in the $t_{\perp}\rightarrow 0$ limit.

\subsubsection{Summary}

We have shown, in agreement with the arguments of the previous
section, that a consideration of the exact expressions for the
interliquid hopping rate presented
above obtained via spectral function methods, leads to the conclusion that
there is a {\em nontrivial incoherent phase\/} in the small $t_{\perp}$ limit
for all $\alpha_c < \alpha < 1/2$. In general, $\alpha_c < 1/4$ 
and is a decreasing
function of $\Delta v/{\bar{v}}$. Open questions are whether $\alpha_c$
reaches $0$ for some physically accessible value of $\Delta v/{\bar{v}}$,
and also the nature of the phase boundary in the $(t_{\perp}, \alpha)$
plane separating a coherent (``deconfined'') 2D state from the 
state of ``confined coherence'', (1D)$_{\rm NFL}\otimes$(1D)$_{\rm incoh}$.

%% file: interp.sec
\section{Interpretation of the Incoherent Response}
\label{sec:interp}

We are really interested in the nature of the
true groundstate of the strongly
interacting problem at strong coupling and finite
$t_{\perp}$.  Our dynamical calculation is meant
to tell us something about this with as few assumptions
and as much sensitivity to the possible breakdown
of coherence as possible.  This is necessary since the
true ground state is not accessible by any
controlled calculation at this time \cite{controlfootnote}
and,
as we have seen, more conventional calculations are
difficult to interpret for questions of coherence.

We believe that the coherence question is crucial for
the simple reason that the splitting of the Fermi surface
for two chains or the formation of a two dimensional
Fermi surface for infinitely many chains are specific manifestations of
interference effects between histories which involve
particles hopping between chains.  This is most easily seen
for the case of free particles where  calculations can be
carried out exactly and the single electron self-energy
responsible for splitting the Fermi surface
clearly relies on interference between paths
that hop between different chains.
On the other hand, our {\it definition}
of a phase with  incoherent hopping  is that such effects 
are {\it completely non-existent} in such a phase
on all energy scales small compared to the high
energy cut-off of the theory. 
There can therefore not be any split Fermi surface
(for two chains) or two dimensional Fermi surface
(for many chains) in this case, and,  since
the shape of the Fermi surface is a 
low energy property of the system, we
are indisputably dealing with a different 
fixed point from any phase in which
split Fermi surfaces are expected.
Coherence, or its absence, is crucial to the
discussion since {\it previous proposals for fixed
points with finite $t_{\perp}$ implicitly assume it to be present}
by taking the dominant action of the
single particle hopping to be the formation
of symmetric and antisymmetric fermion operators,
with corresponding Fermi surfaces.
There has been, up to now, no attempt to justify this assumption,
and in fact our work is really an attempt
to provide that justification {\it and to attach proper limits
to it}.

The approach to coherence we have followed parallels
that in the TLS \cite{TLS} problem; we have 
looked for interference effects in a particularly
natural 
dynamical calculation where they would manifest themselves
as damped oscillations in $\langle \delta N(t)\rangle$.
If they are {\it completely} absent there,
as they are for $\langle \sigma^z(t) \rangle$ at $\alpha = 1/2$
in the TLS,
then we are interpreting this as demonstrating that they
are completely absent in general.
The assumption
that the vanishing of the interference in the quantity
we calculate signifies the loss of coherence and the inability
to observe interference as defined by 
other measures, such as the presence of
split Fermi surfaces, is a plausible assumption, which
is identical to the usual assumptions
made about macroscopic quantum coherence
\cite{TLS}.  Further, there is no evidence of any kind 
which supports the contrary assumption that,
for reasons unstated, a relevant $t_{\perp}$ should
always be taken to act coherently and allow
interference effects. In fact, on the experimental side,
as we will see in Section \ref{sec:organics},
there is extremely strong experimental evidence for
a low temperature phase in which no interference effects
for out of {\it plane} paths
can be observed in the magnetotransport measurements 
on the organic conductor (TMTSF)$_2$PF$_6$.
In a similar way, as discussed in the next Section,
experimental evidence points to a phase (albeit bounded
below in temperature by $T_c$) in the cuprate superconductors
where interplanar transport is incoherent in a nontrivial way
({\em i.e.\/} the incoherence is not the result of inelastic
scattering of any conventional type).
These observations fit in perfectly with a trivial
two dimensional generalization of our 
proposal, while it
is difficult to see how they
could be reconciled with any other
currently existing proposal for
transport in anisotropic systems.

Returning to the specific interpretation of our calculation, 
we first note that for all the models which we understand
(coupled free particles or Fermi liquids)
and which have coherent ground states
(the states are built up out of quasiparticle
operators that have a well defined $k_{\perp}$
and energies which are functions of $k_{\perp}$),
our dynamical calculation turns up a
coherent response with oscillation frequency
essentially
equal (for two chains) to the splitting of the symmetric and antisymmetric
quasiparticle combinations.  Consequently,
we propose to interpret the frequency of any
oscillations which are present as 
the effective splitting of the symmetric
and antisymmetric Fermi surfaces, i.e. the transverse
bandwidth.
We have already seen in Sections \ref{sec:fermions} and
\ref{sec:luttliq} that
our approach, with this interpretation, correctly
describes all of the models where coherence is
known to be present.
Also, for models where $t_{\perp}$ is renormalization
group irrelevant,
we find no oscillations 
{\it and} a convergent long time response,
and no oscillations, so that
again our calculation  and interpretation capture
the correct physics.  

Significantly, our calculation
also exhibits a third regime directly analogous
to the incoherent regime of the TLS.  We interpret
this regime, in which there is no sign
of oscillations in $\langle \delta N(t) \rangle$
but $t_{\perp}$ is still relevant,
as one of incoherence, in the
sense that interference effects
cannot be observed between histories
in which $t_{\perp}$ acts.  This parallels the
interpretation of the corresponding TLS phase
in the problem of macroscopic quantum coherence.
This incoherence results because the only term in our Hamiltonian
which connects states with different numbers of particles in
a given liquid acts in a non-degenerate way: it connects
predominantly states with energies differing by amounts 
large (in fact of order the cut-off in the theory)
compared to the matrix element to those states.
Under these circumstances, it should be impossible to have
coherence in the sense of measurable interference
between histories in which $t_{\perp}$ acts.
A split Fermi surface should therefore not occur and
the system should not have a groundstate
which is built up out of symmetric and antisymmetric
combinations of
fermion operators.
At the very least,
no energy would be gained by forming such 
combinations since the matrix elements to the states
with low enough energies to be mixed in
coherently are vanishingly small (otherwise we
should have seen {\it finite}, if heavily damped 
oscillations, in $\delta N(t)$ \cite{carefulnote}, signaled
in our calculation by short time coherence).
Further, such states are defined in terms of phases which are
only meaningful if 
interference effects of the unobservable type
are considered. We therefore regard the action
of $t_{\perp}$ as qualitatively different in this phase
than in the coherent phase.

Parenthetically, we should note that the physics we are interested
in is the coherence of $t_{\perp}$. Our discussion throughout
has neglected and will neglect operators such as interchain magnetic
interactions generated by the renormalization of $t_{\perp}$.
In particular models, these operators may in the end dominate
the physics and determine the low
energy behavior of the systems.  However, we are 
certainly free to consider sufficiently general
models that the coefficients of these other operators
can be tuned so as to make them unimportant. From this
consideration, it is
clear that {\it in principle}  these other operators
are unimportant 
for the question of  the existence or non-existence of the 
incoherent fixed point we are advocating.  The question
of whether they are in practice important for any
specific model is beyond the scope of this paper,
except for the brief discussion of Sec. \ref{sec:otheroperators}. 

Returning to our main theme,
we believe that dynamical incoherence excludes
the notion of Bloch states with a definite
$k_{\perp}$.  The reason for this is that, if
dynamical incoherence implies a more general
incoherence such that interference effects
are absent between histories that involve the action of
$t_{\perp}$, then
we have no way of comparing the phases of creation operators
acting in different liquids and the whole notion of
a Bloch state is meaningless for a discussion of the dynamics.
In fact, Bloch states are exactly analogous to Schr\"odinger's
cat: there is no point in writing down
$|{\rm alive} \rangle + e^{i\theta}|{\rm dead} \rangle$
because the alive and dead histories will decohere before any
interference
can be observed, likewise for
$\psi^{\dagger}_1 + e^{i\theta}\psi^{\dagger}_2$ 
the two fermion operators will decohere
too fast for $\theta$, or equivalently, $k_{\perp}$
to have any meaning.

Certainly, if a two chain system were well 
described by nearly non-interacting symmetric and antisymmetric
Bloch states, then coherent dynamics would result.
The absence of such coherence in the dynamics
therefore
excludes this possibility.  It is much more
difficult to make definitive statements
about what the dynamics would be like should
the system be well described by, for example,
symmetric and antisymmetric Luttinger liquids.
Nonetheless, we are unaware of any reason to expect
that the damped oscillations one expects in
the dynamics of $\langle \delta N \rangle$
could in this case completely vanish despite the
manifest coherence of the groundstate.
After all, the oscillations are an
interference effect and such effects ought
to be observable under any circumstances 
where Bloch states are meaningful.
In the absence of coherence 
in $\langle \delta N(t)\rangle$, 
we therefore believe that
the groundstate cannot be thought
of as built up out of operators creating
and destroying Bloch states of quasiparticles
with definite transverse momentum.

Incidentally,  it is  clear 
that no finite amount
of damping implies incoherent transport.
If interference effects are observable
at all,
then there is no barrier to the formation
of Bloch states
with definite transverse momentum
and a higher dimensional Fermi surface
(although the transverse bandwidth may be
more heavily renormalized than suggested by
a naive RG calculation).
If a higher dimensional Fermi surface forms
and the system crosses over to a Fermi liquid,
then there will be coherent transport in the
perpendicular direction because the scattering of
quasiparticles always vanishes as the
Fermi surface is
approached.
The quasiparticles therefore have
well defined $k_{\perp}$ and $v_{\perp}$
and transport can be coherent transverse to the chains.
This should be kept in mind when considering
approximate  approaches to 
calculating the single particle Green's function,
since these, in general, result in predictions
of non-vanishing scattering as the Fermi surface is 
approached; this must be an artifact. Since the
dynamical quantity we are calculating does not
directly involve the true ground state,
this argument does not imply that
the damping of its oscillation 
is an artifact (it isn't), but it does make clear that the
damping, {\it if finite}, is not clearly meaningful 
for transverse transport. On the other hand, 
the smallness of 
the oscillation 
frequency,
and certainly its vanishing, 
are very meaningful.

When we find a coherent response,
describing damped oscillations with some frequency,
it is reasonable to believe that the
true groundstate should be constructed by first
making single particle
Bloch states with definite
$k_{\perp}$ and then considering
the interactions of these states.  In higher dimensions,
this should lead us back to Fermi liquid
theory (or at the very least
a higher dimensional Fermi surface), although
the case of a finite number of coupled liquids
may be subtle (see Finkelshtein and Larkin \cite{Bunch,splitfootnote}).
We believe, as previously stated, that the frequency of the damped
oscillation in $\langle \delta N(t)\rangle$
should  in general be equal to
the transverse bandwidth, up to finite factors
of order one. In fact, we saw
that this was the case for coupled free Fermi gasses
and Fermi liquids.  If Luttinger liquids
were coupled with a large enough $t_{\perp}$ or small
enough $\alpha$, $v_{\rho}-v_{\sigma}$, {\em etc.\/} that
damped oscillations resulted, we 
would identify the  oscillation frequency
with the transverse bandwidth. This 
should be $O(t_{\perp}^R)$ and in that
sense corresponds
closely with the proposals made by other authors
for the transverse bandwidth \cite{Bunch},
however in our case the frequency can reach zero
in spite of the renormalization group relevance
of $t_{\perp}$.  In this case,
the naively $O(1)$ prefactor between 
$t_{\perp}^R \sim t_{\perp}^{1/(1-2\alpha)}$ and
the oscillation frequency is vanishing and,
by implication, there
should be no transverse bandwidth (although particles
can clearly still move incoherently between chains, since
$t_{\perp}$ is not irrelevant).

Since the transverse bandwidth specifies
the splitting (warping) of
the Fermi surface expected for two (many)
chains, we propose that this splitting is in effect
the {\em order parameter\/} for the coherence/incoherence
transition.  It is the most natural
low energy example of an interference effect
between histories involving interliquid hops, and,
since the shape of a Fermi surface is an
infinitesimal energy/infinite time property of
the system,
our proposed interpretation
of the splitting/warping
as an order parameter
implies that the incoherent
phase constitutes a truly distinct fixed point.  
This is different from the 
TLS problem where the coherent and incoherent regimes
need not correspond to different fixed points since
the marked distinctions between their short time
behaviors need not persist to infinite times
\cite{sudiponefp} since there is
no corresponding long time
measure of coherence.  On the other hand, for fermions,
the splitting or warping of the Fermi surface should be
determined by the presence or absence of coherence, since
the coherent oscillation frequency, if any, specifies
the splitting.  This splitting (warping) is a true long
time property  accessible in the measurement of
asymptotic correlation functions, and therefore a
qualitative change here requires a new fixed point.

The transition from coherence to incoherence
is, however, not a thermodynamical one, but
one in transport properties, more closely
analogous to the Anderson metal-insulator transition in
localization than to a thermodynamic phase
transition.  We therefore do not expect experimental
signatures of it in specific heat, etc. but rather
in transport data such as the resistivity. In particular,
a substantial change in the perpendicular conductivity
should occur at the transition, particularly in its
frequency dependence, as we have discussed in
Ref. \cite{prl2}.
%will discuss in Sec.
%\ref{sec:highTcexp}

In conclusion,
the incoherence we are proposing is thus a truly new state
of matter with novel properties arising from the
absence of certain interference effects.
Physically, we expect this to lead to
host of unusual properties, including
incoherent
transport in one direction {\it all the way down to
zero temperature in a pure system}. Fortunately
for our rather radical proposal, it appears
that this state has already been seen
in (TMTSF)$_2$PF$_6$, as we shall shortly discuss.

%% file: otheroperators.sec
\section{The Effects of Other Operators}
\label{sec:otheroperators}

Up to this point we have focussed exclusively on the
nature of the single particle hopping between the
Luttinger liquid chains; 
however, we know that one
of the effects of higher order perturbation theory
in  $t_{\perp}$ is to generate other potentially
relevant operators, such as magnetic superexchange
between the chains and so on.
How important are these effects for our conclusions?

Let us begin with the question of whether the 
dominant instability of the uncoupled chains fixed point
is to single particle hopping or to some other operator
generated by the hopping at higher order. 
We discuss here only the case of fermions with spin.
The lowest order
renormalization group equations for $\lambda_{\rm single}$,
the coupling of the single particle hopping operator,
and the coupling constant, $\lambda_O$,
of a generic operator, $O$,
generated at higher order read:
\begin{eqnarray}
\frac{\partial \lambda_{\rm single}}{\partial \ln\Lambda} &
= & (d_t-2) \lambda_{\rm single} \\
\frac{\partial \lambda_O}{\partial \ln\Lambda} & = & (d_O-2) \lambda_O
+ \alpha_O \lambda_{\rm single}^{p} 
\end{eqnarray}
where $\alpha_O$ is a coefficient of order $\Lambda^{-(p-1)}$
which depends on $O$,
$d_t$ is the scaling dimension of $t_{\perp}$,
$d_O$ is the scaling dimension of $O$ and $p$ is 
the order at which the operator is generated.
In practice, the first operators to supersede the
single particle hopping in importance are two body
operators, for which $p=2$.   For repulsive interactions,
the most relevant of these operators couples the 
2$k_F$ components of the spin  operators on the two
chains and we have:
\begin{eqnarray}
d_t  & =  & \frac{1}{2} + \frac{1}{4}
\left( K_{\rho}^{-1} + K_{\rho} \right) \\
d_O & = & 1 + K_{\rho}
\end{eqnarray}
The spin coupling is more relevant than the single particle
hopping for all $K_{\rho} < 1/3$, however, to determine if
it provides the leading instability we need to consider
the initial conditions for the renormalization group
flows:
\begin{eqnarray}
\lambda_{\rm single}^{\rm initial} & = & t_{\perp}  \\
\lambda_O^{\rm initial} & =  & 0 
\end{eqnarray}
In this
case, the criterion, in the limit of small
$t_{\perp}$, for $O$ to dominate, which one obtains
by integrating the renormalization equations until
one of the couplings is of order the cutoff, $\Lambda$,
is:
\begin{equation}
d_O  <  2(d_t -1) 
\end{equation}
or, equivalently, 
\begin{equation}
K_{\rho}  <  \sqrt{5} - 2
\end{equation}
We have obtained this result independently but it is
equivalent to that of Boies, {\it et al.} \cite{Bunch}.

Recall that, according to our arguments, the
action of $t_{\perp}$ is incoherent for
$\alpha > \alpha_c < 1/4$.  This involves
$K_{\rho} < K_{\rho}^c > 2 - \sqrt{3}$.
Since $ 2 - \sqrt{3} > \sqrt{5} - 2$, there is
guaranteed to be a region where the
single particle hopping is
simultaneously 
incoherent and the leading instability of
the uncoupled chains fixed point.  
There is therefore no obstacle to realizing
incoherence due to the two particle
instabilities of the uncoupled chains fixed point.
In fact, one could consider more general models
of coupled chains where a {\it ferromagnetic}
superexchange was added by hand, and then
tune the coupling of the superexchange to
cancel the coupling generated by
$t_{\perp}$ and thus reach smaller values of
$K_{\rho}$ with the single particle hopping still
dominant.  Consequently, it should be clear that the effects of
these operators are model dependent. However, 
it is worth noting that
even in the
simplest coupled chains model, 
fine tuning is unnecessary and incoherent hopping
is the dominant instability over a finite
region of parameter space.

The dominance of the incoherent, single particle hopping is
not sufficient to establish that a
fixed point will be realized with purely
incoherent hopping between the Luttinger liquids
and no other couplings.  This is because the 
incoherent fixed point itself may be unstable to
some other operators.  The stability analysis 
of the incoherent fixed point is
beyond our present understanding of the
problem, however, there are several points we
wish to emphasize.  First,  the instability
of the {\em uncoupled chains fixed point\/}
to two body operators, which has been emphasized
in many other studies of this problem \cite{Bunch},
is completely irrelevant to the question of
the stability of the {\em incoherent fixed point\/}.
If the leading instability of the uncoupled chains
fixed point is to single particle hopping,
and we are in a regime where the hopping is
incoherent, it is the stability
or instability of the incoherent fixed point
which ultimately determines the
low energy physics, and the fact that the
isolated chains fixed point is always unstable
to some two body operator has no bearing.

Second, while, for any particular model, the stability
of the incoherent fixed point
may be very important, 
unless the incoherent
fixed point is unstable to infinitely many
operators, one can, in principle, find a 
sufficiently general model to tune the
coefficients of all operators relevant
about the incoherent fixed point 
so that these coefficients vanish.
In that case the incoherent fixed point will be realized
in the low energy limit.
In light of this, it seems clear to us
that what we have done is sufficient to
demonstrate that {\it the incoherent fixed point
does exist}; whether it is stable for a
specific model, such as that of two or many chains
coupled only with a transverse hopping, is
a different question, and one which requires
further study.  

Finally, the concept of incoherent hopping clearly
generalizes straightforwardly to the coupling of 
planes of strongly interacting fermions
with non-Fermi liquid groundstates,
and in this case one would expect that the
incoherent fixed point would be substantially
more stable than for the one dimensional case
since two dimensional non-Fermi liquids
need not exhibit the nesting related instabilities
of one dimensional Luttinger liquids. In fact,
the experimental case appears to be that the incoherent
state in (TMTSF)$_2$PF$_6$
is unstable only to superconductivity 
(see Sec. \ref{sec:organics}),
and becomes the true groundstate
in a sufficiently strong magnetic field.  In the cuprates
(see Sec. \ref{sec:caxis}),
there appear to be additional instabilities,
perhaps due to finite strengths for the
intra-bilayer couplings 
in bilayer compounds, and the superconducting
instability is also clearly present.  Nonetheless,
many of the finite temperature and finite frequency
properties appear to be the result of the proximity
of the materials to the incoherent fixed point.

In summary, the inclusion of other operators 
beyond the single particle hopping poses no
problem, in principle or in practice, for our arguments
that the leading instability of the uncoupled Luttinger
liquid fixed point can be to incoherent, single particle
hopping.  In this case, there should, at the very least,
be a sizable region in energy/temperature where the
properties of the system will be dominated by the
underlying incoherent fixed point.  The actual realization
of this fixed point in the truly low energy limit
is clearly possible in principle, while the realization
in practice of the natural higher dimensional
analog has, we believe, been demonstrated experimentally
(see Sec. \ref{sec:organics}).

%% file: caxis.sec
\section{Normal State c-axis Conductivity in the Cuprates}
\label{sec:caxis}

The {\em qualitative\/} anisotropy exhibited by the cuprate superconductors
in the normal state implies that they should be described by some fixed point
of the renormalization group that is itself qualitatively anisotropic.
However, the photoemission data provide strong evidence in favor of the
renormalization group relevance of the interlayer single particle hopping.
This implies that the correct low energy fixed point is in fact three
dimensional, yet the transport data require a qualitative anisotropy.
To reconcile these two features  of the experimental data requires a
non-trivial {\em three dimensional\/} alternative to the usual Fermi
liquid fixed point. Note that the perturbative stability of Fermi liquid
theory in three dimensions does not rule out the existence of such a fixed
point; nor does the instability of the $t_{\perp}=0$ fixed point.
In general, when the weak coupling (here, $t_{\perp}=0$) fixed point is
unstable and there exists a stable strong coupling fixed point (here, three
dimensional Fermi liquid theory with finite $t_{\perp}$), one expects that the 
flows will carry the effective Hamiltonian continuously from the weak to
the strong coupling fixed point, however, this need not necessarily be the
case since alternative, stable, strong coupling fixed points may exist.
The primary goal of this paper is to propose such an alternative fixed point in which
$t_{\perp}$ has not renormalized to zero and yet a qualitative anisotropy
remains at the fixed point. The fixed point we propose is one of ``confined
coherence'', at which the interlayer hopping is totally incoherent.

With this in mind, we will now present a more detailed
%In this section we will 
discussion of  the c-axis transport in the HTSC's.
%in greater detail. 
A comprehensive survey by Cooper and Gray
of the data up to 1993 can be found in Ref. \cite{cooper_gray} to which the 
reader is referred for greater detail and a significantly more complete
bibliography.

\subsection{Boltzmann Theory}

We begin by considering the very simplest estimates
of conductivity in a highly anisotropic (2+1)-D metal with a single conduction
band described by a tight-binding Hamiltonian
\begin{equation}
H=- 2t\,[\cos(k_a a)+\cos(k_b b)]-2t_{\perp}(k_a,k_b)\cos(k_c c)
\end{equation}
We have included ${\bf k}$-dependence of $t_{\perp}$ to mimic
more accurately the results of electronic band structure calculations
({\em e.g.\/} for the cuprates, the form $t_{\perp}(k_a,k_b)
=t_{\perp}[\cos(k_a a)-\cos(k_b b)]^2/4$ has been proposed \cite{science}).
Of course, such a model
omits nontrivial many-body effects,
and is an oversimplification of the band structure, however it will
suffice for the points we wish to make.

When the anisotropy is large, $t_{\perp}/t \ll 1$, the Fermi surface is open.
To simplify the analysis, we suppose that the Fermi surface at constant $k_c$
is approximately circular (this is certainly true if the electron density is not 
too large). The Fermi surface is then a warped cylinder, 
and we 
may linearize the dispersion in the ab-plane, whence
\begin{equation}
E(k,k_c)=\hbar v_F (k-k_F)-2t_{\perp}({\bf k})\cos(k_c c)
\end{equation}
where ${\bf k}=(k_a,k_b)$ and $k=|{\bf k}|$.
The conductivity tensor derived from simple Boltzmann transport
theory may be written as
\begin{eqnarray}
\sigma_{\alpha\beta}&=&\frac{1}{4\pi^3} e^2
\int_{\partial{\cal F}}  [v_{\alpha}v_{\beta}\,\tau]({\bf k}, k_c)
%_{\partial{\cal F}} 
\frac{dS_F}{|\nabla_{\bf k}E|} \nonumber \\
&\approx&\frac{1}{4\pi^3} \frac{e^2}{\hbar} \frac{1}{v_F}
\int_{\partial{\cal F}}  [v_{\alpha}v_{\beta}\,\tau]({\bf k}, k_c)
\,dS_F
\end{eqnarray}
The final line follows from the anisotropy assumption
$t_{\perp}/t \ll 1$, and reflects the fact that on an open Fermi
surface, despite the small dispersion along the c-axis, 
there is no ${\bf k}$-state with ``small'' velocity - all velocities
are of order $v_F$.

The c-axis and a-axis conductivities are therefore
\begin{eqnarray}
\sigma_c &=&\frac{1}{4\pi^3} \frac{e^2}{\hbar} \frac{1}{v_F}
\int_{\partial{\cal F}} [v_{\perp}({\bf k})]^2\,\sin^2(k_c c)\,\tau({\bf k}) \\
\sigma_a&=&\frac{1}{2}\frac{1}{4\pi^3} \frac{e^2}{\hbar} v_F
\int_{\partial{\cal F}}\tau({\bf k})
\end{eqnarray}
where $v_{\perp}({\bf k})\equiv 2t_{\perp}({\bf k}) c$.

In the isotropic-$\tau$ approximation we arrive at
\begin{eqnarray}
\sigma_c &\approx &
\frac{1}{2}\frac{1}{4\pi^3} \frac{e^2}{\hbar} \frac{1}{v_F}
\tau\,(2\pi k_F)\,\frac{2\pi}{c}\, \langle v_{\perp}^2({\bf k}) \rangle\nonumber\\
&=&\frac{1}{2\pi} \frac{e^2}{\hbar} k_F
\frac{\langle v_{\perp}^2({\bf k}) \rangle}{v_F^2}
\left(\frac{v_F \tau}{c}\right)
\end{eqnarray}
and
\begin{equation}
\sigma_a \approx \frac{1}{2\pi} \frac{e^2}{\hbar} k_F
\left(\frac{v_F \tau}{c}\right)
\end{equation}
The relation $\sigma_c / \sigma_a \approx (v_{\perp}/v_F)^2$
is a quite general result for a metal of high anisotropy, whenever
simple Boltzmann transport theory and the isotropic-$\tau$
approximation are valid.

\subsection{Mott minimum metallic conductivity}

A simple argument attributed to Mott \cite{mott}
 (and to Ioffe and Regel \cite{ioffe_regel} in a 
different context) provides an estimate of the minimum conductivity a 
metal must have in order to justify a straightforward application of
Boltzmann transport theory. For a metal with a spherical Fermi surface
the simple result $\sigma=n e^2\tau/m$ may be rewritten as
\begin{equation}
\sigma=\frac{1}{3\pi^2}\frac{e^2}{\hbar} k_F\,(k_F l) \label{eq: Mott}
\end{equation}
using $n=k_F^3/3\pi^2$ and $l\equiv v_F\tau$ is the mean free path.
Putting $k_F\sim \pi/a$, the Mott-Ioffe-Regel limit,
$\sigma^{\rm min}$, is obtained by setting $l/a=1$
\begin{equation}
\sigma^{\rm min}\sim\frac{1}{3}\frac{e^2}{\hbar}\frac{1}{a}
\end{equation}
The physics  of this ``minimum metallic conductivity'' is that
once the point is reached where the electronic mean free path,
$l$, is of the order of an interatomic spacing, $a$, the whole apparatus
of Boltzmann transport theory based upon Bloch states of definite
momentum becomes questionable. 

Mott's argument may be suitably modified for an anisotropic metal
using the results derived above. 
The minimum c-axis conductivity 
is obtained by putting $v_{\perp}\tau=c$ giving
\begin{equation}
\sigma_c^{\rm min}\sim\frac{1}{2\pi}\frac{e^2}{\hbar}k_F
\left(\frac{v_{\perp}}{v_F}\right)
\label{c-axis Mott}
\end{equation}
where $v_{\perp}$ is short for $\langle v_{\perp}^2({\bf k}) \rangle^{1/2}$.
Similarly, $v_F\tau=a$ gives
\begin{equation}
\sigma_a^{\rm min}\sim\frac{1}{2\pi}\frac{e^2}{\hbar}\,k_F\,\frac{a}{c}
\end{equation}

We now apply these results to La$_{2-x}$Sr$_x$CuO$_4$. Using
$a \approx$ 0.4 nm, $c \approx$ 0.8 nm, the band structure estimate
\cite{allen}
$v_{\perp}/v_F \approx 1/5$ for La$_{1.85}$Sr$_{0.15}$CuO$_4$, 
and estimating $k_F$ by $n=k_F^2/(2\pi)$ with $n\approx a^{-2}$,
{\em i.e.\/} one electron per Cu, we find
$\sigma_c^{\rm min}\approx 500\,{\Omega}^{-1}$ cm$^{-1}$
and $\sigma_a^{\rm min}\approx 1200\,{\Omega}^{-1}$ cm$^{-1}$
\cite{fnholes}.
By examining the resistivity data for La$_{2-x}$Sr$_x$CuO$_4$ 
shown in Fig. \ref{fig: uchida_dc} we see that at the lowest normal 
state temperatures
accessible $\sigma_c$ is well below $\sigma_c^{\rm min}$ in all but the
most overdoped samples. In contrast, $\sigma_a$ is always well 
above $\sigma_a^{\rm min}$. Moreover, there is something very wrong
with the estimate $\sigma_a / \sigma_c \approx (v_F/v_{\perp})^2
\approx 25$: for example, for the $x=0.15$ sample, 
$\sigma_a (T)/ \sigma_c (T) = \rho_c(T)/\rho_a(T)$ 
{\em increases\/} monotonically as $T$ decreases, with
$\sigma_a (T_c)/ \sigma_c (T_c) \approx 700$!

At this point some remarks are in order. The most important is that
standard Fermi liquid quasiparticle renormalizations do not alter the
above results. To see why, consider that $k_F$ is an invariant under
such renormalizations, a consequence of Luttinger's theorem.
At sufficiently low temperature, scattering is dominated by
impurities, in which case $v_F\tau=l_a$ is an invariant. Finally, 
$v_{\perp}/v_F$ is an invariant because both $v_{\perp}$ and $v_F$
are renormalized by the same quasiparticle renormalization factor, $Z$.
Hence, both $\sigma_c$ and $\sigma_a$ are invariants. This is the
generalization of the isotropic result (\ref{eq: Mott}) which again
is easily seen to be robust to even the most severe Fermi liquid
renormalizations. Quite generally, velocity reducing
factors are exactly cancelled by density of state enhancement.
It is this fact which makes the Mott argument so
powerful. Although Mott originally developed the argument in the
context of localization theory,
we emphasize that it is a general argument indicating how large the 
conductivity of an electronic Fermi liquid {\em has\/} to be in order to
justify the use of quasiparticle Boltzmann transport theory. 

The second remark concerns the use of the isotropic-$\tau$ approximation.
One could envisage a situation in which $\tau({\bf k})$ was a strongly
${\bf k}$-dependent function, such that 
$\tau(k_c\approx 0)\gg \tau(k_c\neq 0)$, in which case $\sigma_c$
could be significantly further reduced. Such a scenario could arise
if planar defects, parallel to the ab-planes, were the dominant source of
scattering \cite{koshelev}. In the case of La$_{2-x}$Sr$_x$CuO$_4$,
however, there is no experimental evidence for such a scenario:
instead, one would expect the dominant source of scattering to be
from isolated Sr atoms which are randomly arranged throughout the 
sample and reside approximately 0.4 nm  from the CuO$_2$ planes,
giving at most a weakly ${\bf k}$-dependent mean free path.

Thirdly, it should be emphasized that, as discussed in the Introduction,
the above theoretical considerations
have been verified in the highly anisotropic (2+1)-D metal Sr$_2$RuO$_4$
\cite{maeno}.
The low temperature in-plane and c-axis resistivities, plotted as a function
of $T^2$, are shown in Fig. \ref{fig: ruthenate}. The behavior is FL-like in all directions. A
crude estimate of the anisotropy from band structure calculations 
\cite{oguchi} gives $\rho_c/\rho_{ab}\approx 100$ in reasonable 
agreement with the data which give 
$\rho_c(T\rightarrow T_c^+)/\rho_{ab}(T\rightarrow T_c^+)\approx 500$.
Of crucial import is the fact that 
$\sigma_c(T\rightarrow T_c^+)\approx 2000\,{\Omega}^{-1}$ cm$^{-1}$,
much larger than the Mott limit 
$\sigma_c^{\rm min}\sim$
a few hundred
%\approx 250\,
${\Omega}^{-1}$ cm$^{-1}$. The conductivity
at low temperatures is truly metallic in all directions. We believe that 
Sr$_2$RuO$_4$ is the most anisotropic Fermi liquid known to date.

A comparison of the dc-conductivity of La$_{2-x}$Sr$_x$CuO$_4$ and
Sr$_2$RuO$_4$ therefore leads us to the conclusion that anisotropy
alone cannot account for the anomalous c-axis transport in 
La$_{2-x}$Sr$_x$CuO$_4$. The behavior of $\rho_c(T)$ cannot be accounted
for within a Fermi liquid framework; there is no plausible mechanism to
lead to a c-axis scattering rate larger than $t_{\perp}$. We are therefore
led to what we believe is the {\em only\/} way to understand the c-axis
transport in the normal state of many of the HTSC's. This is to associate 
the anomalous c-axis conductivity with the fact that the in-plane electron liquid
is not a Fermi liquid.

\subsection{Further Empirical Evidence for Incoherence}

There are other, equally compelling, experimental facts which 
%concur with
%the conclusions we come to on the basis of dc-resistivity. 
are at odds with any Fermi liquid description of c-axis
transport in the cuprates. The first is the 
frequency-dependent conductivity, $\sigma_c(\omega)$,  data for which
are shown in Fig. \ref{fig: uchida_ac}
 for La$_{2-x}$Sr$_x$CuO$_4$ \cite{uchida_ac}. For all but perhaps the most
overdoped sample ($x=0.3$) there is no plausible way to argue for a Drude,
or generalized-Drude, contribution. The same is true for 
YBa$_2$Cu$_3$O$_{6+x}$ (Figs. \ref{fig: ybco_ac(cooper)}, \ref{fig: ybco_ac}), 
even the ``optimally doped''
YBa$_2$Cu$_3$O$_7$ where the ``linear-T'' \cite{fnaffine}
 behavior of $\rho_c$ has often
been interpreted as a sign of metallic c-axis conduction. As the data of
Sch\"{u}tzmann {\em et al.\/} \cite{schutzmann} show (Fig. \ref{fig: schutz}), 
a fit of a Drude term $\sigma_c^D$ to the low frequency
$\sigma_c(\omega)$,
\[
\sigma_c^D(\omega)=\left(\frac{ne^2}{m_b}\right)
\frac{\Gamma(\omega)}{\Gamma^2(\omega)+\omega^2
(m^*(\omega)/m_b)^2}
\]
leads to a width $\Gamma(\omega\rightarrow 0)\sim 1500$K at T=100 K, 
much too large to
account for either in a conventional way or in a way connected to the
anomalous in-plane relaxation rate $\Gamma_{ab}\sim$ T.
In fact, the raw data are very flat, or perhaps weakly rising with
frequency, at all but the lowest and highest (of order an $eV$) frequencies.

Note that in discussing the dc-conductivity in the previous section, it was necessary
to use estimates of the c-axis bandwidth from band structure calculations
in order to estimate, for example, $\sigma_c^{\rm min}$. Within a Drude
framework, the dc-conductivity is 
$\sigma_c^D(\omega=0)=ne^2/(m_b\Gamma)$, so that we only obtain
direct information about the combination $m_b\Gamma$.
The important advantage of ac-conductivity experiments is that
they enable one to disentangle the mass of the carriers from
the relaxation rate. As a result, any attempt to explain away the poor
dc c-axis conductivity in terms of a very small c-axis bandwidth is at
odds with the $\sigma_c(\omega)$ data which exhibit spectral weight
over a very wide energy range. Put another way, $\sigma_c(\omega)$
{\em cannot\/} be characterized by a narrow Drude conductivity of
small weight. Were one willing to ignore the band theory estimates of
$t_{\perp}$ one could have proposed such a scenario on the basis of
the dc-conductivity data alone. 

Again, the c-axis conductivity observed in Sr$_2$RuO$_4$ provides a stark contrast 
to the cuprates. Recent measurements by Katsufuji {\em et al.\/} 
\cite{katsufuji} exhibit
very beautifully the development of a Drude peak in $\sigma_c(\omega)$
at temperatures below 100 K or so. Thus, Sr$_2$RuO$_4$ provides a very
important control system to test predictions for transport in a highly
anisotropic Fermi liquid.

Further experimental evidence in favor of c-axis incoherence can be
found by considering Bi$_2$Sr$_2$CaCu$_2$O$_8$. This is a bilayer HTSC
in which the {\em inter\/}bilayer hopping, $t_{\perp}^{\rm inter}$ is 
significantly smaller
than the {\em intra\/}bilayer hopping, $t_{\perp}^{\rm intra}$.
Reflectivity experiments show no indication of any c-axis
Bloch-Boltzmann quasiparticles and the plasma edge is below
30 cm$^{-1}$, corresponding to an effective mass anisotropy 
$m_c^*/m_{ab}^*\approx 100$. On the other hand, band structure 
calculations estimate $t_{\perp}^{\rm intra}$ to be about 100 $meV$.
Quite generally, if $t_{\perp}^{\rm inter}/t_{\perp}^{\rm intra}\ll 1$,
Fermi liquid theory predicts {\em two\/} Fermi surfaces
split in energy by $\sim 2 t_{\perp}^{\rm intra}$, with each exhibiting a c-axis
dispersion $\sim t_{\perp}^{\rm inter}$. Therefore, even though the 
dc- and low-frequency conductivity may be poor, a strong absorption signal
(albeit broadened by anisotropy of $t_{\perp}$ and lifetime effects)
in $\sigma_c(\omega)$ should be observable at frequencies around 
$\sim 2 t_{\perp}^{\rm intra}$, which corresponds to an inverse wavelength
in the 1000-2000 cm$^{-1}$ range. However, reflectivity data 
\cite{tajima_ac} for 
Bi$_2$Sr$_2$CaCu$_2$O$_8$ provide no evidence of such absorption.
The same considerations may be applied to other bilayer HTSC's, 
YBa$_2$Cu$_3$O$_{6+x}$ for example. Again, there is no evidence of an
intrabilayer splitting.

In principle, the most direct probe to locate Fermi surfaces is photoemission.
The most recent photoemission experiments \cite{ding} on Bi$_2$Sr$_2$CaCu$_2$O$_8$
exhibit just a {\em single\/} Fermi surface, within an experimental resolution
of $\sim 10\,meV$. This correlates completely with the absence of any
resonance in $\sigma_c(\omega)$ in the 1000 cm$^{-1}$ range. 
%The conclusion is that
Electrons are apparently unable to tunnel coherently from plane to plane
even {\em within\/} a bilayer. 

This is consistent with the situation 
{\em vis a vis\/} the c-axis conductivity in the single layer material
La$_{2-x}$Sr$_x$CuO$_4$. The conclusion to draw is that in all
superconducting samples of HTSC's, c-axis transport is an incoherent
process and that there is no physically plausible way to modify FLT
to account for this incoherence. To emphasize the latter point, we 
now briefly discuss the various scenarios for c-axis transport 
which have previously been proposed.

\subsection{Theoretical Implications of Incoherence}

We have demonstrated two key aspects of normal state c-axis
transport in the cuprates: (1) it is incoherent (hence non-metallic), as
evidenced by experimental data; (2) taking band structure estimates for
anisotropy, such incoherence is incompatible with a Fermi liquid description.
Apart from the suggestion of anisotropic localization (discussed and criticized
below), there are really only two ways to attempt to theoretically understand 
this situation,
the first being to retain a FL-like description, and effectively ignore
the band theory estimates of anisotropy, the second being to ascribe the
anomalous c-axis transport to a non-Fermi liquid origin.

We begin with the former. These ``conventional'' attempts \cite{conventional,zha}
at explaining the incoherent c-axis
conduction in the HTSC's assume, either explicitly or implicitly, that
the interplane hopping rate is sufficiently small that the Mott limit 
(\ref{c-axis Mott}) is {\em not\/} violated at low temperatures. The
zero temperature conductivity is then very small, but nonetheless
metallic. One can then advocate various types of inelastic interplanar
hopping processes, which are additive to the conductivity, to generate
a $\rho_c(T)$ which is a decreasing function of temperature. 
The problem with such an approach, though, is that it fails to address
its very starting point, namely why the effective $t_{\perp}$ is so much 
smaller than the band theory prediction (in, say, La$_{2-x}$Sr$_x$CuO$_4$).
Again, one cannot use standard band-narrowing effects from FLT, for these
are {\em isotropic\/} in the sense that both $t$ and $t_{\perp}$ are 
renormalized by the same factor, which therefore leaves 
$\sigma_c^{\rm min}(\omega)$ unchanged. In essence, these proposals
demand some mechanism for an {\em anisotropic\/} downward renormalization
of the bandwidths, the downward renormalization being required to be most
severe in $t_{\perp}$. However, there is no such mechanism within FLT.
From the experimental point of view, we again emphasize that the frequency
dependent conductivity, $\sigma_c(\omega)$, has weight over a {\em very\/}
broad frequency range, and is generically an {\em increasing\/} function
of frequency, which at low temperature is completely incompatible with these 
conventional 
approaches.

If one accepts that the band theory estimates of $t_{\perp}$ 
are not wildly inaccurate, then
one is forced
%, at the very least
%for La$_{2-x}$Sr$_x$CuO$_4$ and YBa$_2$Cu$_3$O$_{6+x}$, 
to look 
elsewhere for an explanation of the poor c-axis conductivity. Within a 
simple Bloch-Boltzmann Fermi liquid framework there is really only one other 
parameter to play with, namely the relaxation time, $\tau({\bf k})$. 
We have already mentioned above that we do not believe there is a
plausible way to argue for a strongly anisotropic $\tau({\bf k})$
(such a strong anisotropy is absolutely essential in the scenarios of Refs. 
\cite{koshelev} and \cite{zha}).
A related proposal, however, has been put forward by Kotliar {\em et al.\/}
\cite{kotliar}.
These authors advocate a regime of ``anisotropic localization'' where the
localization length $\xi$ is sufficiently anisotropic that (for all temperatures
above T$_c$) $\xi^{ab} > l_i^{ab}$ but $\xi^{c} < l_i^{c}$. Here $l_i$ is
the inelastic scattering length. Since effects of weak localization are only
seen when $\xi < l$, the claim is that in such a regime, localization
behavior will be observed in $\rho_c$ but not in $\rho_{ab}$.

From a theoretical point of view, the problem with this suggestion is that
one again must advocate a highly anisotropic random potential. The extreme case
is that of unidirectional randomness, {\em i.e.\/} planar disorder, which
will lead to states which are localized along the c-axis (since a 1D random
potential will always localize) but extended in the ab-plane \cite{koshelev}.
But, as pointed out earlier, in La$_{2-x}$Sr$_x$CuO$_4$ for example, there is
no reason to believe that the disorder introduced by the Sr ions is 
anything like planar. Rather, it is almost isotropic in which case the 
appropriate localization problem to consider is that of quasiparticles with
an anisotropic mass moving in an isotropically random potential. This problem
has been studied by several authors 
\cite{apel,wolfle} with the conclusion that localization
occurs {\em simultaneously} in all directions. At non-zero temperature, any
inelastic processes which cut off localization do so in a proportional way,
in the sense that $\delta\sigma_{ab}/\sigma_{ab}^D=\delta\sigma_{c}/\sigma_{c}^D$
where $\delta\sigma$ denotes the departure from the Drude result $\sigma^D$.
In particular, the temperature dependence of the conductivity will be the 
same in all directions, and one will not observe anisotropic localization.
As Anderson has remarked \cite{pwa_science_92}, the physics here is that 
weak localization requires
coherent backscattering of quasiparticles, and loss of coherence due to 
inelastic scattering in any one direction implies loss of coherence in all
\cite{fnkotliar}.
%[footnote: To be fair, Kotliar {\em et al.\/} do not advocate a FL in-plane
%state, so that direct application of weak localization arguments in FL's 
%may not be appropriate.
%In the absence of any real theory of localization in NFL's, however, we still do
%not see any physical justification for the anisotropic localization scenario.]

From an experimental point of view the anisotropic localization scenario is
even less tenable because the virtually flat low frequency behavior of 
$\sigma_c(\omega)$ is completely at odds with weak localization.

A somewhat different scenario has been proposed by 
 Kumar and Jayannavar \cite{kumar} to account for HTSC
samples which exhibit a $\rho_c(T)$ which {\em increases\/} with temperature.
At temperatures such that the in-plane transport rate 
$\tau_{ab}^{-1}\sim k_B T$ is larger than $t_{\perp}$, one expects to enter a regime 
where c-axis transport is rendered incoherent due to in-plane ``dephasing''.
Here, the in-plane scattering is predominantly inelastic so that
the phase coherence of an electron is destroyed 
at a rate faster than the rate
of interplane hopping. In this regime $\rho_c/\rho_{ab}$ can be
arbitrarily large, but temperature independent. As such, this picture will
only work at high temperatures, the lower temperature cutoff being 
roughly $t_{\perp}$.
For La$_{2-x}$Sr$_x$CuO$_4$ (except highly overdoped) the $\rho_c(T)$ data set
this cutoff at a few hundred degrees, in agreement with the band structure estimate
$t_{\perp}\sim 500$ K.
This scenario {\em cannot\/} account for a monotonically
increasing c-axis resistivity (as temperature
is lowered) coexisting with a monotonically decreasing ab-plane resistivity.

Having ruled out a FL description of the c-axis transport, we are
forced to consider a NFL one. Unlike in the FL case where the interliquid 
hopping operator $t_{\perp}$ is always a relevant operator, in a NFL
$t_{\perp}$ can be relevant, marginal or irrelevant depending upon the
specifics of the NFL. However, as already discussed in the Introduction,
the observation of sharply peaked spectral functions in ARPES experiments
on Bi$_2$Sr$_2$CaCu$_2$O$_8$ and YBa$_2$Cu$_3$O$_{6+x}$ is {\em incompatible
with an irrelevant\/} $t_{\perp}$. It is this fact which rules out a purely
2D description of the cuprates. 
%Such a description has been suggested,
%for example, by
%Nagaosa \cite{nagaosa} within a gauge theory calculation in the uniform
%RVB state in which it is proposed that $t_{\perp}$ is a marginal operator.
%However, it is simply not possible to have a marginal $t_{\perp}$ coexisting
%with a sharply peaked spectral function
%({\em i.e.\/} a spectral function of width much less than $t_{\perp}$)
%The claim in \cite{nagaosa} of a marginal
%$t_{\perp}$ follows only at sufficiently high temperatures (above the 
%holon condensation temperature) where the electron may be treated as 
%a convolution of essentially free holons and spinons. In this regime,
%the electron spectral function is essentially incoherent anyway.
%Unfortunately, at the present time there is really no reliable way 
%within the gauge models to
%calculate either the electron spectral function, or the effect of interlayer
%hopping, at low temperatures.
To our knowledge, the only NFL approach to c-axis transport other than ours 
is that based on 
t-J model gauge theories. The most recent is the work of Nagaosa \cite{nagaosa},
and Lee and Wiegmann \cite{lee_wiegmann}, where it has been argued that at
temperatures $T\gg T_{\rm BE}$ (where $T_{\rm BE}$ is the holon condensation 
termperature)
c-axis conduction is incoherent, and has temperature dependence 
$\sigma_c\propto x T^{1/2}$. In this regime, the electron may be treated as 
a convolution of essentially free holons and spinons, with the result that
$t_{\perp}$ is a marginal operator. While the electron spectral function is
dominated by its incoherent part, there is a ``quasiparticle'' part of
weight $x$ (the hole concentration), and width 
$\Gamma_{\rm qp} \sim(J\,T)^{1/2}$. The Schrieffer tunneling formula
is used to calculate $\sigma_c$, which is a self-consistent procedure
if $t_{\perp}~^{<}_{\sim}\Gamma_{\rm qp}$, {\em i.e.\/} 
$T~^{>}_{\sim} t_{\perp}^2/J$. The source of the incoherent $\sigma_c$
is therefore simply the large $\Gamma_{\rm qp}$, which is similar in
flavor to the phenomenology of Kumar and Jayannavar \cite{kumar}.
Of crucial import, however, is that the gauge model calculation is 
only valid for $T~^{>}_{\sim} T_{\rm BE}\sim xJ$, where $J$ is the Cu-Cu
antiferromagnetic superexchange, $J\sim 1500 K$ in the cuprates.
In fact, as explicitly shown in \cite{nagaosa}, there is a quasiparticle 
contribution to $\sigma_c$ of the form $\sigma_c^{\rm qp}\sim x^2\,T^{-1/2}$
which starts to dominate over the incoherent contribution at a temperature
$T\sim T_{\rm BE}$, and if this result was naively extended to the low temperature
limit it would ultimately lead to a metallic c-axis conductivity 
$\sigma_c\sim x^2\,T^{-1/2}$. However, the calculation 
is only justified  for $T~^{>}_{\sim}T_{\rm BE}$, and in the gauge model $T_{\rm BE}$
is supposed to correspond to the superconducting transition temperature $T_c$.
The claim then is that the metallic $\sigma_c$ is not observed experimentally
because of the onset of superconductivity. For the same reason, the Drude term
predicted in $\sigma_c(\omega)$ (coming from the quasiparticle contribution)
would not be resolvable above $T_c$ due to the broadening from $\Gamma_{\rm qp}$.
To summarize, the picture is that in the normal state the quasiparticle width
due to thermal broadening is sufficiently large compared to the interlayer
hopping rate as to render interplanar transport incoherent. Unlike the
phenomenology of Kumar and Jayannavar, however, the gauge model is capable
of producing a c-axis resistivity that increases as the temperature is lowered
(provided $T>T_{\rm BE}\sim T_c$). 

There are several criticisms of this approach. They all essentially stem from
the same point, namely the validity of application of the Schrieffer tunneling 
formula. As mentioned above, one expects this to be valid if 
$t_{\perp}~^{<}_{\sim}\Gamma_{\rm qp}$, {\em i.e.\/} 
$T~^{>}_{\sim} T_{\rm tunn}=t_{\perp}^2/J$.
In  La$_{2-x}$Sr$_x$CuO$_4$ the band theory estimate of $t_{\perp}\sim 500$ K
gives $T_{\rm tunn}\sim 200$ K, well above $T_c$. On the other hand, one might
argue that the true regime of validity is nontrivial to determine, that
$t_{\perp}~^{<}_{\sim}\Gamma_{\rm qp}$ is a conservative overestimate,
and the true regime may extend to somewhat lower temperatures. Nevertheless,
the criterion for validity of the tunneling formula raises its head when
one tries to compare the calculation of $\sigma_c$ with data. For simplicity,
we shall confine ourselves again to La$_{2-x}$Sr$_x$CuO$_4$. Consider the
result of \cite{nagaosa}:
\begin{equation}
\sigma_c\sim e^2\left(\frac{t_{\perp}}{J}\right)^2 
\left[ ax\left(\frac{J}{T}\right)^{1/2}+b \left(\frac{T}{J}\right)^{1/2}
\right]
\label{gauge_sigma_c}
\end{equation}
While we appreciate that it is unreasonable to demand a detailed agreement
of Eqn. (\ref{gauge_sigma_c})
with experimental data, there are several points worth noting.
Firstly, at high temperature Eqn. (\ref{gauge_sigma_c}) predicts 
$\sigma_c\sim T^{1/2}$, yet the experimental behavior at all doping levels
is that $\sigma_c$ {\em decreases\/} with $T$. Secondly, and of greater import, the
predicted doping dependence is {\em much weaker\/} than that exhibited by the
data. For example,  Eqn. (\ref{gauge_sigma_c}) predicts at most a factor of
4 ratio of the dc conductivity of $x=0.1$ and $x=0.2$ La$_{2-x}$Sr$_x$CuO$_4$,
while the data exhibit a ratio as high as 40. In order to avoid this
situation, Lee and Wiegmann \cite{lee_wiegmann} introduce a strongly
doping dependent hopping rate, {\em i.e.\/} $t_{\perp}=t_{\perp}(x)$.
For example, they take $t_{\perp}(x=0.16)\sim 30$ K and $t_{\perp}(x=0.3)\sim 200$ K.
The argument is circular, however, since the $t_{\perp}$ estimates are taken
from the {\em data\/}, whereas the $t_{\perp}$ in (\ref{gauge_sigma_c})
should be taken from the band theory, in which case the doping dependence would 
be nowhere near as large. Moreover, Lee and Wiegmann argue that the tunneling
formula is invalid in the overdoped regime, except at high temperature
($T~^{>}_{\sim} 500$ K), which again raises the issue of the estimate of
$T_{\rm tunn}$, and whether the use of the tunneling formula is ever valid
in the range where experiments are performed, given the large band theory
estimate of $t_{\perp}$. We note that the use of a strongly doping dependent
$t_{\perp}$ is common to {\em all\/} of the theoretical models for 
c-axis transport discussed above, because the data require a large variation of 
$\sigma_c$ with $x$. The only exception is the scenario we propose, where there
is potential for a strong variation in $\sigma_c$ with doping due to the 
complete lack of interplanar coherence even at $T=0$.

To conclude, while the gauge model calculation and our proposal of
``confined coherence'' have common elements, most notably the use of a NFL
in-plane state and the fact that incoherence results from a ``broad''
spectral function, we emphasize that in the gauge model picture the broadening is
thermally induced, while in our picture incoherence is a property of a truly 
$T=0$ state, due to a  ``broadening'' of the spectral function which is not a thermal,
nor an impurity, broadening. In fact, in our picture the ``broadening''
is not even of Lorentzian type.
These two theoretical approaches
to c-axis transport are therefore logically distinct.

%We therefore believe that, 
Taken together, the ARPES data and the 
c-axis transport data {\em require a theoretical framework in which
interplanar hopping is relevant but incoherent\/}. This is precisely 
the physics of ``confined coherence'', a specific example of which is
afforded, as shown in the previous sections, within the context of weakly
coupled Luttinger liquids. We believe that this is the most
promising framework within which to understand the anomalous c-axis
transport in the cuprates.

\subsection{Summary}
Upon examining the very large amount of data now available on c-axis transport
in the cuprates and comparing it to theoretical expectations one is forced
to accept that there is a very real anomaly. Moreover, there is really
nothing physically reasonable one can do within a FL framework to account for the 
anomalous behavior. We have emphasized the importance of transport measurements
in the highly anisotropic (2+1)-D metal Sr$_2$RuO$_4$ which are in accord
with what is expected from the application of FLT to the predicted band
structure. There is a striking difference between $\sigma_c$ in Sr$_2$RuO$_4$
and $\sigma_c$ in La$_{2-x}$Sr$_x$CuO$_4$. Our central point is that the one
ingredient missing from conventional attempts to understand c-axis transport in
the cuprates is the simple fact that they are {\em not\/} conventional metals:
the in-plane liquid is {\em not\/} a FL. Disorder might be a 
complication, but it alone cannot account for the anomalous physics.
In the absence of any plausible alternative capable of accounting for the
universal fact of incoherent c-axis transport, 
the simplest problem relevant to the issue is that of coherence/incoherence of 
single particle hopping between NFL's (without disorder).
We believe that the physics of ``confined coherence'' exhibited in the model of
coupled 1D Luttinger liquids, when suitably generalized to coupled 2D NFL's,
is the correct paradigm within which to be able to even qualitatively
understand the anomalous c-axis transport in the cuprates.
A first attempt at making semi-quantitative contact with experiments
has been previously presented \cite{prl2}.

%Anderson has also
%pointed out that it 
%is also not correct to
%advocate a regime of anisotropic localization (unless one has good reason
%to believe that scattering is dominated by planar defects).
%In particular,  Kotliar {\em et al.\/} have suggested that in a system with
%an anisotropic localization length, one could be in a regime at finite temperature
%where ab-plane localization was prevented by inelastic scattering, but c-axis
%localization was not. But localization in {\em any\/} direction requires coherent
%backscattering so that

%To see why, one 
%can consider both the strong- and weak-localization limits. In the former,
%it is clear that the effect of introducing an isotropically random static potential 
%will be to localize 

%% file: organics.sec
\section{Incoherence in the Organic Conductor (TMTSF)$_2$PF$_6$}
\label{sec:organics}

The original motivation for the proposal of the incoherent 
$t_{\perp}$ fixed point was the experimental data on the
cuprate superconductors which conflicted with the predictions
of both three 
dimensional Fermi liquid theory and theories in which the
transverse hopping was irrelevant,
however the theoretical arguments we have advanced for
the potential incoherence of hopping between Luttinger
liquids are quite general and other
strongly  interacting, highly anisotropic materials
are also candidates for incoherent behavior.
In fact, we believe that the organic conductor
(TMTSF)$_2$PF$_6$ is very close to having incoherent
single particle hopping in one direction
and can be driven by an appropriate external perturbation
into the incoherent regime. The experimental
evidence supporting this contention is both
extensive and strong as we will now discuss.

We begin with a brief discussion of the 
properties of the relevant material.
(TMTSF)$_2$PF$_6$ is an organic conductor composed of
long conducting chains of the organic molecule 
TMTSF stacked into planes which are separated by 
inorganic PF$_6$ anions.
The resulting three dimensional structure 
is triclinic  and its transport properties are highly
anisotropic with tight binding bandwidths
of $\sim 1 ~eV$ in the main chain direction (hereafter $a$),
$\sim 0.1 ~eV$ in the other in-plane direction (hereafter $b$)
and $\sim 0.003 ~eV$ in the out-of-plane direction (hereafter $c$).
These hopping integrals are so anisotropic
that the material has
a quasi-one dimensional Fermi surface with a pair of
slightly warped Fermi sheets for its single conduction band.   
This band would be quarter filled
except for a dimerization along the main chain direction 
which results in (TMTSF)$_2$PF$_6$ having a
half filled conduction band.  
The effects of interactions should be substantial
in this material
as naive estimates of the on site Coulomb interaction
repulsion energy is of order the largest bandwidth
and the conduction band is half-filled; 
the material
is therefore both strongly correlated and highly anisotropic.

(TMTSF)$_2$PF$_6$ exhibits an extraordinarily
rich low temperature phase diagram as a function of pressure
and applied magnetic field, including spin-density wave 
and
superconducting phases as well as a ``normal'' metal phase,
confirming that correlation effects 
are strong and observable in this material.
Here we concern ourselves with the metallic phase which
occurs under a few kilobars of pressure for temperatures 
high enough or
magnetic fields strong enough to destroy the superconducting
state.  In this phase the magnetoresistance of the material
is highly anomalous both in magnitude and in its dependence 
on the direction of the field. The unusual behavior of the
magnetoresistance provides the main evidence
for the realization of incoherence in this material and
will be central to our discussion.

Figure \ref{fig:woowon} shows the magnetoresistance
data of Kang {\it et al.} \cite{woowon}. 
In these experiments the current is applied along
the $\hat{a}$ direction (the most conducting
direction and the direction in which the bandwidth
is largest) while magnetic fields of various strengths
are rotated in the $bc$ plane.
 The magnetoresistance
is generally increasing as the field is tilted
away from the $b$ direction, but sharp dips in the
magnetoresistance occur for field orientations where the
field parallels a real space lattice vector.  These orientations
of the field
are referred to in the literature as ``Lebed magic angles''
after a proposal by Lebed \cite{fisdwmagiclebed} that 
there should be features associated with the field induced spin
density wave state for magnetic fields directed along real
space lattice directions.
There have been
previous attempts to account for the features
in the magnetoresistance at these angles  based on
commensurability effects
\cite{magicothers} and there are in general many reasons
why there might be features in the magnetoresistance at
these angles. However, as we will discuss,
the magnetoresistance data have features which
can only be accounted for by the transition of
(TMTSF)$_2$PF$_6$ to a state in which coherence is
lost for single particle motion out of the $ab$ planes.

Noting that (TMTSF)$_2$PF$_6$ is strongly correlated
and highly anisotropic, one might expect that the single
particle motion in this material might be incoherent 
in either the $b$ and $c$ directions or only in the $c$ direction.  
The interchain analysis of
Sec. \ref{sec:fermions} and \ref{sec:luttliq} 
would apply directly
to the former case, and the latter case is possible if
the ground state of a single $ab$ plane of (TMTSF)$_2$PF$_6$
had a non-Fermi liquid ground state with properties
similar to those of a one dimensional Luttinger
liquid (e.g. anomalous exponents, power law singularity
in $n_k$ at $k_F$ rather than a step function, different
spin and charge velocities).  Experimentally,  the evidence
is that {\it in the absence of a magnetic field with
a sufficient projection onto  the direction perpendicular
to the $a$ and $c$ lattice directions} the material is three
dimensionally coherent; we will discuss the evidence for
this coherence later.  For
now, we take this as given, but consider the possibility
that the single particle hopping in the $c$ direction
is very nearly incoherent.  In this case 
a magnetic field along the direction perpendicular
to the $a$ and $c$ lattice directions should 
have very novel effects.  

The reason for this is that
a field in this direction introduces inelasticity into
the hopping of electrons in the $c$ direction. After
making a Peierls substitution one obtains that an
electron in one plane with crystal momentum $k$ hops
into a state with  crystal momentum $k+ \hat{a} e  B l_c$ 
when hopping one lattice spacing, $l_c$, in the $c$ direction.
Here
$e$ is the electron charge,  $B$ is the magnetic and
$l_c$ is the $c$-axis lattice spacing. This momentum
shift implies an energy shift $\Delta E \sim v_F e  B l_c$
because the highly anisotropic nature of (TMTSF)$_2$PF$_6$
gives it a well defined Fermi velocity in the $a$ direction.
As we have argued, the question of
whether or not $t_{\perp}$ acts coherently
or incoherently is essentially the question, for small
$t_{\perp}$, of whether or not perturbation theory in
$t_{\perp}$ is degenerate or non-degenerate and
the added inelasticity helps to further lift the
degeneracy of the states connected by the $t_{\perp}$
operator, enhancing the possibility for incoherence.
This is analogous to the increase in incoherence
predicted for the two level system problem in the
case of non-degenerate levels \cite{TLS}.
The reason for commensurability effects in the magnitude
resistance is clear in this scenario:
if the field is directed along a real space lattice direction
out of the $ab$ plane then, for hops in that
direction, there is no 
change in crystal momentum thus  no
inelasticity induced by the field and no enhancement of the
incoherence.  The
material should then retain three dimensional coherence.  In general,
this should result in a decrease in the electrical resistivity
compared to cases where $B$ is not parallel to a real space lattice vector,
since the higher dimensional state should have substantially
less
scattering.  The changes in the resistance associated with this
transition should be large (of order the resistance itself)
as the transition is a qualitative one.
The dips should be particularly pronounced for $c$-axis transport
data such as those of Figure \ref{fig:guybc}, in which data of
Danner {\it et al.} \cite{guynfl} are shown for magnetoresistance
for currents perpendicular to the $ab$ plane
as fields of various strength are rotated in the
$bc$ plane.

While incoherence offers a natural explanation for the
dip features in the magnetoresistance and can account for
the anomalous magnitude of the magnetoresistance itself,
other explanations for some of the behaviors have been proposed
\cite{magicothers}.  Fortunately, there are a number of truly 
unique predictions of the incoherence theory which
offer genuine and stringent tests.
Additionally, there are several qualitative features of
the data which are inconsistent with all
other proposed theoretical explanations 
but completely compatible with the incoherence theory.

We first discuss the four novel
predictions of the incoherence explanation for
the magic angle effects
which are distinct from other theories
(some of these were discussed in \cite{magic}).
The first of these
is that there is a hierarchy among the magic angle dips.  For a dip
in the $a$ axis magnetoresistance
to occur for an orientation corresponding to a given real
space lattice direction, three dimensional coherence must rely on 
hopping in that direction. Since the coherence of
hopping in any direction out of the $ab$ plane implies
three dimensional coherence, this means that to observe a
dip hops in all other directions out of the $ab$ plane must
be incoherent; the central dip for fields nearly in the
$\hat{c}$ direction is therefore primary.  If the field
is too weak to disrupt the $c$ axis hopping sufficiently
strongly for higher order hopping integrals like that in the
$\hat{b} + \hat{c}$ direction to become important, then
no feature will be seen in the vicinity of the
other magic angles.  This is born out experimentally as shown by 
Figure \ref{fig:diphierarchy}
 which depicts the field dependence of the magnetoresistance
in the vicinity of the $\hat{c}$ and $\hat{b}+\hat{c}$ magic
angles.

For all fields strong enough to destroy the superconducting
state, there is a sharp angular dependence near $\hat{c}$
but the angular dependence near $\hat{b}+\hat{c}$ turns on
after the field has reached a strength of about $1$ Tesla.

The second prediction of the incoherence theory 
is that the feature at $\hat{b}$ is {\it not a magic angle
dip}.   No transition to three dimensional coherence occurs
when the field parallels $\hat{b}$ (in fact quite the
contrary-two dimensionality is maximal).
The shape and field dependence of this feature should
therefore be completely different from those occuring at the
other magic angles. 
This difference between this feature and those
occuring at the magic angles is very clear in the data
of Figure \ref{fig:woowon} but has been ignored by all previous theories.

The third prediction of the incoherence theory concerns the shape
of the magnetoresistance curve everywhere away from the
magic angle dips.  Our proposal is that for fields
with a
large projection along the direction perpendicular
to $\hat{a}$ and $\hat{c}$ and with large projections
along the directions perpendicular
to $\hat{a}$ and $n \hat{b} + \hat{c}$ for small $n$,
the field can combine with the interaction effects to
render all single particle hopping out of the $ab$ plane
completely incoherent.  If this occurs, then
in a path integral calculation of any
physical quantity, the phase
associated with taking an electron around a closed
loop that does not lie entirely in a single $ab$ plane
will be randomized by the incoherence of
the interplane hopping. Thus only loops lying entirely in
a single $ab$ plane have well defined phases, and since
the only effect of a magnetic field 
(at the level of a Peierls substitution where we have
ignored the effects of the field on the Wannier functions
themselves on which the tight binding description is based)
is to a add a phase 
proportional to the flux enclosed to the weight of such
paths, the magnetic field will effect only these paths.
Since only the component of the magnetic field  that is
perpendicular to the $ab$ plane contributes to the
fluxes such paths enclose, 
{\it all physical properties
will be independent of the other two
components of the magnetic field}.
In particular the magnetoresistance will depend only on 
the component of the field out of the $ab$ plane.
It is already clear from Figure \ref{fig:woowon}
 that the magnetoresistance
for a field along $\hat{b}$ is independent of field strength
for fields larger than $0.8$ Tesla, satisfying this prediction.
In Figure \ref{fig:scalingxx}, we
replot the data of Figure \ref{fig:woowon}  as
$R_{xx}(H) - R_{xx}(H \cdot \hat{c} = 0)$ versus
$\vec{H} \cdot \hat{n}_{ab}$ to demonstrate the more general
scaling:
{\it the magnetorestance is only a function of the 
field perpendicular to the $ab$ plane}.

Not only is $R_{xx}(H) - R_{xx}(H \cdot \hat{c}=0) = f(\vec{H} \cdot
\hat{n}_{ab})$, but $R_{xx}(H) - R_{xx}(H \cdot \hat{c}) 
\propto \sqrt{\vec{H} \cdot \hat{n}_{ab}}~$!  The fact there
there should exist some scaling function $f$ so that
$R_{xx}(H) - R_{xx}(H \cdot \hat{c}=0) = f(\vec{H} \cdot
\hat{n}_{ab})$ is a unique prediction of the
incoherence theory  which the data dramatically confirm.
The only alternative explanation for this would be
for the single particle hopping out of the $ab$ plane
to be an irrelevant operator.   This cannot be reconciled
with the fact that the  conductivity out of the $ab$ plane is 
not insulating
for an 18 Tesla field directed
along $\hat{b}$
down to temperatures of 50 mK \cite{chaikinunpub}
and the fact that the magnetoresistance for currents
out of the $ab$ plane 
can, depending on pressure,
change by less than a factor of two 
as the transition to planes without
coherent coupling is made. 
Further, there exists no proposal for a mechanism
by which a magnetic field along 
the direction perpendicular to $\hat{a}$ and $\hat{c}$ should
enhance the irrelevance of hopping out of the $ab$ plane
in the metallic state of (TMTSF)$_2$PF$_6$.
Thus the simple irrelevance of $t_{\perp}$ 
appears thoroughly implausible as an explanation for the observed
scaling behavior.

The fact that the scaling function for $R_{xx}$ is a square root
is not at present understood and is not a prediction of the
incoherence theory.  It is compatible with the theory and
totally incompatible with any three dimensional Fermi liquid
explanation for the magnetoresistance of (TMTSF)$_2$PF$_6$.
Likewise, the data shown if Figures \ref{fig:guybc} 
and \ref{fig:scalingzz} for the
magnetoresistance out of the $ab$ plane, while not
directly predicted by the incoherence theory, support
it very strongly.  The magnetoresistance in this direction 
also scales, depending only on $\vec{H} \cdot\hat{n}_{ab}$,
being proportional to 
$\left(\vec{H} \cdot\hat{n}_{ab} \right)^{x}$
where $ x \approx \frac{3}{2}$.

The anomalous power law requires a non-Fermi liquid
state,
while the scaling requires
a two and excludes a three dimensionally coherent state. It is particularly
striking that, except for the magic angle behavior, the
magnetoresistance is largest when the component of the
magnetic field out of the $ab$ plane is largest
{\it even though the current is in that direction}
while the magnetoresistance saturates to complete field independence
for fields along $\hat{b}$ once incoherence has been achieved
{\it even though this field is perpendicular to the current}.
Semiclassically, no saturation of the magnetoresistance would
occur for this second field-current configuration.

The final prediction of the incoherence theory 
for magnetoresistance experiments for fields
in the $bc$ plane is that
for weak magnetic fields the magnetoresistance should
exhibit a maximum for fields along the direction normal
to $\hat{a}$ and $\hat{c}$, rather than
the minimum observed  at $\hat{b}$ at higher fields.
This should occur as a result of the central dip expanding
as the field is reduced until the data have smoothly crossed
over to a form appropriate for a Fermi liquid 
with an open Fermi surface.
This is exactly what occurs \cite{chaikinunpub}. 
This result is particularly 
dramatic since it demonstrates that 
there really is a transition from  a state
with essentially the expected Fermi liquid
properties to a state with exactly the expected
properties for a state with ``confined coherence''.
This makes it clear that (1) it is the field
which induces the incoherence, not disorder or
finite temperature and (2) the behaviors in the
``confined coherence'' region really are incompatible
with Fermi liquid theory, it is not that we have somehow
considered an inappropriate Fermi liquid model.
%For the data taken at $0.5$ Kelvin this can not be
%checked easily because the fields required to destroy the 
%superconductivity are too large to leave any trace of
%coherence.  Above the superconducting transition
%temperature the magic angle effects are still observed
%and it is should be simple
%to
%test this prediction but this has not yet
%been done \cite{lowfieldfootnote}.  

The incoherence theory thus offers a natural, simple explanation
for the magic angle magnetoresistance effects in (TMTSF)$_2$PF$_6$,
making {\it four unique, strikingly confirmed predictions }.
Moreover, the theory is  consistent with all aspects of the data where
no direct prediction is yet possible,
even those aspects of the data which are incompatible
with other theories proposed to date.  
Further, as we now discuss,
an independently proposed test of the theory has been
carried out by Danner, {\it et al.} \cite{guynfl}
which probes the qualitative nature of c-axis transport directly.

In investigating the magnetoresistance
of
(TMTSF)$_2$ClO$_4$,
a similar material to (TMTSF)$_2$PF$_6$,
for currents perpendicular to the
$ab$ plane and fields in the $ac$ plane 
close to the $\hat{a}$ direction, 
Danner {\it et al.} found sharp magnetoresistance 
resonances as shown in
Figure \ref{fig:aclebed}.

They accounted for these effects with a quasiclassical
theory in which the averaging over quasiclassical
orbits of the velocity
out of the $ab$ plane is more effective for
certain field orientations than for others \cite{batman}
and argued that these resonances can
be used 
to probe the bandwidth of (TMTSF)$_2$ClO$_4$ in the $\hat{b}$ 
and $\hat{c}$ directions.  
Their explanation agrees quite well with the data
and some of the results of their simulations
are plotted in Figure \ref{fig:crvr}

It is possible 
to probe the bandwidth of (TMTSF)$_2$ClO$_4$ in the $\hat{b}$
and $\hat{c}$ directions because the shape of the resonances 
in the semiclassical theory is a calculable function of
these bandwidths and is sensitive to both of their magnitudes.
The shape and therefore the test is most sensitive to the
value of $t_b$, but also somewhat to the value of $t_c$.
Importantly, however, the test is
{\it entirely dependent on the existence of coherent
motion out of the $ab$ plane}.  Without such
coherence there is no notion of a
quasi-classical velocity or momentum transverse
to the $ab$ planes and the averaging effect simply
should not exist.   The natural prediction for (TMTSF)$_2$PF$_6$
of the incoherence theory  is that these  same resonances should
be present for that material
if and only if the component of field out of
the $ac$ plane is not sufficiently large for 
complete incoherence
to have set in.  If it is
large enough, then the behavior should cross over to a
broad background with a magnetoresistance depending only
on $\vec{B} \cdot \hat{n}_{ab}$ 
(in fact the background $\Delta R_{zz}$
should be $ \sim (\vec{B} \cdot \hat{n}_{ab})^x$
where $x \sim 3/2$), and 
the background should have dips superimposed where
the component of the field perpendicular to the $\hat{a}$
direction points along a real space lattice direction.
This occurs because, for these fields, 
hops along the field have no associated
$\Delta k_a$ and therefore no significant inelasticity;
these field are not  effective
in inducing incoherence.  A magnetoresistance with
almost exactly this structure is seen \cite{guynfl}.

For fields with no component out of the $ac$ plane, there
are features in the PF$_6$ magnetoresistance data which are 
analogous to those seen in ClO$_4$.  The features for PF$_6$
are much weaker and there is a large background magnetoresistance
compared to ClO$_4$ for fields tilted away from the $\hat{a}$
direction (see Figure \ref{fig:rzpvr})

In fact,
the identification of the features in the
PF$_6$ magnetoresistance plot with those present for
ClO$_4$ is not in itself compelling, however, the
similarity  is much clearer if the second
derivative with respect to angle is plotted instead
\cite{guynfl} as in Figure \ref{fig:dbatmandtheta},
which clearly demonstrates that the same resonances are
present in PF$_6$ as in ClO$_4$.

As the field is tilted out of the $ac$ plane we expect the
$c$-axis hopping to become incoherent and the resonance features
should vanish.  As the data of Danner {\it et al.} show,
the resonances are gone or nearly gone already for fields a field with a
$0.2$ Tesla component out of the $ac$ plane
 As the resonances
disappear the background magnetoresistance crosses over to 
the $(\vec{H} \cdot \hat{n}_{ab})^{\frac{3}{2}}$ for fields with
projections onto the $bc$ plane which are away from magic angle
directions; meanwhile the dips at Lebed magic angles appear as
the component of the field out of the $ac$ plane grows, becoming
clear when the field strength out of the $ac$ plane reaches $1$
Tesla.
These effects are  
as shown in Figures \ref{fig:rzpvrhy} and
\ref{fig:drphy}.

Notice that the quasiclassical explanation of \cite{batman}
can be extended to include fields out of the $ac$ plane
as shown in Figure \ref{fig:cryvr}.

The results of the experiment are in exact agreement with the predictions of
the incoherence theory with the exception of the large background present
in PF$_6$ for fields in or nearly  in the $ac$ plane.  The background
crosses over to the two dimensional magnetoresistance of the incoherent
state smoothly and is clearly a precursor of that resistance, however
that resistance is itself not well understood at the present time.  Since
it is a property of the isolated planes of the incoherent state, which
must have a non-Fermi liquid character for the theory to make sense
(and of course to have any chance of accounting for the magnetoresistance
itself),  we need to understand this non-Fermi liquid state well enough
to study the effects of magnetic field on it, which has not been possible
to this time.  

We believe that the degree to which the incoherence theory
predicts and explains the experimental data is compelling. 
Moreover, those
features of the data
which are as yet inexplicable are in no way in conflict with 
the theory, whereas they are in fundamental
conflict with  a three dimensionally
coherent picture for (TMTSF)$_2$PF$_6$.
Less exotic sources of incoherence than our proposal,
e.g. disorder and finite temperature effects,
are also effectively ruled out since the
relevant experimental anomalies are clearest at
the lowest temperatures and in the cleanest samples;
we believe them to be the result of an interaction
induced confinement of coherence even in
 the pure
system, zero temperature limit.

Parenthetically, we should remark that the experimental data
on (TMTSF)$_2$PF$_6$ place some very interesting constraints on
the breakdown of Fermi liquid theory occurring in the incoherent
phase.  First, further investigations \cite{chaikinkatyaunpub}
of the magnetoresistance
have revealed that the 3/2 power law
seen in the $R_{zz}$ magnetoresistance is {\it nonuniversal}
and strongly pressure dependent.  The non-Fermi liquid state
underlying this behavior must therefore support non-universal
power laws, a property reminiscent of the Luttinger liquid
ideas advocated by Anderson \cite{philshift}.  Second,
at the temperatures where the non-Fermi liquid behavior occurs,
the conduction band of (TMTSF)$_2$PF$_6$
is half-filled as the result of a dimerization transition.
A successful theory must therefore be consistent with
metallic behavior at half-filling in the non-Fermi liquid
state.  It is also interesting to note that
the field required to destroy the coherence and
stabilize the non-Fermi liquid state is smaller than the critical
field in the $b$ direction.  The superconductivity 
in this material is therefore exotic in the sense
that the superconducting state can be entered directly from a
state which is clearly non-Fermi liquid.
It seems likely that the non-Fermi liquid physics is central to
the occurrence of superconductivity at reasonably high
temperature in (TMTSF)$_2$PF$_6$, when all other
measured properties are most 
consistent with repulsive interactions.

In summary, when all of the data on (TMTSF)$_2$PF$_6$ are
taken together,
we regard them as providing an
experimental demonstration
of the existence of an alternative finite $t_{\perp}$ fixed point at which
there is no three dimensional coherence  even though $t_{\perp}$ is
not irrelevant.  These are exactly the essential, novel features
of our proposal and we regard the experiments as demonstrating the
existence of a phase of confined coherence, although the 
question of the precise connection
to the Luttinger  calculation we have done is still open.

%% file: concl.sec
\section{Conclusion}
\label{sec:concl}

Our essential claims in this paper are that there are 
sound theoretical reasons for believing in the existence
of a previously undiscovered state of matter in which
transport in one direction has been rendered incoherent 
by interaction effects, and that, 
most importantly, there exist experiments
which  demonstrate the existence of this phase
beyond reasonable doubt.  

Our initial reason for believing that such a state
might exist was the irreconcilability, discussed in
the Introduction 
and in Sec. \ref{sec:caxis},
of the experimental
data on the high temperature superconductors with
the various existing theoretical proposals.
This led us to propose as a natural alternative
to previous theoretical suggestions a
state in which three dimensional coherence was lost
through interactions. We believe
that this is a separate question from
the renormalization
group relevance of $t_{\perp}$ and
that,  while the two effects are related,
incoherence in general sets in before irrelevance.

Our proposal is based
on a close analogy between the incoherence found
in the two level system problem (discussed in section
\ref{sec:TLS}) and single particle hopping between Luttinger
liquids (discussed in Sec. \ref{sec:fermions}
 and \ref{sec:luttliq}).  This
analogy suggests that the oscillation frequency of
the number difference between two coupled Luttinger liquids
can 
be identified with the magnitude of the
coherent, single particle  hopping between
the liquids.  If this oscillation frequency vanishes there
is no coherent hopping \cite{general}.
In Sec. \ref{sec:interp}
we have discussed in detail how the dynamical
calculations of Sec. \ref{sec:TLS}, \ref{sec:fermions}
 and \ref{sec:luttliq}
can thus be reliably interpreted to reveal the nature 
(i.e. coherence or incoherence) of the
$t_{\perp} \neq 0$ groundstate.

For the special case of
coupled Luttinger liquids which we have considered in
detail theoretically,  this leads us to propose
a schematic renormalization group diagram 
for the two and many chain problems
very different from previous ideas
\cite{Bunch}.

Our picture differs from that reached by
other
studies of the same problem 
in that the possibility of an incoherent
fixed point for intermediate interactions
is included.
In general, other
works postulated that the
system either
crosses over to a two dimensional Fermi liquid
or has a renormalization group irrelevant
$t_{\perp}$, resulting in a trivial confinement of
the electrons to their original chains. In
our case, there exists an additional intervening
phase which we refer to as a phase of ``confined
coherence'' since, in this phase, coherent transport
(though not the electrons) is confined to the
chains.

The change from coherent transport in all directions
to confined coherence should be thought of as a
true transition as discussed in Sec. \ref{sec:interp}.  
The  natural order parameter
is the shape of the Fermi surface, specifically
the $k_{\perp}$ dependence of the Fermi surface, since
the phase with confined coherence has no higher
dimensional Fermi surface and hence no 
dependence at all on $k_{\perp}$, while the
fully coherent phase has a Fermi surface with
warping categorized by some renormalized, coherent
$t_{\perp}$ which is non-zero.  

Having introduced this transition, we are
able to explain the apparent contradictions
in the experimental data on the cuprate superconductors
which originally motivated our work 
(see Sec. \ref{sec:caxis} and \cite{prl2}).
This strongly supports our theoretical
proposal, however, the clearest experimental evidence in favor
of the existence of the new phase
comes from experiments on
(TMTSF)$_2$PF$_6$, where magnetoresistance
data demonstrate that the material can undergo
a transition from a three dimensional Fermi liquid
to a state with no coherent
three dimensional transport 
\cite{guynfl}. This demonstrates a state
in which the coherent single particle 
hopping is indeed vanishing, yet
in this state conductivity
in
the perpendicular direction is not insulating.
This rules out the renormalization group irrelevance
of $t_{\perp}$ and thus requires a state of
exactly the
sort we are proposing. The material is also categorized by
an extremely long mean free path,
while the relevant experiments are conducted at
0.5 K, smaller than any of the relevant energy scales in the
problem. Further, the effects of interest are 
increasingly pronounced with cleaner samples
and lower temperatures, so that
neither disorder
effects nor finite temperature can explain the results.
This was discussed in detail in  Sec. \ref{sec:organics}
along with various
other experimental results on (TMTSF)$_2$PF$_6$
that
are uniquely understandable in our theory.

In summary, we have made a theoretical proposal for
a new state in order to reconcile features of the
experimental data on the high $T_c$ cuprates.  Our
best  calculation suggests that the proposal is
viable for sufficiently strongly correlated, anisotropic
systems and there exists experiments on the
material (TMTSF)$_2$PF$_6$ that are uniquely
understandable in terms of a state in which 
exactly the proposed incoherence obtains.

%% file: appendix.sec
\section{Appendix}
\label{sec:appendix}

\subsection{Interliquid Hopping Rate}

To $O(t_{\perp}^2)$,
\begin{equation}
P(t)\equiv\mid\langle O_1O_2\mid e^{iH_0 t}e^{-iHt}\mid O_1O_2\rangle\mid^2
\end{equation}
is given by 
\begin{equation}
1-P(t)=2t_{\perp}^2L{\rm Re}\int_0^t dt_1 \int_0^{t_1} dt_2
\int dx\left\{\langle c^{(1)}(x,t_1)c^{(1)\dag}(0,t_2)\rangle
\langle c^{(2)\dag}(x,t_1)c^{(2)}(0,t_2)\rangle +(1\leftrightarrow 2)
\right\}
\end{equation}
where the superscripts on the electron operators label the liquid in
which the operator acts. In obvious notation, the interliquid hopping
rates $\Gamma_{ij}$ defined by 
\begin{equation}
\Gamma(t)=\Gamma_{12}(t)+\Gamma_{12}(t)\equiv -\frac{dP(t)}{dt}
\end{equation}
are then given by
\begin{eqnarray}
\Gamma_{12}(t)&=&2t_{\perp}^2L{\rm Re}\int_0^t dt'
\int\frac{dk}{2\pi}\langle c^{(1)}(k,t')c^{(1)\dag}(k,0)\rangle
\langle c^{(2)\dag}(k,t')c^{(2)}(k,0)\rangle \nonumber\\
&=&2t_{\perp}^2L{\rm Re}\int_{-\infty}^{\infty} dt'\,\theta(t')\,\theta(t-t')
\int\frac{dk}{2\pi}{\cal J}_1^{(1)}(k,t'){\cal J}_2^{(2)}(k,-t')
\end{eqnarray}
and similarly for $\Gamma_{21}$. Here
we have introduced the Green's functions
\begin{eqnarray*}
{\cal J}_1(k,t')&\equiv&\langle c(k,t')c^{\dag}(k,0)\rangle\\
{\cal J}_2(k,t')&\equiv&\langle c^{\dag}(k,0)c(k,t')\rangle
\end{eqnarray*}
Upon Fourier transforming we obtain
\begin{equation}
\Gamma_{12}(t)=2t_{\perp}^2L\int\frac{d\omega}{2\pi}
\frac{\sin\omega t}{\omega}\int\frac{d\omega'}{2\pi}
\int\frac{dk}{2\pi}{\cal J}_1^{(1)}(k,\omega'){\cal J}_2^{(2)}(k,\omega'-\omega)
\end{equation}
which, in the notation introduced in the text, may be written
\begin{equation}
\Gamma_{12}(t)=2t_{\perp}^2L\int\frac{d\omega}{2\pi}
\frac{\sin\omega t}{\omega}A_{12}(\omega)
\end{equation}
where
\begin{equation}
A_{ij}(\omega)=
\int\frac{d\omega'}{2\pi}
\int\frac{dk}{2\pi}{\cal J}_1^{(i)}(k,\omega'){\cal J}_2^{(j)}(k,\omega'-\omega)
\end{equation}
Note that ${\cal J}_{1,2}$ can be expressed in terms of the electron 
spectral function via
\begin{equation}
{\cal J}_{1,2}(k,\omega')=\frac{\rho(k,\omega'-\mu)}
{1+e^{\mp\beta(\omega'-\mu)}}
\end{equation}
which reduces in the zero temperature limit to
\begin{equation}
{\cal J}_{1,2}(k,\omega')=
\theta_{\pm}(\omega'-\mu)\rho(k,\omega'-\mu)
\end{equation}

\subsection{Coupled Fermi Liquids}

At low enough energies, the Fermi liquid spectral function may be taken to be
\[
\rho(k,\omega)=\frac{2Z^2\gamma\omega^2}
{(\omega-E_k)^2+Z^2\gamma^2\omega^4}
\]
We present the calculation for $A_{12}(\omega)$, which is the interliquid hopping
spectral function with potential coherence. We have
\begin{equation}
A_{12}(\omega)=\theta_+(\omega+\mu_2-\mu_1)
\int_{\mu_1}^{\mu_2+\omega}\frac{d\omega'}{2\pi}
\int\frac{dk}{2\pi}\rho(k,\omega'-\mu_1)\rho(k,\omega'-\omega-\mu_2)
\end{equation}
where we have taken $\Delta\mu\equiv\mu_2-\mu_1 > 0$.
For sufficiently small $\omega$, one can approximate one of the $\rho$'s by a 
$\delta$-function to obtain
\begin{eqnarray}
A_{12}(\omega)&=&\theta_+(\omega+\mu_2-\mu_1)
\int_{0}^{\omega+\Delta\mu}\frac{d\omega'}{2\pi}
\int\frac{dk}{2\pi}\rho(k,\omega')\rho(k,\omega'-\omega-\Delta\mu) \nonumber\\
&\sim&\theta_+(\omega+\mu_2-\mu_1)
\int_{0}^{\omega+\Delta\mu}\frac{d\omega'}{2\pi}
\int\frac{dk}{2\pi}2\pi Z\,\delta(\omega'-vk+\mu_1)\rho(k,\omega'-\omega-\Delta\mu)
 \nonumber\\
&=&\theta_+(\omega+\Delta\mu)
\frac{1}{v}\int_{0}^{\omega+\Delta\mu}\frac{d\omega'}{2\pi}
Z\,\rho((\omega'+\mu_1)/v,\omega'-\omega-\Delta\mu)
 \nonumber\\
&=&\theta_+(\omega+\Delta\mu)
\frac{1}{v}\int_{0}^{\omega+\Delta\mu}\frac{d\omega'}{2\pi}
\frac{2 Z^3 \gamma (\omega'-\omega-\Delta\mu)^2}
{\omega^2+Z^2\gamma^2(\omega'-\omega-\Delta\mu)^4}
\end{eqnarray}
Shifting the integration variable gives
\begin{eqnarray}
A_{12}(\omega)&\sim&\theta_+(\omega+\Delta\mu)
\frac{1}{v}\int_{-\Delta\mu}^{\omega}\frac{d\omega'}{2\pi}
\frac{2 Z^3 \gamma (\omega'-\omega)^2}
{\omega^2+Z^2 \gamma^2(\omega'-\omega)^4} \nonumber\\
&=&\theta_+(\omega+\Delta\mu)\frac{1}{v}
\left\{\int_{0}^{\omega}+\int_{-\Delta\mu}^{0}\right\}
\end{eqnarray}
For low energies, $\omega^2\gg Z^2 \gamma^2\omega^4$, the first integral 
evaluates to $(3\pi v)^{-1} Z^3\gamma\,\omega$.
The second integral needs care to evaluate, since
it is formally divergent at $\omega=0$. It can be rewritten as 
\begin{eqnarray}
&&Z^2\int_{0}^{\Delta\mu}\frac{d\omega'}{2\pi}
\frac{2Z\gamma(\omega'+\omega)^2}
{\omega^2+Z^2\gamma^2(\omega'+\omega)^4} \nonumber\\
&=&Z^2\Delta\mu\lim_{\Omega\rightarrow 0}\frac{Z\gamma}{\pi}
\frac{(\omega+\Omega)^2}{\omega^2+Z^2\gamma^2(\omega+\Omega)^4}\nonumber\\
&=&Z^2\Delta\mu\delta(\omega)
\end{eqnarray}
in the limit of small $\Delta\mu$. 

We therefore find
\begin{equation}
A_{12}(\omega)\sim\theta_+(\omega+\Delta\mu)\frac{1}{v}\left\{
Z^2\Delta\mu\delta(\omega)+\frac{Z^3\gamma}{3\pi}\omega\right\}
\end{equation}
the $\delta(\omega)$ piece representing a coherent term, while the
term linear in $\omega$ is marginal.

\subsection{Coupled Luttinger Liquids}

We now consider the case of coupled Luttinger liquids. In order to
proceed, we need Luttinger liquid spectral functions. A simple method
of obtaining these is presented below.

\subsubsection{Calculation of Luttinger Liquid Spectral Functions}
\underline{Fracton Convolution Formula}

The key to using spectral function methods for Luttinger liquids
is to write the Luttinger liquid electron spectral function
as a convolution of ``fracton'' spectral functions, the latter being
sharp $\delta$-functions. The fractons have a precise relation to the
charge and spin degrees of freedom and therefore represent in a
certain  well-defined sense ``holons'' and ``spinons''. 

We begin with some quite general observations. 
%One way to think about
The method of bosonization allows one to express fermionic fields as
exponentials of (essentially free) bosonic fields. For definiteness,
suppose one may write the electron destruction operator as
\begin{equation}
\psi(x,t)=\lambda e^{i\phi_a(x,t)} e^{i\phi_b(x,t)}
\end{equation}
where $\phi_a$, $\phi_b$ are bosonic fields chosen such that the electron
Hamiltonian may be rewritten to be harmonic in these fields, $\phi_a$ and $\phi_b$
being decoupled at the Hamiltonian level.
$\lambda$ is a (real) parameter chosen
to get the dimensions of correlation functions correct.
We refer to the operators $e^{i\phi_a}$ and $e^{i\phi_b}$
as ``fractons'', the name being suggestive of the fact that
a real electron is made by ``glueing'' fractons together via
multiplication at the same space-time point.

We now write
\begin{eqnarray}
{\cal J}_1^{(e)}(k,\omega)&=&\frac{\lambda^2}{Z}\sum_{n,m}\int_{-\infty}^{\infty} dt
e^{i\omega t} \int dx e^{-ikx} 
\langle n|\psi(x)|m \rangle\langle m|\psi^{\dag}(0)|n \rangle
e^{-i(E_m-E_n)t} e^{-\beta E_n} \nonumber \\
&=&\frac{2\pi\lambda^2}{Z}
\sum_{n,m} e^{-\beta E_n} \int dx e^{-ikx}
\langle n|\psi(x)|m \rangle\langle m|\psi^{\dag}(0)|n \rangle
\delta(E_n-E_m+\omega) 
\end{eqnarray}
Since $\phi_a$ and $\phi_b$ are independent, all exact eigenstates $|m \rangle$
are of the form $|m_a,m_b \rangle$. Moreover, translation invariance allows
one to choose $|m_{a,b}\rangle$ to have definite momentum $k_{a,b}$, so that
(using $Z=Z_a Z_b$)
\begin{eqnarray}
{\cal J}_1^{(e)}(k,\omega)&=&\lambda^2
\left\{\frac{1}{Z_a}\sum_{n_a,m_a}e^{-\beta E_{n_a}}
\left|\langle n_a|e^{i\phi_a(0)}|m_a\rangle\right|^2\right\}
\left\{\frac{1}{Z_b}\sum_{n_b,m_b}e^{-\beta E_{n_b}}
\left|\langle n_b|e^{i\phi_b(0)}|m_b\rangle\right|^2\right\} \nonumber \\
&&2\pi\delta(\omega+E_{n_a}+E_{n_b}-E_{m_a}-E_{m_b})\;
2\pi\delta(k-k_{m_a}-k_{m_b})
\end{eqnarray}
But
\begin{eqnarray}
{\cal J}_1^{(a)}(k,\omega)&\equiv&\int_{-\infty}^{\infty} dt
e^{i\omega t} \int dx e^{-ikx} 
\langle e^{i\phi_a(x,t)} e^{-i\phi_a(0,0)}\rangle \nonumber \\
&=&\frac{1}{Z_a}\sum_{n_a,m_a}e^{-\beta E_{n_a}}
\left|\langle n_a|e^{i\phi_a(0)}|m_a\rangle\right|^2
2\pi\delta(k-k_{m_a})\;2\pi\delta(\omega+E_{n_a}-E_{m_a})
\end{eqnarray}
from which we obtain the convolution formula
\begin{eqnarray}
{\cal J}_1^{(e)}(k,\omega)&=&\frac{\lambda^2}{(2\pi)^2}
\int dk_1 \int dk_2 \int d\omega_1 \int d\omega_2 \nonumber\\
&&{\cal J}_1^{(a)}(k_1,\omega_1) {\cal J}_1^{(b)}(k_2,\omega_2)\;
\delta(\omega-\omega_1-\omega_2)\;\delta(k-k_1-k_2)
\end{eqnarray}

The generalization to three or more fracton fields is straightforward
and in fact necessary for dealing with spin-charge separated Luttinger liquids.

\underline{Fracton Spectral Functions}

It remains to give expressions for the fracton spectral functions,
from which the electron spectral function may be constructed using 
the above convolution formula. The space-time correlation function
for a fracton of weight $p$ is
\begin{equation}
\langle e^{i\phi(x,t)} e^{i\phi(0,0)}\rangle
\sim\left(\frac{i\pi a}{v\beta}\right)^p
\left\{\sinh\left[\frac{\pi(x+ia-vt)}{v\beta}\right]\right\}^{-p}
\end{equation}
from which one obtains
\begin{equation}
{\cal J}_1^{(p)}(k,\omega)=\left(\frac{i\pi a}{v\beta}\right)^p
\;2\pi\delta(\omega-vk)\int_{-\infty}^{\infty}
\frac{dz\;e^{-ikz}}{\left[\sinh\left(\frac{\pi(z+ia)}{v\beta}\right)\right]^p}
\end{equation}
for the fracton spectral function. Here, $\beta$ is inverse temperature,
$v$ the velocity of the relevant bosonic excitations, $\phi$, and $a$ is a
short-distance cutoff.

For our purposes, we shall only need the zero-temperature limit:
\begin{eqnarray}
{\cal J}_1^{(p)}(k,\omega)
&\stackrel{\beta\rightarrow\infty}{\sim}&
\left(\frac{i\pi a}{v\beta}\right)^p
\;2\pi\delta(\omega-vk)\int_{-\infty}^{\infty}
\frac{dz\;e^{-ikz}(v\beta)^p}{[\pi(z+ia)]^p} \nonumber \\
&=& (ia)^p\;2\pi\delta(\omega-vk)\int_{-\infty}^{\infty}
\frac{dz\;e^{-ikz}}{(z+ia)^p}
\end{eqnarray}
Choosing the cut for $z^{-p}$ to be along the positive real axis, we
may deform the contour of integration to obtain
\begin{equation}
\int_{-\infty}^{\infty}\frac{dz\;e^{-ikz}}{(z+ia)^p}
=(1-e^{-i2\pi p})\;\theta_+(k)\,k^{p-1} \int_{0}^{\infty}d\theta\;
e^{-i\theta}\theta^{-p}
\end{equation}
The $\theta$-integral may be performed by deforming the contour to
run up the negative imaginary axis. A change of variable to $x=i\theta$
gives
\begin{eqnarray}
{\cal J}_1^{(p)}(k,\omega)&=& a^p 2\pi\delta(\omega-vk)(1-e^{-i2\pi p})\;
\theta_+(k)\,k^{p-1}\;e^{i\pi p}(-i)\int_{0}^{\infty}dx\;e^{-x} x^{-p} \nonumber \\
&=& 4\pi\;a^p \sin(\pi p)\;\Gamma(1-p)\;k^{p-1}
\theta_+(\omega)\,\delta(\omega-vk)
\nonumber \\
&=& \frac{4\pi^2}{\Gamma(p)}a^p\;k^{p-1}\theta_+(\omega)\delta(\omega-vk)
\end{eqnarray}
the final line following from the identity $\Gamma(x)\Gamma(1-x)=
\pi/\sin(\pi x)$.

\subsubsection{Calculation of Interliquid Hopping Spectral Functions}
\underline{Spinless Luttinger liquid}

 From the previous section, we may write
\begin{eqnarray}
{\cal J}_1(k,\omega)&\propto&\int dk_1 dk_2\int d\omega_1 d\omega_2
\delta(\omega-\mu-\omega_1-\omega_2)\delta(k-k_F-k_1-k_2)\nonumber\\
&&\theta_+(\omega_1)(\omega_1/v_c)^p\delta(\omega_1-v_ck_1)
\theta_+(\omega_2)(\omega_2/v_c)^{p-1}\delta(\omega_2+v_ck_2)
\end{eqnarray}
A similar expression for ${\cal J}_2$ holds.
Then
\begin{eqnarray}
A_{12}(\omega)&\propto&\int\frac{d\omega'}{2\pi}\int\frac{dk}{2\pi}
\int_0^{\infty} d\omega_1 d\omega_2
\delta(\omega'-\mu_1-\omega_1-\omega_2)
\delta(k-k_F^{(1)}-\frac{\omega_1}{v_c}+\frac{\omega_2}{v_c})
(\omega_1/v_c)^p
(\omega_2/v_c)^{p-1} \nonumber\\
&&\int_{-\infty}^0 d\omega_1' d\omega_2'
\delta(\omega'-\omega-\mu_2-\omega_1'-\omega_2')
\delta(k-k_F^{(2)}-\frac{\omega_1'}{v_c}+\frac{\omega_2'}{v_c})
(\omega_1'/v_c)^p
(\omega_2'/v_c)^{p-1} \nonumber \\
&=&v_c^{3-4p}\int_0^{\infty} d\omega_1 d\omega_2 d\omega_1' d\omega_2'
\delta(\omega_1+\omega_2+\omega_1'+\omega_2'-(\omega+\Delta\mu)) \nonumber\\
&&\delta(\omega_1+\omega_1'-\omega_2-\omega_2'-v_c\Delta k)
\omega_1^p \omega_2^{p-1}
(\omega_1')^p(\omega_2')^{p-1}
\end{eqnarray}
To get some idea of how to evaluate these integrals, let us first consider 
the case $\Delta\mu=0$:
\begin{eqnarray}
A_{12}(\omega)_{\Delta\mu=0}
&\propto&\int_0^{\infty} d\omega_1 d\omega_2 d\omega_1' d\omega_2'
\delta(\omega_1+\omega_2+\omega_1'+\omega_2'-\omega) \nonumber\\
&&\delta(\omega_1+\omega_1'-\omega_2-\omega_2')
\omega_1^p \omega_2^{p-1}
(\omega_1')^p(\omega_2')^{p-1}
\end{eqnarray}
Rescaling of the variables of integration, $\omega_i=\omega x_i$ etc.,
immediately yields $A_{12}(\omega)_{\Delta\mu=0}\propto\omega^{4p}$.
However, it is instructive to be more explicit:
\begin{eqnarray}
A_{12}(\omega)_{\Delta\mu=0}
&\propto&\int_0^{\infty} d\omega_2 d\omega_1' d\omega_2'
\delta(2(\omega_2+\omega_2')-\omega)(\omega_2+\omega_2'-\omega_1')^p
\theta_+(\omega_2+\omega_2'-\omega_1')
\omega_2^{p-1}
(\omega_1')^p(\omega_2')^{p-1} \nonumber\\
&=&\int_0^{\infty} d\omega_1' d\omega_2'
(\omega/2-\omega_1')^p
(\omega/2-\omega_2')^{p-1}
(\omega_1')^p(\omega_2')^{p-1}
\theta_+(\omega/2-\omega_1')\theta_+(\omega/2-\omega_2') \nonumber\\
&=&\theta_+(\omega)\left\{\int_0^{\omega/2} d\omega_1'
(\omega/2-\omega_1')^p(\omega_1')^p \right\}
\left\{\int_0^{\omega/2} d\omega_2'
(\omega/2-\omega_2')^{p-1}(\omega_2')^{p-1} \right\}
 \nonumber\\
&\propto& \theta_+(\omega)\omega^{2p+1}\omega^{2p-1}
=\theta_+(\omega)\omega^{4p}
\end{eqnarray}

For $\Delta\mu >0$ write (defining $\Delta\equiv\Delta\mu-v_c\Delta k
\equiv (v-v_c)\Delta k =(1-v/v_c)\Delta k$)
\begin{eqnarray}
A_{12}(\omega)
&\propto&\int_0^{\infty} d\omega_1 d\omega_2 d\omega_1' d\omega_2'
\delta(\omega_1+\omega_2+\omega_1'+\omega_2'-(\omega+\Delta\mu)) \nonumber\\
&&\delta(\omega_1+\omega_1'-\omega_2-\omega_2'-v_c\Delta k)
\omega_1^p \omega_2^{p-1}
(\omega_1')^p(\omega_2')^{p-1} \nonumber\\
&=&\int_{-\Delta\mu}^{\infty} d\omega_1 
\int_0^{\infty}d\omega_2 d\omega_1' d\omega_2'
\delta(\omega_1+\omega_2+\omega_1'+\omega_2'-\omega) \nonumber\\
&&\delta(\omega_1+\omega_1'-\omega_2-\omega_2'+\Delta)
(\omega_1+\Delta\mu)^p \omega_2^{p-1}
(\omega_1')^p(\omega_2')^{p-1} \nonumber \\
&=&\int_{-\Delta\mu}^{\infty} d\omega_1 
\int_0^{\infty}d\omega_2 d\omega_2'
\delta(2(\omega_2+\omega_2')-(\omega+\Delta)) 
(\omega_1+\Delta\mu)^p \omega_2^{p-1}
(\omega_2')^{p-1}\nonumber\\
&&(\omega_2+\omega_2'-\omega_1-\Delta)^p\theta_+(\omega_2+\omega_2'-\omega_1-\Delta)
 \nonumber \\
&=&\int_{-\Delta\mu}^{\infty} d\omega_1 
\int_0^{\infty} d\omega_2'
(\omega_1+\Delta\mu)^p (\omega_2')^{p-1}\nonumber\\
&&\left(\frac{(\omega+\Delta)}{2}-\omega_2'\right)^{p-1}
\theta_+\left(\frac{(\omega+\Delta)}{2}-\omega_2'\right)
\left(\frac{(\omega-\Delta)}{2}-\omega_1\right)^{p}
\theta_+\left(\frac{(\omega-\Delta)}{2}-\omega_1\right)
 \nonumber \\
&=&\left\{\int_{0}^{\infty} d\omega_1 \omega_1^p
\left(\frac{(\omega-\Delta)}{2}+\Delta\mu-\omega_1\right)^{p}
\theta_+\left(\frac{(\omega-\Delta)}{2}+\Delta\mu-\omega_1\right)\right\}
 \nonumber\\
&&\left\{\int_0^{\infty} d\omega_2' (\omega_2')^{p-1}
\left(\frac{(\omega+\Delta)}{2}-\omega_2'\right)^{p-1}
\theta_+\left(\frac{(\omega+\Delta)}{2}-\omega_2'\right)\right\}
\end{eqnarray}
Using $\Delta=(1-v/v_c)\Delta k$ this simplifies to
\begin{eqnarray}
A_{12}(\omega)\propto\theta_+\left[\omega+\left(1-\frac{v_c}{v}\right)\Delta\mu\right]
\theta_+\left[\omega+\left(1+\frac{v_c}{v}\right)\Delta\mu\right]\nonumber\\
\left(\omega+\left(1-\frac{v_c}{v}\right)\Delta\mu\right)^{2p-1}
\left(\omega+\left(1+\frac{v_c}{v}\right)\Delta\mu\right)^{2p+1}
\end{eqnarray}

It is straightforward to show that 
$A_{21}(\Delta\mu,\omega)=A_{12}(-\Delta\mu,\omega)$. Without loss of generality,
we may take $v-v_c < 0$, so that
\begin{equation}
A_{12}(\omega)\propto\theta_+\left[\omega-\left(\frac{v_c}{v}-1\right)\Delta\mu\right]
\left(\omega-\left(\frac{v_c}{v}-1\right)\Delta\mu\right)^{2p-1}
\left(\omega+\left(\frac{v_c}{v}+1\right)\Delta\mu\right)^{2p+1}
\end{equation}
and
\begin{equation}
A_{21}(\omega)\propto
\theta_+\left[\omega-\left(\frac{v_c}{v}+1\right)\Delta\mu\right]
\left(\omega+\left(\frac{v_c}{v}-1\right)\Delta\mu\right)^{2p-1}
\left(\omega+\left(\frac{v_c}{v}+1\right)\Delta\mu\right)^{2p+1}
\end{equation}

\underline{Chiral Luttinger liquid}

Here we have charge- and spin-excitations with different velocities, but
no anomalous exponent. The prescription for glueing spin- and
charge-fractons together to give the electron spectral function is given by
\begin{eqnarray}
{\cal J}_{(1,2)}(k,\omega)=\frac{1}{(2\pi)^2}&&\int dk_1 \int dk_2
\int d\omega_1 \int d\omega_2 
{\cal J}_{(1,2)}^{(c)}(k_1,\omega_1){\cal J}_{(1,2)}^{(s)}(k_2,\omega_2)
\nonumber\\
&&\delta(\omega-\mu-\omega_1-\omega_2)\delta(k-k_F-k_1-k_2)
\end{eqnarray}
where
\begin{eqnarray}
{\cal J}_1^{(c)}(k,\omega)\propto\theta_+(\omega)\delta(\omega-v_c k)
\left(\frac{\omega}{v_c}\right)^{-1/2} \\
{\cal J}_1^{(s)}(k,\omega)\propto\theta_+(\omega)\delta(\omega-v_s k)
\left(\frac{\omega}{v_s}\right)^{-1/2} \\
{\cal J}_2^{(c)}(k,\omega)\propto\theta_-(\omega)\delta(\omega-v_c k)
\left(\frac{-\omega}{v_c}\right)^{-1/2} \\
{\cal J}_2^{(s)}(k,\omega)\propto\theta_-(\omega)\delta(\omega-v_s k)
\left(\frac{-\omega}{v_s}\right)^{-1/2} 
\end{eqnarray}

It is simple to show that if $\Delta\mu=0$ then $A_{12}(\omega)=A_{21}(\omega)=0$,
so we begin with the general case of $\Delta\mu >0$:
\begin{eqnarray}
A_{12}(\omega)&=&\int\frac{d\omega'}{2\pi}\int\frac{dk}{2\pi} 
{\cal J}_1^{(1)}(k,\omega'){\cal J}_2^{(2)}(k,\omega'-\omega) \nonumber \\
&\propto&\int\frac{d\omega'}{2\pi}\int\frac{dk}{2\pi}  \nonumber \\
&&\int_0^{\infty}d\omega_1 d\omega_2 \delta(\omega'-\mu_1-\omega_1-\omega_2)
\delta(k-k_F^{(1)}-\frac{\omega_1}{v_c}-\frac{\omega_2}{v_s})
(\omega_1/v_c)^{-1/2}
(\omega_2/v_s)^{-1/2} \nonumber\\
&&\int_0^{\infty} d\omega_1' d\omega_2'
\delta(\omega'-\omega-\mu_2+\omega_1'+\omega_2')
\delta(k-k_F^{(2)}+\frac{\omega_1'}{v_c}+\frac{\omega_2'}{v_s})
(\omega_1'/v_c)^{-1/2}
(\omega_2'/v_s)^{-1/2} \nonumber \\
&\propto&\int_0^{\infty}d\omega_1 d\omega_2 d\omega_1' d\omega_2'
\delta(\omega_1+\omega_2+\omega_1'+\omega_2'-(\omega +\Delta\mu))\nonumber\\
&&\delta\left(\frac{(\omega_1+\omega_1')}{v_c}+\frac{(\omega_2+\omega_2')}{v_s}
-\Delta k\right)(\omega_1/v_c)^{-1/2}(\omega_2/v_s)^{-1/2}
(\omega_1'/v_c)^{-1/2}(\omega_2'/v_s)^{-1/2} \nonumber \\
&\propto&\int_0^{\infty} d\omega_2 d\omega_1' d\omega_2'
\delta(-\Delta-(\omega_2+\omega_2')\Delta v/v_s -\omega)
(\omega_2/v_s)^{-1/2}
(\omega_1'/v_c)^{-1/2}(\omega_2'/v_s)^{-1/2}\nonumber\\
&&\left(\Delta k-\frac{(\omega_2+\omega_2')}{v_s}-\frac{\omega_1'}{v_c}\right)^{-1/2}
\theta_+\left(\Delta k-\frac{(\omega_2+\omega_2')}{v_s}-\frac{\omega_1'}{v_c}\right)
\nonumber \\
&\propto&\frac{v_s}{\Delta v}\left\{
\int_0^{\infty} d\omega_1'
(\omega_1'/v_c)^{-1/2}
\left(\Delta k+\frac{(\omega+\Delta)}{\Delta v}-\frac{\omega_1'}{v_c}\right)^{-1/2}
\theta_+
\left(\Delta k+\frac{(\omega+\Delta)}{\Delta v}-\frac{\omega_1'}{v_c}\right)\right\}
\nonumber\\
&&\left\{\int_0^{\infty} d\omega_2'
(\omega_2'/v_s)^{-1/2}
\left(-\frac{(\omega+\Delta)}{\Delta v}-\frac{\omega_2'}{v_s}\right)^{-1/2}
\theta_+
\left(-\frac{(\omega+\Delta)}{\Delta v}-\frac{\omega_2'}{v_s}\right)\right\}
\end{eqnarray}
Using $\int_0^{x}dt\;t^{-1/2}(x-t)^{-1/2}=\int_0^{1}dt\;t^{-1/2}(1-t)^{-1/2}$
(i.e. independent of $x$), we obtain
\begin{eqnarray}
A_{12}(\omega)&\propto&\frac{1}{\Delta v}\theta_+(v_c\Delta k-\Delta\mu-\omega)
\theta_+(\omega+\Delta\mu-v_s\Delta k) \\
A_{21}(\omega)&=& 0
\end{eqnarray}
that is, $A_{12}(\omega)$ has constant non-vanishing weight in the interval
$\omega\in[v_s\Delta k-\Delta\mu,v_c\Delta k-\Delta\mu]$. Note that, in the
limit $\Delta v\rightarrow 0$, this step function of width $\Delta v$, height
$\propto 1/\Delta v$, goes over to a $\delta$-function, as it should, for
the limit $\Delta v\rightarrow 0$ is the limit of free electrons.

\underline{Spinny Luttinger liquid}

This case has aspects of both the spinless Luttinger liquid, and the
chiral Luttinger liquid. We shall see that the high frequency behavior
is essentially the same as that of the spinless case, while the effect of
spin-charge separation is to destroy the low frequency divergence in $A_{12}(\omega)$
which was present in the spinless case.

For the electron spectral function we have the convolution
\begin{eqnarray}
{\cal J}_1&(k,\omega)&\propto\int dk_1 dk_2 dk_3
\int d\omega_1 d\omega_2 d\omega_3
\delta(\omega-\mu-\sum_i\omega_i)\delta(k-k_F-\sum_i k_i)\nonumber\\
&&\delta(\omega_1-v_ck_1)\theta_+(\omega_1)(\omega_1/v_c)^{\alpha-1/2}
\delta(\omega_2-v_sk_2)\theta_+(\omega_2)(\omega_2/v_s)^{-1/2}
\delta(\omega_3+v_ck_3)\theta_+(\omega_3)(\omega_3/v_c)^{\alpha-1}
\nonumber\\
&\propto&
\int_0^{\infty} d\omega_1 d\omega_2 d\omega_3
\delta(\omega-\mu-\sum_i\omega_i)
\delta\left(k-k_F-\frac{(\omega_1-\omega_3)}{v_c}-\frac{\omega_2}{v_s}\right)
\nonumber\\
&&(\omega_1/v_c)^{\alpha-1/2}
(\omega_2/v_s)^{-1/2}
(\omega_3/v_c)^{\alpha-1}
\end{eqnarray}
Similarly
\begin{eqnarray}
{\cal J}_2(k,\omega)&\propto&
\int_0^{\infty} d\omega_1' d\omega_2' d\omega_3'
\delta(\omega-\mu+\sum_i\omega_i')
\delta\left(k-k_F+\frac{(\omega_1'-\omega_3')}{v_c}+\frac{\omega_2'}{v_s}\right)
\nonumber\\
&&(\omega_1'/v_c)^{\alpha-1/2}
(\omega_2'/v_s)^{-1/2}
(\omega_3'/v_c)^{\alpha-1}
\end{eqnarray}

For $\Delta\mu >0$ we calculate $A_{12}(\omega)$ as follows:
\begin{eqnarray}
A_{12}(\omega)&\propto&\int\frac{d\omega'}{2\pi}\int\frac{dk}{2\pi}
\int_0^{\infty} d\omega_1 d\omega_2 d\omega_3
\delta(\omega'-\sum_i\omega_i)
\delta\left(k-k_F-\frac{(\omega_1-\omega_3)}{v_c}-\frac{\omega_2}{v_s}\right)
\omega_1^{\alpha-1/2}
\omega_2^{-1/2}
\omega_3^{\alpha-1}
\nonumber\\
&&\int_0^{\infty}d\omega_1' d\omega_2' d\omega_3'
\delta(\omega-\omega'-\Delta\mu+\sum_i\omega_i')
\delta\left(k-k_F+\frac{(\omega_1'-\omega_3')}{v_c}+\frac{\omega_2'}{v_s}\right)
(\omega_1')^{\alpha-1/2}
(\omega_2')^{-1/2}
(\omega_3')^{\alpha-1} \nonumber\\
&\propto&\int_{-\Delta\mu}^{\infty}d\omega_1
\int_0^{\infty}d\omega_2 d\omega_3 d\omega_1' d\omega_2' d\omega_3'
\nonumber\\
&&
\delta(\sum_i(\omega_i+\omega_i')-\omega)
\delta((\omega_1+\omega_1')-(\omega_3+\omega_3')+(v_c/v_s)(\omega_2+\omega_2')
+\Delta)\nonumber\\
&&(\omega_1+\Delta\mu)^{\alpha-1/2}\omega_2^{-1/2}\omega_3^{\alpha-1}
(\omega_1')^{\alpha-1/2}(\omega_2')^{-1/2}(\omega_3')^{\alpha-1} \nonumber\\
&\propto&\int_{-\Delta\mu}^{\infty}d\omega_1
\int_0^{\infty}d\omega_2 d\omega_3 d\omega_2' d\omega_3'
\delta\left(2(\omega_3+\omega_3')+\left(1-\frac{v_c}{v_s}\right)(\omega_2+\omega_2')
-(\omega+\Delta)\right)\nonumber\\
&&(\omega_1+\Delta\mu)^{\alpha-1/2}\omega_2^{-1/2}\omega_3^{\alpha-1}
(\omega_2')^{-1/2}(\omega_3')^{\alpha-1} \nonumber\\
&&\left((\omega_3+\omega_3')-\frac{v_c}{v_s}(\omega_2+\omega_2')
-(\omega_1+\Delta)\right)^{\alpha-1/2}
\theta_+\left((\omega_3+\omega_3')-\frac{v_c}{v_s}(\omega_2+\omega_2')
-(\omega_1+\Delta)\right)\nonumber\\
&=&\int_{0}^{\infty}d\omega_1
d\omega_2 d\omega_2' d\omega_3'
\omega_1^{\alpha-1/2}\omega_2^{-1/2}
(\omega_2')^{-1/2}(\omega_3')^{\alpha-1}\nonumber\\
&&\left(\frac{1}{2}\left[(\omega+\Delta)-\left(1-\frac{v_c}{v_s}\right)
(\omega_2+\omega_2')\right]-\omega_3'\right)^{\alpha-1}
\theta_+\left(\frac{1}{2}\left[(\omega_1+\Delta)-\left(1-\frac{v_c}{v_s}\right)
(\omega_2+\omega_2')\right]-\omega_3'\right)\nonumber\\
&&\left(\frac{1}{2}\left[(\omega-\Delta)-\left(1+\frac{v_c}{v_s}\right)
(\omega_2+\omega_2')+2\Delta\mu\right]-\omega_1\right)^{\alpha-1/2}
\nonumber \\
&&
\theta_+\left(\frac{1}{2}\left[(\omega-\Delta)-\left(1+\frac{v_c}{v_s}\right)
(\omega_2+\omega_2')+2\Delta\mu\right]-\omega_1\right)\nonumber\\
&\propto&\int_{0}^{\infty} d\omega_2 d\omega_2' 
\omega_2^{-1/2}(\omega_2')^{-1/2}
\left((\omega+\Delta)-\left(1-\frac{v_c}{v_s}\right)
(\omega_2+\omega_2')\right)^{2\alpha-1}
\nonumber \\
&&
\theta_+\left((\omega+\Delta)-\left(1-\frac{v_c}{v_s}\right)
(\omega_2+\omega_2')\right)\nonumber\\
&&\left((\omega-\Delta)-\left(1+\frac{v_c}{v_s}\right)
(\omega_2+\omega_2')+2\Delta\mu\right)^{2\alpha}
\theta_+\left((\omega-\Delta)-\left(1+\frac{v_c}{v_s}\right)
(\omega_2+\omega_2')+2\Delta\mu\right)\nonumber\\
&=&\int_{0}^{\infty} d\omega_2 d\omega_2' 
\omega_2^{-1/2}(\omega_2')^{-1/2}
\left((\omega+\Delta)+\frac{\Delta v}{v_s}
(\omega_2+\omega_2')\right)^{2\alpha-1}
\theta_+\left((\omega+\Delta)+\frac{\Delta v}{v_s}
(\omega_2+\omega_2')\right)\nonumber\\
&&\left((\omega-\Delta+2\Delta\mu)-\frac{2\bar{v}}{v_s}
(\omega_2+\omega_2')\right)^{2\alpha}
\theta_+\left((\omega-\Delta+2\Delta\mu)-\frac{2\bar{v}}{v_s}
(\omega_2+\omega_2')\right)
\end{eqnarray}

To proceed further we need to calculate the generic integral
\[
I(a,b)=\int_0^{\infty}dx\int_0^{\infty}dy x^{-1/2}y^{-1/2}
(a-(x+y))^{2\alpha}(b+(x+y))^{2\alpha-1}\theta_+(a-(x+y))\theta_+(b+(x+y))
\]
This double integral can be reduced to a single integral via change
of variables
\begin{eqnarray}
\eta=\frac{1}{\sqrt{2}}(x+y)\nonumber\\
\xi=\frac{1}{\sqrt{2}}(x-y)\nonumber
\end{eqnarray}
giving
\begin{eqnarray}
I(a,b)&=&\sqrt{2}\int_0^{\infty}d\eta\int_{-\eta}^{\eta}d\xi
\frac{(a-\sqrt{2}\eta)^{2\alpha}(b+\sqrt{2}\eta)^{2\alpha-1}}{(\eta^2-\xi^2)}^{1/2}
\theta_+(a-\sqrt{2}\eta)\theta_+(b+\sqrt{2}\eta)\nonumber\\
&=&\sqrt{2}\int_0^{\infty}d\eta
(a-\sqrt{2}\eta)^{2\alpha}(b+\sqrt{2}\eta)^{2\alpha-1}
\theta_+(a-\sqrt{2}\eta)\theta_+(b+\sqrt{2}\eta)
\left\{\int_{-1}^1\frac{d(\xi/\eta)}{1-(\xi/\eta)^2)^{1/2}}\right\}\nonumber\\
&=&\pi\int_0^{\infty}d\eta
(a-\eta)^{2\alpha}(b+\eta)^{2\alpha-1}
\theta_+(a-\eta)\theta_+(b+\eta)
\end{eqnarray}
We may then write
\[
A_{12}(\omega)\propto\left(\frac{\bar{v}}{v_s}\right)^{2\alpha}
\left(\frac{\Delta v}{v_s}\right)^{2\alpha-1}\;I(a,b)
\]
with
\begin{eqnarray}
a&=&\frac{v_s}{\bar{v}}(\omega+\Delta\mu+v_c\Delta k)\nonumber\\
b&=&\frac{v_s}{\Delta v}(\omega+\Delta\mu-v_c\Delta k)\nonumber
\end{eqnarray}

We can reduce $I(a,b)$ further. There are two cases (from
here on we write $v\Delta k$ for $\Delta\mu$):

(i) $b>0$ ({\em i.e.\/} $\omega > (v_c-v)\Delta k$)

\begin{eqnarray}
I(a,b)&=&\pi\int_{0}^{a}dx\; (a-x)^{2\alpha}(b+x)^{2\alpha-1} \nonumber\\
&=&\frac{\pi}{(1+2\alpha)}\; a^{2\alpha+1} b^{2\alpha-1}
~_2F_1\left(1,1-2\alpha;2+2\alpha;-\frac{a}{b}\right)
\end{eqnarray}

(ii) $b<0, a+b>0$ ({\em i.e.\/} $(v_s-v)\Delta k < \omega < (v_c-v)\Delta k$)

\begin{eqnarray}
I(a,b)&=&\pi\int_{-b}^{a}dx\; (a-x)^{2\alpha}(b+x)^{2\alpha-1} \nonumber\\
&=&\pi\int_{0}^{a+b}dy\; (a+b-y)^{2\alpha}y^{2\alpha-1} \nonumber \\
&=& (a+b)^{4\alpha}\int_{0}^{1}dt\; t^{2\alpha-1}(1-t)^{2\alpha} \nonumber\\
&=& (a+b)^{4\alpha}\frac{\Gamma(2\alpha)\Gamma(1+2\alpha)}{\Gamma(1+4\alpha)}
\end{eqnarray}

Upon substituting for $a$ and $b$, and reinserting
the appropriate prefactors, one obtains the results for the
interliquid hopping spectral function 
presented in the main body of the paper.

%% file: long.bbl
\begin{references}
\bibitem{fncrossover} Note that in a Fermi
liquid, the peak contains weight $Z$, where $Z$ is $O(1)$,
and the crossover scale at which $t_{\perp}$ asserts
its relevance is $Z\,t_{\perp}^{bare}$.  For non-Fermi
liquid models, the weight in the peak can be made to vanish
and the crossover scale cannot be determined unambiguosly
from experiment since the photoemission data only
determine $\rho(\vec{k},\omega)$ up to an unknown constant.
\bibitem {fn_P(t)} To avoid possible confusion, note that `$P(t)$' is 
actually used in \cite{TLS} 
to refer to $\langle\sigma_z(t)\rangle$.
\bibitem{divnote} This division is actually highly conservative
since the oscillation frequency, if any, can be much smaller
than $\Delta^R$, but never significantly larger.
\bibitem{fkkworry} The quasiparticle feature of the
spectral function will approach a delta functions only as
$k \rightarrow k_F$, however since we are perturbing in
$t_{\perp}$, which is arbitrarily small, the width
is unimportant.
\bibitem{fermipoint}  The operators are taken to be near to
the right Fermi point here and non-universal,
cutoff dependent features of the spectral functions
are neglected.  For simplicity, we hereafter
work only with the right Fermi
point and universal aspects of the spectral 
functions unless otherwise stated.
\bibitem{vJfootnote} Here we considered a
$\delta N$ at only the right Fermi point,
if we add equal numbers at both Fermi points then
$E_F^1 - E_F^2 = (k_F^1 -k_F^2) v_N$, which is again in
general different from
$v_S (k_F^1 - k_F^2) = \sqrt{v_J v_N} (k_F^1 - k_F^2)$
but may be smaller if the interaction is attractive.
\bibitem{energyfootnote} If we instead choose
the definition of high energy to
be energies above those making a significant
contribution to $\langle \delta N(t) \rangle$,
the result is $O(1)$
%We find approximately
%$\frac{4-7 \alpha}{3(2- 4 \alpha)}$.
for all $\alpha$  so that it appears reasonable
that the incoherent hops could completely wipe out
the coherence for any $\alpha$ for one extra particle.
\bibitem{wenfootnote}
We emphasize that this is an interpretation of the results obtained.
In general,  at least half of the poles obtained in the Green's function
occurs at
complex frequency and the resulting Green's function is incompatible
with a Fermi liquid groundstate. The calculation requires 
interpretation, and while that of the author, that the splitting of
the poles and the renormalization relevance of $t_{\perp}$
indicate flow to a higher dimensionally coherent state, is reasonable,
other possibilities are not excluded.
\bibitem{fn:P(t)} In general, one would expect that in the thermodynamic
limit $P(t)$ would rapidly vanish; if one wished to use $P(t)$ to
study the coherence issue to all orders in $t_{\perp}$, 
and in particular if using numerical methods,
one would need to consider something like $[P(t)]^{\frac{a}{L}}$.
\bibitem{controlfootnote}
In high dimensions, Fermi liquid theory is stable with 
respect to weak interactions, however we are dealing here
with strong interactions and strong anisotropy and nothing
controlled can be said.  Given this, there are a number of
advantages to our dynamical calculation. Specifically,
we have not had to assume anything about the nature of
the Fermi surface or of the true ground state to pose
our dynamical calculation, our approach is directly 
connected to the TLS and is therefore interpretable in
terms of coherence, and finally, we are calculating a
well defined physical response.
\bibitem{carefulnote}  While we have not actually proven it, 
we believe it is extremely reasonable to expect that, if the
states connected to the $t_{\perp} = 0$ groundstates (at fixed
$N$ and $J$) have
enegies effectively non-degenerate with those groundstates,
then the states connected by  to the true
$t_{\perp} \neq 0$ groundstatte by $t_{\perp}$ 
should also have this non-degeneracy.  This is essentially
the assumption made in the TLS problem in  \cite{TLS}.
\bibitem{splitfootnote}  Here we do not consider the possibility
that operators generated at higher order by the renormalization
of $t_{\perp}$ may be relevant and themselves destroy the
Fermi surface.
\bibitem{fnaffine} The term ``linear-T'' is highly misleading, suggesting a
zero temperature intercept of $\rho_c$ close to zero; in fact, if the normal
state data are extrapolated to zero temperature, the intercept is not at all
small, but typically of order $1 m\Omega\,cm$.
\bibitem{general} The
notion of intrinsic incoherent transport resulting
from interactions is clearly more general than the
model we have considered, and our calculation is really meant
to establish that such incoherence is possible.
More generally, we can expect it to occur
for sufficiently strong
interactions if the system, for $t_{\perp} = 0$,
(1) is gapless, 
(2) lacks  the quasiparticle structure  of a Fermi liquid
and (3) lacks a sharp Fermi surface (not necessarily
implied by 2).
\end{references}
